\documentclass[a4paper,fleqn,usenatbib]{mnras}

\usepackage[T1]{fontenc}
\usepackage{ae,aecompl}

\usepackage{amssymb}	
\usepackage{amsmath}
\usepackage{graphicx}
\usepackage[usenames]{color}
\usepackage{bm} 

\usepackage{xspace}

\usepackage{multirow}

\usepackage{lineno}

\newcommand{\MpcOh}{ \,  \mathrm{Mpc}  \, h^{-1} }

\newcommand{\nn}{ \nonumber }

\newcommand{\beq}{\begin{equation}}
\newcommand{\eeq}{\end{equation}}
\newcommand{\beqa}{\begin{eqnarray}}
\newcommand{\eeqa}{\end{eqnarray}}

\newcommand{\change}[1]{{\textcolor{black}{#1}}\xspace}

\newif\ifcomments
\commentstrue   

\usepackage{eso-pic}

\AddToShipoutPictureBG*{%
  \AtPageUpperLeft{%
    \hspace{0.75\paperwidth}%
    \raisebox{-3.5\baselineskip}{%
      \makebox[0pt][l]{\textnormal{DES 2017-0306}}
}}}%

\AddToShipoutPictureBG*{%
  \AtPageUpperLeft{%
    \hspace{0.75\paperwidth}%
    \raisebox{-4.5\baselineskip}{%
      \makebox[0pt][l]{\textnormal{Fermilab-PUB-17-590}}
}}}%

\title[BAO from angular correlation function]{BAO from angular clustering: optimization and mitigation of theoretical systematics}

\author[K.~C.~Chan et al]{
\parbox{\textwidth}{
\Large
K.~C.~Chan$^{1,2,3} $\thanks{E-mail: chankc@mail.sysu.edu.cn (KCC)},
M.~Crocce$^{2,3}$,
A.~J.~Ross$^{4}$,
S.~Avila$^{5,6}$,
J.~Elvin-Poole$^{7}$,
M.~Manera$^{8}$,
W.~J.~Percival$^{5}$,
R.~Rosenfeld$^{9,10}$,
T.~M.~C.~Abbott$^{11}$,
F.~B.~Abdalla$^{8,12}$,
S.~Allam$^{13}$,
E.~Bertin$^{14,15}$,
D.~Brooks$^{8}$,
D.~L.~Burke$^{16,17}$,
A.~Carnero~Rosell$^{10,18}$,
M.~Carrasco~Kind$^{19,20}$,
J.~Carretero$^{21}$,
F.~J.~Castander$^{2,3}$,
C.~E.~Cunha$^{16}$,
C.~B.~D'Andrea$^{22}$,
L.~N.~da Costa$^{10,18}$,
C.~Davis$^{16}$,
J.~De~Vicente$^{23}$,
T.~F.~Eifler$^{24,25}$,
J.~Estrada$^{13}$,
B.~Flaugher$^{13}$,
P.~Fosalba$^{2,3}$,
J.~Frieman$^{13,26}$,
J.~Garc\'ia-Bellido$^{6}$,
E.~Gaztanaga$^{2,3}$,
D.~W.~Gerdes$^{27,28}$,
D.~Gruen$^{16,17}$,
R.~A.~Gruendl$^{19,20}$,
J.~Gschwend$^{10,18}$,
G.~Gutierrez$^{13}$,
W.~G.~Hartley$^{8,29}$,
K.~Honscheid$^{4,30}$,
B.~Hoyle$^{31,32}$,
D.~J.~James$^{33}$,
E.~Krause$^{24,25}$,
K.~Kuehn$^{34}$,
O.~Lahav$^{8}$,
M.~Lima$^{35,10}$,
M.~March$^{22}$,
F.~Menanteau$^{19,20}$,
C.~J.~Miller$^{27,28}$,
R.~Miquel$^{36,21}$,
A.~A.~Plazas$^{25}$,
K.~Reil$^{17}$,
A.~Roodman$^{16,17}$,
E.~Sanchez$^{23}$,
V.~Scarpine$^{13}$,
I.~Sevilla-Noarbe$^{23}$,
M.~Smith$^{37}$,
M.~Soares-Santos$^{13}$,
F.~Sobreira$^{38,10}$,
E.~Suchyta$^{39}$,
M.~E.~C.~Swanson$^{20}$,
G.~Tarle$^{28}$,
D.~Thomas$^{5}$,
A.~R.~Walker$^{11}$
\begin{center} (DES Collaboration) \end{center}
}
\vspace{0.4cm}
\\
\parbox{\textwidth}{
$^{1}$ School of Physics and Astronomy, Sun Yat-Sen University, Guangzhou 510275, China\\
$^{2}$ Institut d'Estudis Espacials de Catalunya (IEEC), 08193 Barcelona, Spain\\
$^{3}$ Institute of Space Sciences (ICE, CSIC),  Campus UAB, Carrer de Can Magrans, s/n,  08193 Barcelona, Spain\\
$^{4}$ Center for Cosmology and Astro-Particle Physics, The Ohio State University, Columbus, OH 43210, USA\\
$^{5}$ Institute of Cosmology \& Gravitation, University of Portsmouth, Portsmouth, PO1 3FX, UK\\
$^{6}$ Instituto de Fisica Teorica UAM/CSIC, Universidad Autonoma de Madrid, 28049 Madrid, Spain\\
$^{7}$ Jodrell Bank Center for Astrophysics, School of Physics and Astronomy, University of Manchester, Oxford Road, Manchester, M13 9PL, UK\\
$^{8}$ Department of Physics \& Astronomy, University College London, Gower Street, London, WC1E 6BT, UK\\
$^{9}$ ICTP South American Institute for Fundamental Research\\ Instituto de F\'{\i}sica Te\'orica, Universidade Estadual Paulista, S\~ao Paulo, Brazil\\
$^{10}$ Laborat\'orio Interinstitucional de e-Astronomia - LIneA, Rua Gal. Jos\'e Cristino 77, Rio de Janeiro, RJ - 20921-400, Brazil\\
$^{11}$ Cerro Tololo Inter-American Observatory, National Optical Astronomy Observatory, Casilla 603, La Serena, Chile\\
$^{12}$ Department of Physics and Electronics, Rhodes University, PO Box 94, Grahamstown, 6140, South Africa\\
$^{13}$ Fermi National Accelerator Laboratory, P. O. Box 500, Batavia, IL 60510, USA\\
$^{14}$ CNRS, UMR 7095, Institut d'Astrophysique de Paris, F-75014, Paris, France\\
$^{15}$ Sorbonne Universit\'es, UPMC Univ Paris 06, UMR 7095, Institut d'Astrophysique de Paris, F-75014, Paris, France\\
$^{16}$ Kavli Institute for Particle Astrophysics \& Cosmology, P. O. Box 2450, Stanford University, Stanford, CA 94305, USA\\
$^{17}$ SLAC National Accelerator Laboratory, Menlo Park, CA 94025, USA\\
$^{18}$ Observat\'orio Nacional, Rua Gal. Jos\'e Cristino 77, Rio de Janeiro, RJ - 20921-400, Brazil\\
$^{19}$ Department of Astronomy, University of Illinois at Urbana-Champaign, 1002 W. Green Street, Urbana, IL 61801, USA\\
$^{20}$ National Center for Supercomputing Applications, 1205 West Clark St., Urbana, IL 61801, USA\\
$^{21}$ Institut de F\'{\i}sica d'Altes Energies (IFAE), The Barcelona Institute of Science and Technology, Campus UAB, 08193 Bellaterra (Barcelona) Spain\\
$^{22}$ Department of Physics and Astronomy, University of Pennsylvania, Philadelphia, PA 19104, USA\\
$^{23}$ Centro de Investigaciones Energ\'eticas, Medioambientales y Tecnol\'ogicas (CIEMAT), Madrid, Spain\\
$^{24}$ Department of Astronomy/Steward Observatory, 933 North Cherry Avenue, Tucson, AZ 85721-0065, USA\\
$^{25}$ Jet Propulsion Laboratory, California Institute of Technology, 4800 Oak Grove Dr., Pasadena, CA 91109, USA\\
$^{26}$ Kavli Institute for Cosmological Physics, University of Chicago, Chicago, IL 60637, USA\\
$^{27}$ Department of Astronomy, University of Michigan, Ann Arbor, MI 48109, USA\\
$^{28}$ Department of Physics, University of Michigan, Ann Arbor, MI 48109, USA\\
$^{29}$ Department of Physics, ETH Zurich, Wolfgang-Pauli-Strasse 16, CH-8093 Zurich, Switzerland\\
$^{30}$ Department of Physics, The Ohio State University, Columbus, OH 43210, USA\\
$^{31}$ Max Planck Institute for Extraterrestrial Physics, Giessenbachstrasse, 85748 Garching, Germany\\
$^{32}$ Universit\"ats-Sternwarte, Fakult\"at f\"ur Physik, Ludwig-Maximilians Universit\"at M\"unchen, Scheinerstr. 1, 81679 M\"unchen, Germany\\
$^{33}$ Harvard-Smithsonian Center for Astrophysics, Cambridge, MA 02138, USA\\
$^{34}$ Australian Astronomical Observatory, North Ryde, NSW 2113, Australia\\
$^{35}$ Departamento de F\'isica Matem\'atica, Instituto de F\'isica, Universidade de S\~ao Paulo, CP 66318, S\~ao Paulo, SP, 05314-970, Brazil\\
$^{36}$ Instituci\'o Catalana de Recerca i Estudis Avan\c{c}ats, E-08010 Barcelona, Spain\\
$^{37}$ School of Physics and Astronomy, University of Southampton,  Southampton, SO17 1BJ, UK\\
$^{38}$ Instituto de F\'isica Gleb Wataghin, Universidade Estadual de Campinas, 13083-859, Campinas, SP, Brazil\\
$^{39}$ Computer Science and Mathematics Division, Oak Ridge National Laboratory, Oak Ridge, TN 37831\\
}
}

\date{Accepted XXX. Received YYY; in original form ZZZ}

\pubyear{2018}

\begin{document}
\label{firstpage}
\pagerange{\pageref{firstpage}--\pageref{lastpage}}
\maketitle

\begin{abstract}
  
 \change{We study the methodology and potential theoretical
   systematics of measuring Baryon Acoustic Oscillations (BAO) using
   the angular correlation functions in tomographic bins.  We calibrate
   and optimize the pipeline for the Dark Energy Survey Year 1 dataset
   using 1800 mocks.   We compare the BAO fitting results obtained
   with three estimators: the Maximum Likelihood Estimator (MLE),
   Profile Likelihood, and Markov Chain Monte Carlo. The fit results
   from the MLE are the least biased and their derived 1-$\sigma$ error
   bar are closest to the Gaussian distribution value  after removing
   the extreme mocks with non-detected BAO signal.   
   We show that incorrect assumptions in constructing the template,
   such as mismatches from the cosmology of the mocks or the
   underlying photo-$z$ errors, can lead to BAO angular shifts. We
   find that MLE is
   the method that best traces this systematic biases, allowing to
   recover the true angular distance values. 
In a real survey analysis, it may happen that the final data sample properties are slightly different from those of the mock catalog.  We show that the effect on the mock covariance due to the sample differences can be corrected with the help of the Gaussian covariance matrix or more effectively using the eigenmode expansion of the mock covariance. In the eigenmode expansion, the eigenmodes are provided by some proxy covariance matrix.  The eigenmode expansion is significantly less susceptible to statistical fluctuations relative to the direct measurements of the covariance matrix because of the number of free parameters is substantially reduced. }

\end{abstract}

\begin{keywords}
  cosmology: observations - (cosmology:) large-scale structure of Universe
\end{keywords}



\section{Introduction}

\change{ Baryon Acoustic Oscillations (BAO), generated in the early universe, leave their imprint in the distribution of galaxies \citep{PeeblesYu1970,SunyaevZeldovich1970}. }  At early times ($z\gtrsim 1100$),  photons and baryons form a tightly coupled plasma, and sound waves propagate in this plasma. After the recombination of hydrogen, photons can free stream in the universe. The acoustic wave pattern remains frozen in the baryon distribution. The sound horizon at the drag epoch is close to 150 Mpc. \change{ The formation of the BAO in the early universe is governed by the well-understood linear physics, see \citet{BondEfstathiou1984,BondEfstathiou1987,HuSugiyama1996,HuSugiyamaSilk1997,Dodelson_2003} for the details of the cosmic microwave background physics.}  Given that this scale is relatively large, it is less susceptible to astrophysical contamination and other nonlinear effects. Another reason for its robustness is that it exhibits  sharp features in 2-point correlations, while nonlinearity tends to produce changes in the broad band power.\change{ The BAO can serve as a standard ruler \citep{EisensteinHu1998,MeiksinWhitePeacock1999,EisensteinSeoWhite2007}.}  Observations of the BAO feature in the distribution of galaxies has been recognized as one of the most important cosmological probes that enables us to measure the Hubble parameter and the equation of state of the dark energy [see e.g.~\cite{WeinbergMortonson_etal2013,Aubourg:2014yra}]. The potential distortions of BAO due to nonlinear evolution and galaxy bias are less than 0.5\% \citep{CrocceScoccimarro_2008,PadmanabhanWhite_2009}.

Using spectroscopic data, the BAO was first clearly detected in SDSS \citep{Eisenstein_etal2005} and 2dFGS \citep{Cole_etal2005}, and subsequently in numerous studies, e.g.~\cite{Gaztanaga:2008xz,Percival_etal2010,Beutler_etal2011, Kazin_etal2014, Ross_etal2015,Alam_etal2017,Bautista_etal2017,Ata_etal2017}.  Spectroscopic data give precise redshift information, but such surveys are relatively expensive, as they require spectroscopic observations of galaxies targeted from existing imaging surveys. \change{ On the other hand, multi-band imaging surveys relying on the use of photometric redshift (photo-$z$)  for radial information  \citep{Koo1985}  can cheaply survey large volumes.}  There are several (generally weaker) detections of the BAO feature in the galaxy distribution using photometric data \citep{Padmanabhan_etal2007,EstradaSefusattiFrieman2009, Hutsi2010, Seo_etal2012,Carnero_etal2012,deSimoni_etal2013}. The Dark Energy Survey (DES) and future surveys such as LSST \citep{LSSTScienceBook} will deliver an enormous amount of photometric data with well calibrated photo-$z$'s, thus we expect that accurate BAO measurements will be achieved from these surveys.

In this work we investigate the BAO detection using the angular correlation function of a galaxy sample optimally selected from the first year of DES data (DES Y1) \citep{LSSsample}. DES is one of the largest ongoing galaxy surveys, and its goal is to reveal the nature of the dark energy. One of the routes to achieve this goal is to accurately measure the BAO scale in the distribution of galaxies as a function of redshift. As it is a photometric survey, its redshift information is not so precise, but it can cover a large volume.  This is advantageous to the BAO measurement:  the sound horizon scale is large, it requires large survey volume to get good statistics. The BAO sample \citep{LSSsample} derived from the first year DES data \citep{Drlica-Wagner:2017tkk} already consists of about 1.3 million galaxies covering more than 1318 deg$^2$. This is only 1\% of the total number of galaxies identified in DES Y1, which can be used for science analyses such as \cite{Abbott:2017wau}.

A measurement of the BAO at the effective redshift (or mean redshift) of the survey, $z_{\rm eff} = 0.8 $  was presented in \cite{BAOmain} using this sample. This effective redshift is less explored by other existing surveys.  In \cite{BAOmain}, three statistics: the angular correlation function $w$, the angular power spectrum $C_{\ell}$, and the 3D correlation function $\xi$ were used. See \cite{Clpaper} and \citet{Ross:2017emc} for the details on the  $C_{\ell}$ and  $\xi$ analysis. These statistics are sensitive to different systematics and they provide important cross checks for the analyses.  By measuring angular correlation $w$ from the data divided into a number of redshift bins (or tomography), no precise redshift information is required, thus this statistics is  well-suited for extracting BAO information from the photometric sample. In the current paper, we present the details of the calibrations and optimization applied when using $w$ to measure the BAO.  Although the fiducial setup is tailored to DES Y1, the analysis and apparatus developed here will be useful for upcoming DES data and other large scale imaging surveys.  

\change{This paper is organized as follows. In Sec.~\ref{sec:theory}, we introduce the theory for the BAO modelling  and for the Gaussian covariance matrix. We also describe the mock catalogs used to test the pipeline. 
In Sec.~\ref{sec:fitting_methods} we discuss the extraction of the angular diameter distance using BAO template fitting methods, and we present three different estimators for such procedure: Maximum Likelihood Estimator (MLE), Profile Likelihood (PL), and Markov Chain Monte Carlo (MCMC). Potential systematics errors in the angular diameter distance scale due to the assumed template used for BAO extraction are studied in  Sec.~\ref{sec:template_systematics}, such as the BAO damping scale, the assumed cosmological model or the propagation of photo-$z$ errors. In Sec.~\ref{sec:optimizing_analysis} we discuss various optimizations for the analysis. Sec.~\ref{sec:Covariance} is devoted to issues related to the covariance; in particular we present an eigenmode expansion that allows to adapt the covariance to changes in the underlying template or sample assumptions. We present our conclusions in Sec.~\ref{sec:Conclusion}.  In Appendix~\ref{sec:appendix}, we further compare the results obtained with the three estimators.}

\section{ Theory and Mock Catalogs}
\label{sec:theory}
\change{To detect the BAO signal in the data we employ a template fitting method.  A template encodes the expected shape and amplitude of the BAO feature. It is computed using the expected properties, e.g.~the survey and galaxy sample charactereistics.  A large set of mock catalogs are constructed to mimick those detailed characteristics. The template is then fitted to the correlation functions measured on the mock catalogs to extract the BAO distance scale, and to study different systematic and statistical effects.}

In this section we discuss how the template is constructed.  We also introduce a Gaussian theory covariance that is employed at different moments in the paper.  The mock catalogs used in this study are briefly described in \ref{sec:halogen}, with full details given in \cite{Halogen}.


\subsection{The BAO template} 
The angular correlation function $w(\theta)$ measures the correlation between two points separated by an angle $\theta $.  We use linear theory for the angular correlation function, except that we include the effect where nonlinear structure growth smooths the BAO feature by including one additional ``BAO damping'' factor.  This is sufficient for the angular scales we consider, $\theta > 0.5^{\circ} $. For example, we have checked that  it can fit the mean result from the mocks well. See Sec.~\ref{sec:DampingScale} for more details. In this work the template is computed in configuration space directly as \citep{CrocceCabreGazta_2011}
\beq
w(\theta) = \int d z_1  \int dz_2 g(z_1) g(z_2)  \xi_s \big( s(z_1,z_2,\theta), \hat{\bm{s}} \cdot \bm{\hat{l}} \big), 
\eeq
with $g(z)  \equiv D(z) \phi(z) $, $\phi(z)$ is the redshift distribution of the sample, and $D$ is the linear growth factor.  The redshift space correlation function $ \xi_s $ depends on the separation vector $\bm{s} $  in redshift space  and the dot  product between the line of sight direction $\bm{\hat{l}}  $  and $ \hat{\bm{s}}$, which is given by
\beq
 \hat{\bm{s}} \cdot \bm{\hat{l}} = \frac{r(z_2) - r(z_1)   }{ s } \cos \frac{\theta}{2}, 
 \eeq
 with $r(z)$ being the comoving distance to redshift $z$.  \change{In the linear regime, the redshift-space correlation function is related to the real space power spectrum as \citep{Kaiser87,Hamilton1992,ColeFisherWeinberg1994} }
\beq
\xi_s( \bm{s}  ) = \sum_{\ell=0,2,4} i^\ell  A_\ell P_\ell ( \hat{\bm{s}} \cdot  \hat{\bm{l}} ) \int \frac{ dk k^2 }{ 2 \pi^2 } j_\ell(ks)  P_{\rm m} (k), 
\eeq
where $P_\ell$ is the Legendre polynomial of order $\ell$ and  $j_\ell$ the spherical Bessel function. In our model, the power spectrum $P_{\rm m} (k)$ including the BAO damping is parametrized as 
\beq
P_{\rm m} (k) =   ( P_{\rm lin} - P_{\rm nw} ) e^{- k^2 \Sigma^2 } + P_{\rm nw}, 
\eeq
where $P_{\rm lin}$  is the linear matter power spectrum computed by {\tt camb} \citep{CAMB}  and $ P_{\rm nw}$ is the linear one without BAO wiggles given in \cite{EisensteinHu1998}. \change{The damping scale should be anisotropic; but given the accuracy of our data we take the isotropic average $ \Sigma = 5.2 \MpcOh$. In Sec.~\ref{sec:DampingScale} we show how this value is determined.   The multipole coefficients $A_\ell $ are given by
\begin{equation}
A_\ell = \left\{
\begin{array}{cl}
 b^2 + \frac{2}{3} bf + \frac{1}{5} f^2   & \text{for } \ell =0,\\
\frac{4}{3} bf  + \frac{4}{7} f^2   & \text{for } \ell = 2,\\
\frac{8  }{35  } f^2  & \text{for } \ell=4,
\end{array} \right.
\end{equation}
where  $b$ is the linear galaxy bias and  $f \equiv d \ln D / d \ln a $ with $a$ being the scale factor. We use  $f = \Omega_{\rm m}^{0.55} $, where  $\Omega_{\rm m} $ is the density parameter of matter \citep{PeeblesGroth1975,Linder2005} evaluated at the mean redshift of the photo-$z$ distribution. The bias parameters $b$ are assumed to be constant in each redshift bin, which is sufficient given the narrow bins used here, but they can vary from bin to bin. 
They are determined by fitting, 
for each mock and redshift bin, the model $w$ to the mock correlation function measurement in the $\theta$ range $[0.5^\circ, 2.5^\circ]$. We have checked that using a smaller range of $[0.5^\circ, 1.5^\circ]$ results in less than 0.001 fractional variation of the best fit $\alpha$ (except a few extreme cases). }

A template fitting method is employed to detect the BAO feature in the angular correlation function.  Analogous to that in \cite{Seo_etal2012}, we use the following template 
\beq
\label{eq:BAO_template}
T_\alpha(\theta) = B  \, w( \alpha \theta ) +  A_0   +  \frac{A_1}{\theta} +
\frac{A_2}{\theta^2} , 
\eeq
where $w(\theta)$ is the angular correlation function computed in some cosmology (the fiducial setting is the MICE cosmology, see below). The parameter $\alpha$ gives the shift in the model BAO position relative to the fiducial one. The parameter $B$ allows for a shift in the overall amplitude. Its value is expected to be close to 1 as we have determined the physical bias parameter $b$ by fitting to the data first.  The polynomial in $1/\theta$ gives a smooth contribution and is not expected to lead to strong features in the BAO range.  We will also test the model with varying number of $A_i$:  with $A_0$ only (denoted as $N_p=1$), $A_0$ and $A_1$ ($N_p=2$), and  $A_0$, $A_1$, and $A_2$ ($N_p=3$). We fit the template over a range of angles, from $0.5^{\circ}$  to $5^{\circ}$. 

The MICE cosmology is the reference cosmology adopted in the DES Y1 BAO analysis \citep{BAOmain}, and thus is in this paper as well. \change{ It was chosen primarily because of the MICE simulation set \citep{Fosalba_etal2015}, a large high-resolution galaxy lightcone simulation tailored in part to reproduce DES observables, and accessible to us. In particular, MICE halo catalogs were used to calibrate the Halogen mock catalogs used in DES Y1 BAO analysis \citep{BAOmain} and all the supporting papers including this one}.  In summary, the cosmological parameters in MICE cosmology are $\Omega_{\rm m} = 0.25$, $\Omega_{\Lambda} = 0.75$, $\Omega_{\rm b} =0.044 $, $h=0.7  $, $\sigma_8 =0.8  $, and  $n_{\rm s} = 0.95 $. Such a low matter density is no longer compatible with the current accepted value by Planck [$\Omega_{\rm m} = 0.31$ \citep{Ade:2015xua}]. We investigate  in Sec.~\ref{sec:mismatch_cosmology} how the BAO fit is affected when there is mismatch between the template cosmology and the cosmology of the mocks.

\subsection{Theory covariance matrix}

We consider both the covariance derived from the mock catalogs and an analytic Gaussian covariance model in this work.   Here we derive an expression for the Gaussian covariance matrix for two point function obervables  between different redshift bins, $w_{ij}(\theta)$, accounting for shot-noise and angular binning.  The cross-correlation between bins $i$ and $j$ can be expressed in terms of angular power spectra $C_\ell$
through a Legendre transform as follows \citep{Peebles}
\begin{align}
\label{eq:w_from_Cl}
w_{ij}(\theta) =  \sum_{\ell} \frac{( 2 \ell + 1)}{ 4 \pi } P_\ell(\cos \theta)  C^{ij}_\ell.   
\end{align}
Note we do not include a term $1/n$ because it only contributes to the zero separation limit \citep{2011MNRAS.415.2193R}. In this work we used the public code {\tt camb sources}\footnote{http://camb.info/sources/} to compute the harmonic spectra $C^{ij}_\ell$. The covariance matrix between two point correlations at different pairs of bins is then given by
\begin{align}
 & \mathrm{Cov}[ w_{ij}(\theta),  w _{mn}(\theta')] =  \sum_{\ell_1,\ell_2} \frac{(2 \ell_1 +1)(2 \ell_2 +1) }{(4 \pi)^2}  \nonumber \\
 & \qquad \qquad   \times  P_{\ell_1}(\cos \theta)  P_{\ell_2}(\cos\theta')  \mathrm{Cov}[ C^{ij}_{\ell_1}, C^{mn}_{\ell_2} ],
\end{align}
which under the assumptions that the covariance scales inversely with the sky fraction in consideration ($f_{\rm sky}$) and that in the all-sky limit the spectra band-powers are diagonal, can be further written as \citep{CrocceCabreGazta_2011}
\begin{align}
  \label{eq:cov_mat_crossz}
&\mathrm{Cov}[ w_{ij}(\theta),  w _{mn}(\theta')] =  \sum_{\ell} \frac{(2 \ell +1)}{f_{\rm sky}(4 \pi)^2} P_\ell(\cos \theta) P_{\ell}(\cos\theta') \nonumber \\
& \times \bigg[(C^{im}_{\ell}+ \frac{\delta^{im}_{\rm K}}{\bar{n}_i})   ( C^{jn}_{\ell}+\frac{\delta^{jn}_{\rm K}}{\bar{n}_j}) + (C^{in}_{\ell}+\frac{\delta^{in}_{\rm K}}{\bar{n}_i })(C^{jm}_{\ell}+\frac{\delta^{jm}_{\rm K}}{\bar{n}_j}) \bigg],
\end{align}
where $\delta_{\rm K}^{ab}$ is the Kronecker delta  and ${\bar n}_i$ is the projected galaxy number density in bin $i$, and we have assumed Poisson shot noise.

The fact that $w$ is measured over a finite angular binning can be taken into account if we express the above summations in terms of the  the bin-averaged Legendre polynomial defined as \citep{Salazar-Albornoz:2016psd}
\begin{align}
\label{eq:cov_all_indexes}
  \bar{P}_\ell &= \frac{ \int_{\theta_{-} }^{\theta_{+}}   P_\ell( \cos \theta) \, \sin \theta \, d \theta }{  \int_{ \theta_{-}}^{ \theta_{+}}   \sin \theta \,d \theta}   \nn \\
& =\frac{ P_{\ell+1} ( x_+ ) - P_{\ell + 1} (x_-) - P_{\ell-1}(x_+) + P_{\ell -1 } (x_-) }{ (2\ell +1 ) (x_+ - x_- ) },
\end{align}
where $\theta_+$ and $\theta_-$ ($x_+$ and $x_-$) denotes the upper and lower limit of the bin (the cosine of the upper limit and lower limit) respectively.  It is important to use the bin-averaged $\bar{P}_\ell$ in Eq.~\eqref{eq:cov_mat_crossz}, otherwise the error is overestimated [see similiar arguments in \citet{Cohn:2005ex,Smith:2007gi,Sanchez:2008iw,Salazar-Albornoz:2016psd}].  The bin-averaged $P_\ell$ is substantially smaller than the un-averaged  $P_\ell$ for high $\ell \gtrsim 1000$ for the typical bin width we consider $\Delta \theta \sim 0.1^{\circ} $.  This is because for high $\ell$ multipoles, the variation of $P_\ell$ across the bin width is not negligible.  The effect of the bin-averaging is small for the mean [i.e.~Eq.~\eqref{eq:w_from_Cl}] because although the high $\ell$ part is inaccurate, it is rapidly oscillatory, the net effect is small.  For the diagonal of the covariance, the terms are all positive, and so the effect is large.   

In addition, care must be taken with the shot-noise terms, which are scale independent and hence factor out in the infinite sum over Legendre polynomials. One is left with a sum of the type $\sum_\ell (2\ell+1) P_{\ell}(x) P_\ell(x') = 2 \delta_{\rm D}(x-x')$ which is formally infinite. The fact that these are over averaged multipoles regularizes them to $\sum_\ell (2\ell+1) \bar{P}_{\ell}(x) \bar{P}_\ell(x') = 2 / (x_- - x_+)$ for auto-correlations (same angular bin) and zero otherwise. On the other hand, the cross terms of the form $C_\ell/{\bar n}$ do converge when summed over because $C_\ell$ decays at high $\ell$. Hence we treat the pure noise terms separately and perform the sum analytically. There are only two relevant cases

\begin{align}
& \mathrm{Cov}[ w_{ij}(\theta),  w _{ij}(\theta')] =  \frac{\delta_{\rm K}^{\theta \theta'}}{ 8 \pi^2 f_{\rm sky} {\bar n}_i {\bar n}_j (x_- - x_+ )} + \sum_{\ell} \frac{2 \ell +1}{ (4 \pi)^2  f_{\rm sky}} \nonumber \\  & \quad \times \bar{P}_\ell(\cos \theta) \bar{P}_{\ell}(\cos\theta') 
\bigg(C^{ii}_{\ell}C^{jj}_{\ell}+   + C^{ij}_{\ell}C^{ij}_{\ell} +   \frac{C^{ii}_{\ell}}{{\bar n}_j} + \frac{C^{jj}_{\ell}}{{\bar n}_i} \bigg) 
\end{align}
for  $i \ne j $, 
and, 
\begin{align}
  \label{eq:Gaussian_cov_auto}
\mathrm{Cov}[ w_{ii}(\theta),&  w _{ii}(\theta')] =  \frac{  \delta_{\rm K}^{\theta \theta'}}{  4\pi^2 f_{\rm sky} {\bar n}^2_i ( x_- - x_+ ) } + \sum_{\ell} \frac{2 \ell +1}{ 8 \pi^2  f_{\rm sky}} \nonumber \\  &\times  \bar{P}_\ell(\cos \theta) \bar{P}_{\ell}(\cos\theta') 
\bigg(C^{ii}_{\ell}C^{ii}_{\ell}+2 \,\frac{C^{ii}_{\ell}}{{\bar n}_i} \bigg). 
\end{align}
The remaining combinations of the indices ($i,j,m,n$) in Eq.\eqref{eq:cov_mat_crossz} do not give rise to scale independent shot noise term,  and can therefore be obtained from Eq.~\eqref{eq:cov_mat_crossz} without the need to sum them separately. 

Finally we note that the survey angular geometry mask does not appear explicitly in the Gaussian covariance, only the survey area through $f_{\rm sky}$. In configuration space the effect of the mask is less severe than in Fourier/harmonic space. \change{Nonethelss the geometry of the mask, including the fact that it has holes, makes the number of random pairs as a function of separation not to simply scale with the effective area of the survey. Hence the shot noise term will not exactly follow Eq.~\eqref{eq:cov_mat_crossz} but acquire a scale dependence, see ~\cite{Krause:2017ekm}. We have checked that this effect is negligible for our BAO sample. Furthermore, \citet{Halogen} compared the covariance matrix obtained from mocks to the Gaussian covariance matrix and found agreement to within 10\%. In Sec.~\ref{sec:Covariance} we study this issue in greater detail. }   Another effect, the supersample covariance due to the coupling of the small scale modes inside the survey with the long mode when the window function is present and is only important for small scales \citep{Hamilton:2005dx,Takada:2013bfn,Li:2014sga,Chan:2017fiv}.

\subsection{Mock Catalogs} 
\label{sec:halogen}

We calibrate our methodology using a sample of 1800 Halogen mocks \citep{Halogen} that match the BAO sample of DES Y1. We outline the basic information here, and refer the readers to \cite{Halogen}, and \cite{Avila_etal2015}, for more details.  For each mock realization, the dark matter particle distribution is created using second-order Lagrangian perturbation theory.  Each mock run uses $1280^3$ particles in a box size of $3072  \MpcOh$.  Halos are then placed in the dark matter density field based on the prescriptions described in \citet{Avila_etal2015}.  The halo abundance, bias as a function of halo mass, and  velocity distribution are matched to those in the MICE simulation \citep{Fosalba_etal2015}.   Halos are arranged in the lightcone with the observer placed at one corner of the simulation box. The lightcone is spanned by 12 snapshots from $z=0.3$ to 1.3.  From this octant, the full sky mock is formed by replicating it 8 times with periodic boundary conditions.  Galaxies are placed in the halos using a  hybrid Halo Occupation Distribution--Halo Abundance Matching prescription that allows for galaxy bias and number density evolution.

The mocks match to the properties of the DES Y1 BAO sample. The angular mask and sample properties including the photo-$z$ distribution, the number density, and the galaxy bias are matched. Redshift uncertainties are accurately modeled by fitting a double skewed Gaussian curve to photo-$z$ distribution measured from the data, and this relation is then applied to the mocks. The final mock catalog covers an area of 1318 $\mathrm{deg}^2$ on the sky as in DES Y1. We consider the mock data in the photo-$z$ redshift range [0.6,1], and there are close to 1.3 million galaxies per mock in this range.  In the fiducial setting, the sample is further divided into four redshift bins of width $ \Delta z = 0.1 $. In total, we produce 1800 realizations, and we use them to calibrate the pipeline and estimate the covariance matrix. Unless otherwise stated, the mock covariance is used. 

We measure the angular correlation function from the mocks with the Landy-Szalay estimator \citep{LandySzalay_1993}
\beq
w(\theta) = \frac{ DD(\theta) - 2 DR( \theta) + RR(\theta)    }{ RR(\theta)   }, 
\eeq
where $DD$, $DR$, and $RR$ are the pair counts between the data-data, data-random and random-random catalogs respectively, normalized based on the size of catalog. \change{ The number of objects in the random catalog is 20 times those in the data.} $w$ is computed using the public code CUTE \citep{Alonso_CUTE}.

\section{BAO fitting methods  } 
\label{sec:fitting_methods}

\subsection{Methods overview}

\change{ For high signal-to-noise BAO data one expects to recover Gaussian likelihoods e.g.~in the case of \citet{Alam_etal2017}. } For such data, many methods for the BAO fitting would be expected to yield consistent results. But this might not be our situation, as the expected signal to noise is close to 2. 

Here we compare three methods to derive the BAO angular scale from the data, testing them thoroughly with the mocks: Maximum Likelihood Estimator (MLE), Profile Likelihood (PL), and Markov Chain Monte Carlo (MCMC). We define each below. These methods differ in how they define the probability distribution of the interested variables, and how the best fit values and errors are computed.  For a review of these statistical methods see e.g.~\cite{Press:2007:NRE:1403886,HoggBovyLang2010,Trotta:2017wnx}.

In the BAO fitting even though the full likelihood is multi-dimensional (with all the parameters in the template) we are ultimately interested in only one, the BAO dilation parameter $\alpha  $.  Here we are mostly interested in which estimator gives the most reliable result for $\alpha $. 

Throughout we use  $ \bar{ \alpha }$  and  $ \sigma_{ \alpha } $ to denote the best fit and the error obtained from the fitting method for an individual realization. We will use angular brackets to represent the ensemble average over the mocks: e.g., $\langle \bar{\alpha}\rangle$ is the mean of the best fit distribution, while $ \mathrm{std} (\bar{\alpha} )$ is the standard deviation of  the  $\bar{ \alpha }$  distribution. 

\subsubsection{Maximum Likelihood Estimator (MLE)}
\label{sec:MLE_description}

MLE is a point estimator for some parameters $\bm{\lambda}$, and it seeks the best fit by maximizing the likelihood function $  L( D | \bm{\lambda}) $, where $D$ denotes the dataset and $\bm{\lambda} $ the parameters.  If the data is Gaussian distributed, and we can relate the likelihood to the $\chi^2 $ as \citep{Press:2007:NRE:1403886}
\beq
\label{eq:L_chi2_relation} 
L \propto  \exp( - \chi^2 /2 ), 
\eeq
with the $\chi^2$ defined as
\beq
\chi^2( \bm{\lambda} ) = \sum_{i,j} [ D_i - x_i(\bm{\lambda} ) [ C^{-1}_{\quad ij }  [D_j - x_j(\bm{\lambda}) ],
\eeq
where $C$ is the covariance matrix and $\bm{x}$ the model.  The best fit can be obtained by minimizing the $\chi^2$. MLE itself does not require the likelihood to be Gaussian, and in that case $\chi^2 $ minimization can still be used to find the best fit although its connection to the probability distribution is not direct. 

\change{ For the model Eq.~\eqref{eq:BAO_template},  although there are  large number of nuisance parameters $B$ and $A_i$, they  appear \textit{linearly} and can be fit analytically using the least square fit method.  Suppose that the model is given by
\begin{align}
\bm{x}_i =   \sum_\rho A_{i \rho} \bm{\lambda}_\rho, 
\end{align}
where $A$ is called the design matrix, then the best fit model parameters $ \bm{\lambda}$ obtained by minimizing the $\chi^2$ reads \citep{GCowan}
\beq
\label{eq:lambda_analyticfit}
\bm{\lambda} = ( A^T C^{-1} A )^{-1} A^T C^{-1} \bm{D}.
\eeq  }
Thus in principle, we can end up with only one parameter $\alpha$ and its best-fit can be found by a grid search. In practice, we find that sometimes we get the unphysical result $B<0$. To avoid this unsavory situation, we impose the prior that $B>0$. 
\change{ To do so, the $\chi^2( \alpha, B_i , \bm{A}_i ) $ is first analytically minimized with respect to the parameters $ \bm{A}_i $ using Eq.~\eqref{eq:lambda_analyticfit} to get the best fit  $\bm{A}_i^{\rm bf} $.  We then numerically search for the best fit $B_i^{\rm bf}$ with the prior that  $B_i>0$ such that  $ \chi^2( \alpha, B_i, \bm{A}_i^{\rm bf}  ) $ is minimized with respect to $B_i$. Finally we are left with a one-parameter  function $\chi^2( \alpha, B_i^{\rm bf}, \bm{A}_i^{\rm bf} ) $.  MLE has been adopted as a convenient choice for BAO fits in numerous recent studies [e.g., \citet{Anderson_etal2014}].  }

For MLE, we use the 1-$\sigma$ error bar derived from the deviation from the minimum $\chi_{\rm min}^2 $  by $ \Delta \chi^2 = 1 $ \citep{Lampton_etal1976,Press:2007:NRE:1403886}.  This does assume that the likelihood of $\lambda$ is Gaussian distributed [or $\chi^2( \lambda )  $ is a quadratic function of $\lambda$] and there is only one parameter $\lambda$. \change{ It also applies to our case when the other parameters are maximized, see \cite{Press:2007:NRE:1403886}.  This rule can be obtained as follows.  We can expand the log of the likelihood $ \ln L $ about $\alpha = \alpha_0 $ where the maximum of the likelihood is attained as 
\begin{align}
  \ln \mathcal{L}(\alpha)   
    & \approx  \ln \mathcal{L}_0 +   \frac{ 1}{ 2 }  \frac{ \partial^2  \ln \mathcal{L} }{ \partial \alpha^2   } \Big|_{\alpha_0 }      ( \alpha - \alpha_0 )^2 . 
\end{align}
When the Cram\'er-Rao bound [see \citet{Heavens2009} for a review]  is saturated, the variance of $\alpha $, $\sigma_{\alpha}^2 $  is given by  
\beq
 \frac{ 1}{ \sigma_{\alpha  }^2 } =  -  \frac{ \partial^2  \ln \mathcal{L} }{ \partial \alpha^2   } \Big|_{\alpha_0 }.  
 \eeq
I.e.~the curvature of $\ln L $ encodes the error bar  on the parameter.  Hence at 1-$\sigma$ from  $\alpha_0$,  $ \alpha_0 \pm \sigma_\alpha  $
\beq
\label{eq:lnL_halfdeviation}
\ln \mathcal{L}( \alpha \pm \sigma_{\alpha} ) =   \ln \mathcal{L}_0 - \frac{ 1}{ 2 } .
\eeq
From Eq.~\eqref{eq:L_chi2_relation}, this is equivalent to $\Delta \chi^2 =1 $ rule for the 1-$\sigma$ error bar in MLE.  }

We will further take the symmetric error bar by averaging over the lower and upper bars. In the frequentist's interpretion, because  $\alpha$ is a parameter, we can interpret the error bars only when the experiments are repeated. Suppose $N$ independent measurements are repeated, we expect to have 68\% of the time the 1-$\sigma $ bars enclosing the true value ($\alpha = 1 $ for the unbiased case) for a Gaussian distribution.

We will consider the likelihood in the range of $\alpha \in $ [0.8,1.2], and BAO is regarded as being detected only if the 1-$\sigma $ interval can be constructed within the interval [0.8,1.2].

\subsubsection{Profile Likelihood (PL)}

Instead of only using the information at the maximum of the likelihood, we can compute the weighted mean and the standard deviation using the weight $W$ 
\begin{align}
W( \alpha )  = \frac{ L(D| \alpha) }{ \int d \alpha' \,  L(D| \alpha') }. 
\end{align}
Here the partial likelihood $ L( D| \alpha )$  is obtained by partially maximizing the likelihood with respect to all the other nuisance parameters except  $ \alpha $ following the procedures outlined in Sec.~\ref{sec:MLE_description}.  The mean and variance are given by
\begin{align}
\bar{\alpha} &= \int  d\alpha \,   W(\alpha) \alpha  ,  \\
\sigma_\alpha^2 &= \int  d\alpha \,   W(\alpha) ( \alpha - \bar{\alpha} )^2  . 
\end{align}
This method sits between the MLE and MCMC in concept: it adopts MLE for parameters other than $\alpha $, but performs a full marginalization for $\alpha $ as in the MCMC approach.  The integration range is taken to be [0.8,1.2]. For PL, $\bar{\alpha}$ and $\sigma_\alpha^2 $ can always be defined, and it does not require a minimum in $\chi^2 $ to be found in the $\alpha$ range.  

\subsubsection{ Markov Chain Monte Carlo (MCMC)}

The MCMC has been widely used in cosmological parameter fitting in last couple of  decades, e.g.~\citet{WMAP9Yr2013,Ade:2015xua, Abbott:2017wau}, mainly because it can efficiently sample large number of parameters.  MCMC is based on the Bayes theorem
\beq
P( \bm{\lambda} | D ) = \frac{ P( \bm{\lambda} )  L( D | \bm{\lambda} )   }{ P(D) },   
\eeq
where $ P(\bm{\lambda}) $ is the prior distribution, and $P(D)$ is the probability of the data (often called the evidence). 

In the Bayesian approach, $\alpha $ is a random variable, we can talk about  the chance that the $\alpha$ value lies in the 1-$\sigma $ interval. Strict Bayesians will stop at the posterior distributions as their final product, but to compare with other methods we will deviate from the strict Bayesianism and use the posterior distributions to compute the summary statistics \citep{HoggBovyLang2010}.   If the distribution is Gaussian we expect 68\% of chance. We use the median of the MCMC chain for $\bar{\alpha}$, and $ \sigma_{\alpha }$ is derived from 16 and 84 percentiles of the chain.  Again to facilitate the comparison with other methods, we average over the left and right error bars to get a symmetric one.  Alternatively, we can use the mean and the standard deviation for the best fit and its error bar. We opt for the median and the percentiles because we find that the results encloses $ \langle \bar{ \alpha } \rangle $ closer to the Gaussian expectation.   The prior on $\alpha$  is taken to be [0.6,1.4]. Similar to PL, $\bar{\alpha}$ and $ \sigma_{\alpha }$ can always be defined.

We use the MCMC implementation {\tt emcee} \citep{Emcee_Foreman-Mackey2013}, in which multiple walkers are employed and the correlation among the walkers are reduced by using the information among them.   We use 100 walkers, 3000 burn-in steps, and 2000 steps for the run. In this setting, the MCMC fitting code takes about 50 times longer than the MLE code does. \change{ We have conducted some convergence test on the number of steps required. We took a sample of steps in the range from 1500 to 50000 and find that fluctuations of  $\bar{\alpha}$  are within 0.1\% from the convergent value (assuming convergence attained with 50000 steps) and those of  $\sigma_{\alpha}$ within 2\%. Thus 2000 steps are sufficient to make sure that it does not impact the results later on. We note that the usage of the percentiles of the distribution is much less suspectible to statistical fluctuations than using the mean and the variance.   }

\subsubsection{Comparison criteria}
\label{sec:selection_criterion}

We will test these methods against the mock catalogs.  We will check how stable the fit results are, especially how small a bias (in comparison to the known true value) that each method yields.

The full information is in the likelihood/posterior.  The best fit and its error matter because we want to effectively represent the distribution by these two numbers. Here are some features that the desirable summary statistics should have.   First we want the estimator to yield an unbiased mean result. Second, in order to have the proper probabilistic interpretation of the error bar, it is desirable that the error bar encloses the true answer close to the Gaussian expection, which is 68\% for 1-$\sigma$. Hence we will check the fraction of times that the error bar derived encloses the true answer.
Another useful metric is that the standard deviation of the distribution, $ \mathrm{std}( \bar{ \alpha }  )$  agrees with the mean of the error derived  $ \langle  \sigma_{\alpha} \rangle $. One way to compare the deviation of $\bar{ \alpha } $ from the expected true answer (the spread of the $ \bar{ \alpha }$ distribution) and the error derived  $\sigma_{\alpha} $ is to consider the normalized variable 
\beq
d_{\rm norm } = \frac{ \bar{ \alpha }  - \langle  \bar{ \alpha }  \rangle  }{ \sigma_{\alpha } }. 
\eeq
The distribution of this variable can be compared to a unit normal distribution as a test of the Gaussianity of the recovered results.

\change{ Note that we do not enforce the error distribution to be Gaussian, indeed they are not (see Fig.~\ref{fig:error_distribution_NAP4}). We only want the 1-$\sigma$ error bar to enclose the true answer close to the Guassian expectation.   In principle,  for MLE we can adjust the value of $\Delta \chi^2 $ so that it encloses the true value (e.g.~$\alpha =1 $) 68\% of the time. Similar adjustments can be done for other estimators.  By doing so, the derived error bar yields the desired Gaussian probability expectation. However, the $\Delta \chi^2 =1 $ rule works well for us, and no adjustment is needed.  }



\subsection{Comparison of the BAO fit results by MLE, PL, and MCMC }

\begin{figure}
\begin{center}
\includegraphics[width=\linewidth]{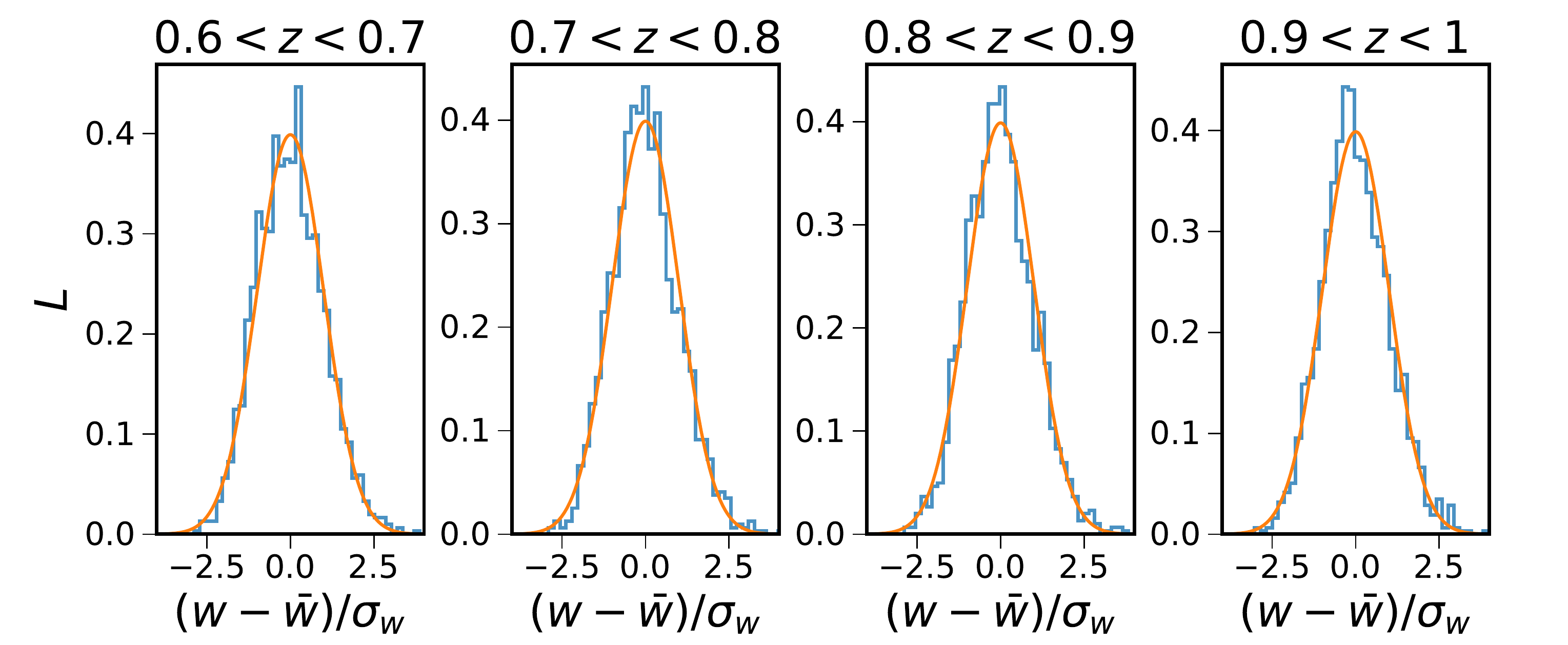}
\caption{  The likelihood distribution of $w$ at the bin $\theta= 2^{\circ  }$ (histogram). The standard normal variable is used. The Gaussian distribution with zero mean and unity variance (solid line) is plotted for comparison.    }
\label{fig:likelihood_1angle}
\end{center}
\end{figure}

\begin{table*}
  \caption{The BAO fit with MLE, PL, and MCMC obtained with two sets of detection criteria (left and right). For Detection Criterion 1, we only consider those mocks whose 1-$\sigma$ interval $\bar{\alpha }  \pm \sigma_\alpha  $ fall within the interval [0.8,1.2]. For Detection Criterion 2, we use the same set of mocks for all three methods, those having 1-$\sigma$ interval falling within  [0.8,1.2] using the MLE estimate of $\bar{\alpha }$ and $\sigma_\alpha$.  For each criterion, the first fraction is normalized with respect to the total number of mocks (1800), while the second fraction is normalized with respect to the number of mocks satisfying the selection criterion. }
\label{tab:data_pruning}
  \begin{tabular}{|l|c|c|c||c|c|c|}
    \hline
    \multirow{2}{*}{} &
      \multicolumn{3}{c}{ Detection Criterion 1} &
      \multicolumn{3}{c}{ Detection Criterion 2} \\
      \hline 
    &  ${\bar \alpha} \pm \sigma_\alpha$ in [0.8,1.2]  &  $\langle {\bar \alpha} \rangle \pm {\rm std}({\bar \alpha})$  & frac. with ${\bar \alpha} \pm \sigma_\alpha$ & ${\bar \alpha} \pm \sigma_\alpha$ in [0.8,1.2] & $\langle {\bar \alpha} \rangle \pm {\rm std}({\bar \alpha})$ & frac. with ${\bar \alpha} \pm \sigma_\alpha$ \\
    &  (frac. selected) & & enclosing $\langle \bar{\alpha} \rangle $ & MLE (frac. selected) &  & enclosing $\langle \bar{\alpha} \rangle $ \\
    \hline
 MLE           & 0.91      & $ 1.001 \pm 0.052 $  &  0.69    &  0.91  & $ 1.001 \pm 0.052$   & 0.69    \\
PL             & 0.99      &  $1.004 \pm 0.049 $  &  0.77    &  0.91  & $ 1.003 \pm 0.046 $  & 0.78     \\
MCMC           & 0.84      &  $1.007 \pm 0.049 $  &  0.74    &  0.91  & $ 1.007 \pm 0.059  $ & 0.74    \\
    \hline
  \end{tabular}
\end{table*}

\begin{figure}
\begin{center}
\includegraphics[width=\linewidth]{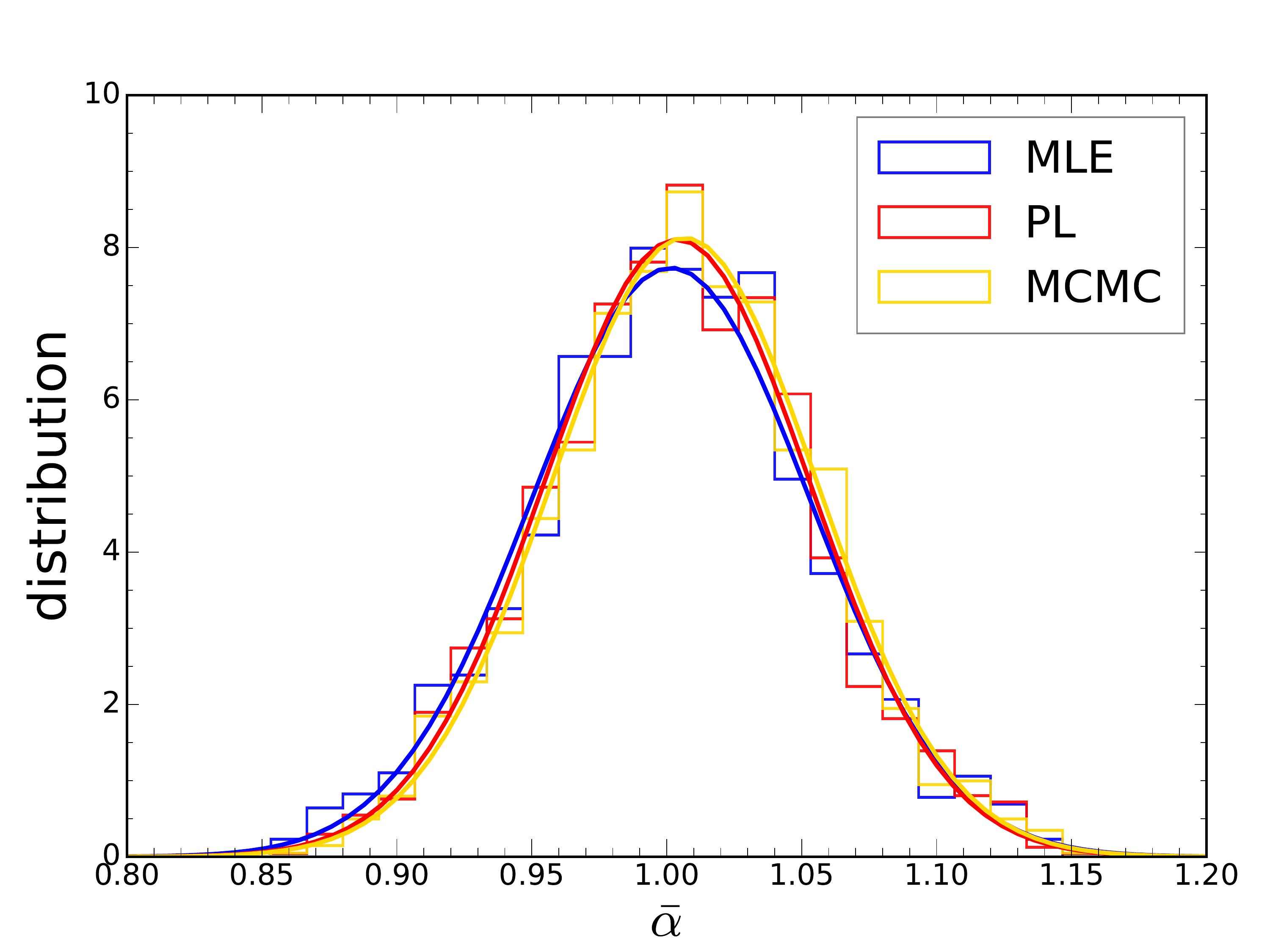}
\caption{  The histograms show the distribution of the best fit $\bar{\alpha} $ obtained using MLE (blue), PL (red), and MCMC (yellow).  The solid lines (blue for MLE, red for PL, and yellow for MCMC) are the Gaussian distributions with the same mean and variance as the corresponding histograms.    }
\label{fig:alpha_distribution_NAP4}
\end{center}
\end{figure}

\begin{figure}
\begin{center}
\includegraphics[width=\linewidth]{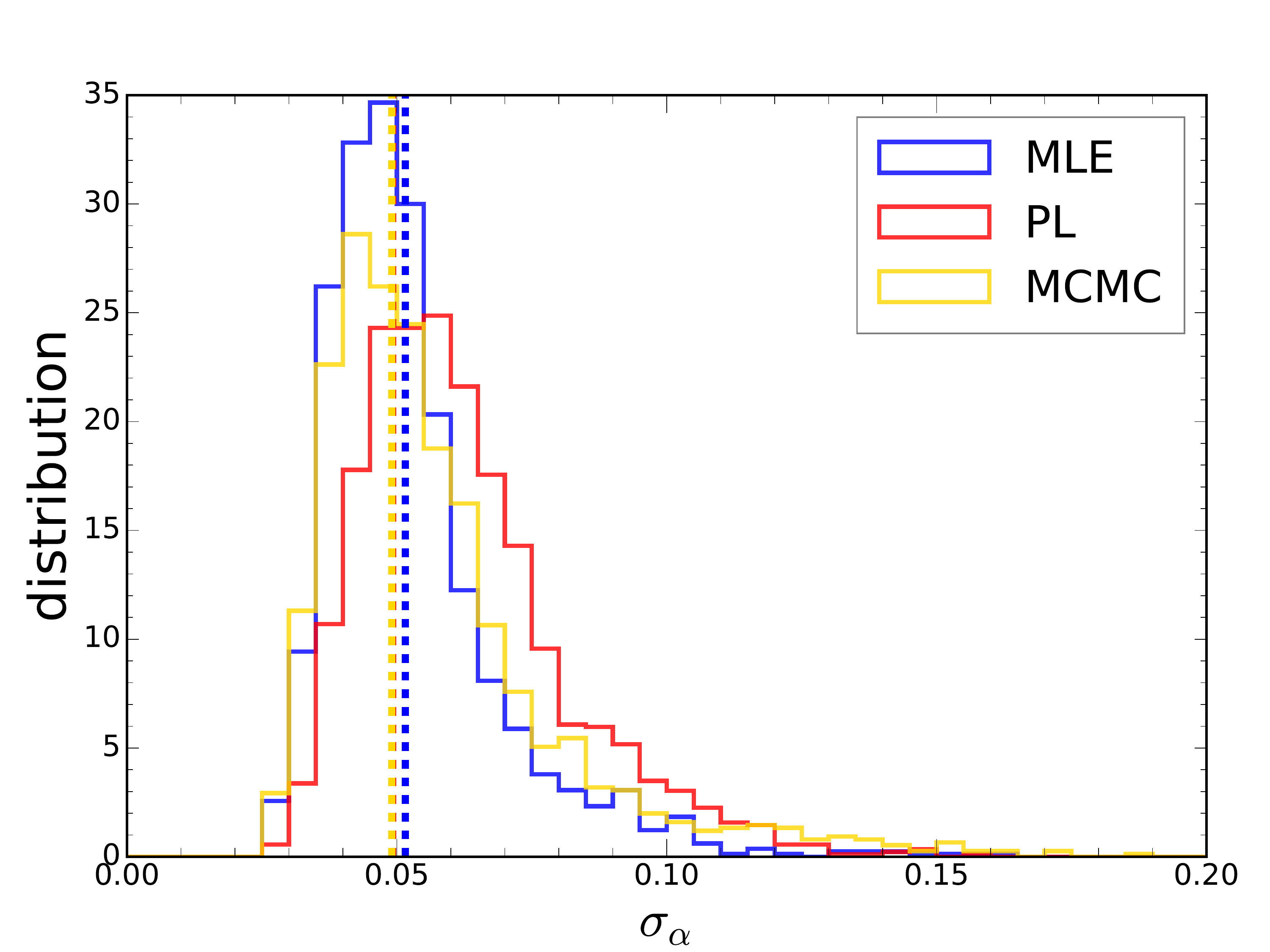}
\caption{The distribution of the error derived from each individual mock (blue for MLE, red for PL, and yellow for MCMC). The vertical dashed lines are the standard deviations of the best fit $\bar{\alpha} $ shown in Fig.~\ref{fig:alpha_distribution_NAP4}. The $\langle \sigma_\alpha  \rangle $ and $\mathrm{std}(\bar{\alpha}  )$ for MLE, PL, and MCMC are (0.053, 0.052), (0.062, 0.0492), and (0.057, 0.049) respectively. While for MLE  $\langle \sigma_\alpha  \rangle $ and $\mathrm{std}(\bar{\alpha}  )$ coincide, PL and MCMC yield larger $\langle \sigma_\alpha  \rangle $.   }
\label{fig:error_distribution_NAP4}
\end{center}
\end{figure}

\begin{figure}
\begin{center}
\includegraphics[width=\linewidth]{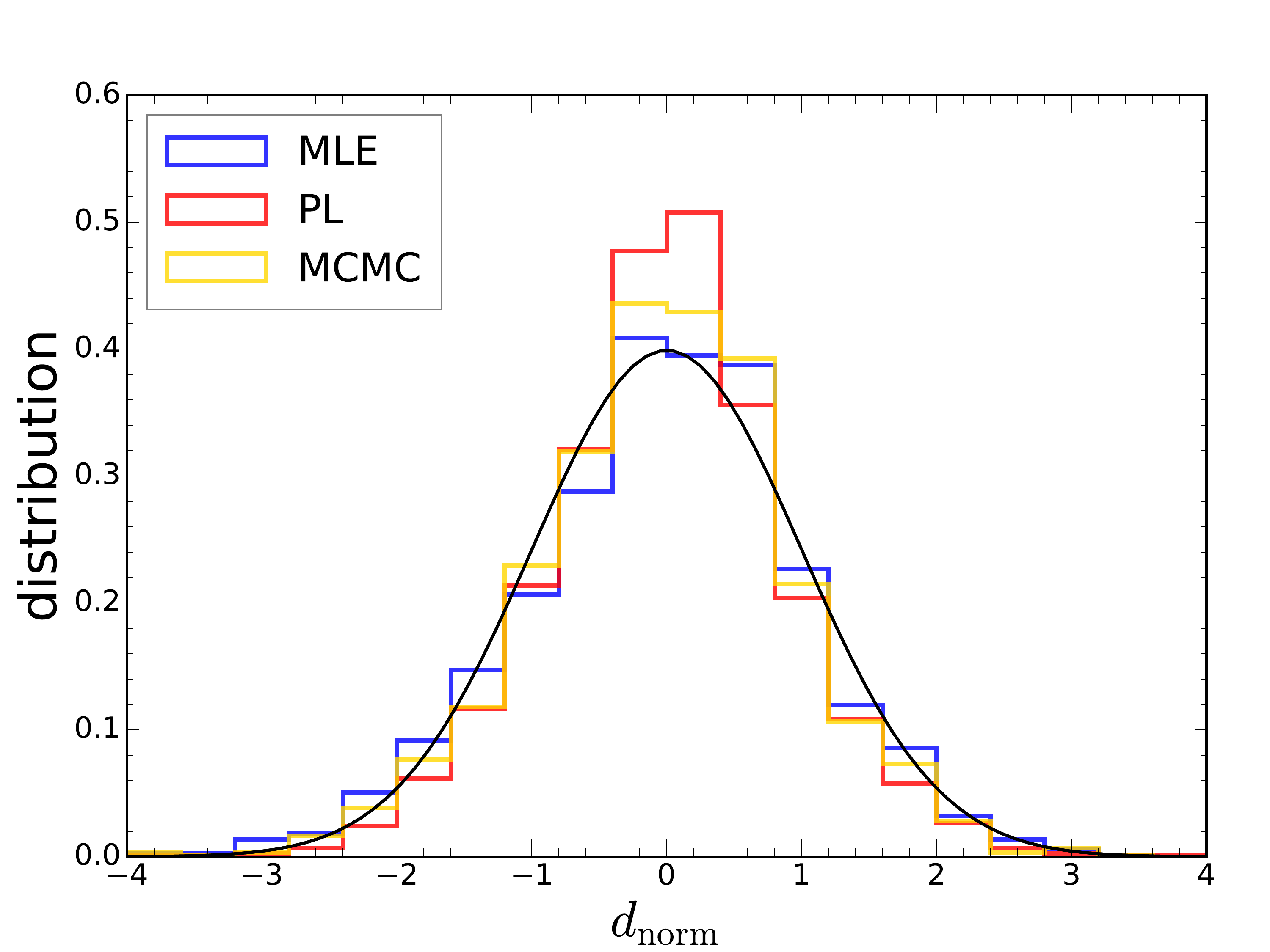}
\caption{ The distribution of the normalized variable $d_{\rm norm} $. The results obtained with MLE (blue), PL (red), and MCMC (yellow) are shown. The Gaussian distribution with zero mean and unity variance (solid, black) are shown for reference.  }
\label{fig:dnorm_fiducial}
\end{center}
\end{figure}

Following \citet{BAOmain}, we consider those fits  with their 1-$\sigma$ intervals of $\alpha$  falling outside the range [0.8,1.2] as non-detections.  These non-detections are poorly fit by our template, and they cause the distribution of $\bar{ \alpha } $  to be highly non-Gaussian. Thus we will remove the non-detection mocks first. We will comment more on this at the end of the section.

\change{ Before proceeding to the comparison, we first verify the Gaussian likelihood assumption Eq.~\eqref{eq:L_chi2_relation} using the mock catalogs. The likelihood tends to be Gaussian distributed thanks to the central limit theorem. Because the Gaussian likelihood assumption is central to the analysis, we need to check it [e.g.~\citet{Scoccimarro2000,Hahn:2018zja}]. In Fig.~\ref{fig:likelihood_1angle}, we show the likelihood distribution of the values of $w$ measured from the mocks. We have shown the results for the bin $\theta = 2^\circ $. We have used the  standard normal variable  $ (w - \bar{w} ) / \sigma_w $, where  $\bar{w}$ and $ \sigma_w $ are the mean and the standard deviation of the distribution of $w$. We find  that the likelihood indeed follows the Gaussian distribution well. }

In Table \ref{tab:data_pruning}, we show the fit with MLE, PL, and MCMC for two detection criteria.  First, for MLE with error derived from $ \Delta \chi^2 =1 $, there are 91\% of the mocks with their 1-$\sigma$ error bars fall within the interval of $[0.8,1.2]$, while for MCMC the fraction is 84\%. For PL, it is also almost 100\%.  

In Fig.~\ref{fig:alpha_distribution_NAP4}, we show the distribution of the best fit $\alpha$ obtained with these three methods after pruning the non-detection mocks. As a comparison we have also plotted the Gaussian distribution with the same mean and variance, we find that $ \bar{\alpha} $  follows the Gaussian distribution well.  The mean of the best fit from MLE is the least biased among the three methods.  The distributions of  $\bar{\alpha} $ from PL and MCMC are quite similar. They tend to be more skewed towards $\bar{\alpha} > 1 $, and this can be seen from their corresponding Gaussian distribution and the $\langle  \bar{\alpha }\rangle  $ shown in  Table \ref{tab:data_pruning}. In particular for MCMC,  $ \langle \bar{\alpha} \rangle $, it is larger than 1 by 0.007.   In Table \ref{tab:data_pruning}, the fraction  of mocks with the 1-$\sigma$ error bar enclosing the $\langle \bar{\alpha } \rangle  $ is also shown. We find that MLE with $\Delta \chi^2 =1  $ prescription encloses  $\langle \bar{\alpha} \rangle $ 0.69 of the time, which is very close to the Gaussian expectation, while PL and MCMC  are higher than the Gaussian expectation by 9\% and 6\% respectively.

In order to see the error derived more clearly, in Fig.~\ref{fig:error_distribution_NAP4}, the distribution of the errors derived from these three methods are shown. As a comparison we also show the $\mathrm{std}(\bar{\alpha })$.  In the caption of Fig.~\ref{fig:error_distribution_NAP4} we have given the numbers for the $\langle \sigma_\alpha  \rangle $ and $\mathrm{std}(\bar{\alpha}  )$ for MLE, PL, and MCMC. The $\langle \sigma_\alpha  \rangle $ obtained from the MLE method coincides with $\mathrm{std}(\bar{\alpha}  )$ well, while MCMC gives slightly larger error $\sigma_{\alpha}$. PL tends to give the largest error with $\langle \sigma_\alpha  \rangle  / \mathrm{std}(\bar{\alpha}  ) = 1.27$.

We plot the distributions of the normalized variable for these methods in Fig.~\ref{fig:dnorm_fiducial}.  By comparing with the Gaussian distribution with zero mean and unity variance, we find that MLE with $\Delta \chi^2 =1 $, $d_{\rm norm } $  is more Gaussian, while MCMC is slightly worse, and PL is the least Gaussian. 

In the detection criterion we have applied, the non-detections are different for each of the methods. We therefore wish to check if the differences we have found in the results are simply due to that.
 We now apply the same selection criterion that the 1-$\sigma$ interval obtained from MLE falls within [0.8,1.2]  to all the three methods. The results are shown in Table \ref{tab:data_pruning} as Detection Criterion 2. We find that the results are essentially the same as the previous one, ruling out that the differences are due to the selection criterion.  Note that for MCMC, the $\mathrm{std}(\bar{\alpha }) $ has increased quite appreciably because using the MLE detection criterion, a large fraction of extreme mocks are retained and they cause a big increase in the $\mathrm{std}(\bar{\alpha })$.  In the rest of the analysis we will stick to the original pruning criterion.   

\change{ The choice of the interval  $[0.8,1.2]$ is somewhat arbitrary, but based on past results [e.g.,~\citet{Anderson_etal2014,Ross:2016gvb}] that suggest it is a reasonable choice (though, admittedly, more a rule of thumb than anything else). To further justify this choice, in the Appendix~\ref{sec:appendix}, we show the comparison of results obtained with a larger interval [0.6,1.4]. We find that by including the extreme mocks, the distribution of the best fit  $\bar{\alpha}$ exhibits strong tails, and it does not agree with the Gaussian distribution  with the same mean and variance.    These extreme mocks not only enlarge the mean of the error bars $\langle \sigma_\alpha \rangle $,  they also cause bias in $\langle  \bar{\alpha} \rangle $. In particular for MLE, $|\langle  \bar{\alpha} \rangle - 1 |$ changes from 0.001 (Table \ref{tab:data_pruning}) to 0.029 (Table \ref{tab:data_almost_no_pruning}). For MCMC, it only changes mildly from 0.007 to 0.010. Overall we find that when the wider interval [0.6,1.4] is adopted, the MCMC is superior to the other two methods because it gets less biasd results, and  the derived 1-$\sigma$ intervals which enclose $\langle \bar{\alpha} \rangle $ is 71\% of the mocks, the closest to the Gaussian expectation. Moreover,  $\langle \sigma_\alpha \rangle  / \mathrm{std}(\bar{\alpha})=0.92$ is close to 1.  }

\change{Because the small fraction of the mocks without a BAO detection highly bias the distributions of the best fit parameters, and since it would be difficult to extract useful BAO information from them,   we adopt a smaller interval to get rid of them. In practice,  if the best fit with its 1-$\sigma$ interval is outside the range [0.8,1.2] in our actual data, the data would be poorly fitted (or the error bar poorly estimated) by our existing methodology and we can hardly claim to detect the BAO signals in the data.  In this case, we may need to change the fiducial cosmological model or wait for additional data.} 

Thus, overall we find that after pruning the non-detections, MLE yields the estimate with the least bias and the error bar using $\Delta \chi^2=1 $ results in the 1-$\sigma$ error closest to the Gaussian expectation.  Therefore if we prune the data, MLE furnishes an effective BAO fitting method.  For the rest of the analysis, we always use the pruned data.

\section {Template systematics }
\label{sec:template_systematics}

In this section, we study the potential systematics associated with the template and investigate how they can affect the BAO fit results.   We will first determine the physical damping scale $\Sigma $ in the template by fitting to the mock catalog.  Although the damping scale does not bias the mean for the BAO fit, it can strongly affect the error bar that is derived.  We then examine how the BAO fit is affected when the template does not coincide precisely with the data, i.e.~the BAO scale in the template is different from that in the data.  The parameter $\alpha$ is introduced to allow the shift in the BAO scale, it is crucial to check how successful it deals with the mismatch.

The BAO angular scale $\theta_{\rm BAO}$ can be estimated as
\beq
\label{eq:theta_BAO}
\theta_{\rm BAO} =  \frac{r_{\rm s}}{ D_{\rm A}},
\eeq
where  $r_{\rm s} $ is the comoving sound horizon at the drag epoch  and $ D_{\rm A} $ denotes the \textit{ comoving} angular diameter distance [as in \citet{WeinbergMortonson_etal2013}] and in a \change{ flat cosmology }  it is given by
\beq
D_{\rm A} = \int_0^z \frac{ c \, dz'  }{H(z') },
\eeq
where $H$ is the Hubble parameter and $c$ is the speed of light. Here we consider two ways that the template mismatches the data. First, the cosmology for the data could be different from that of the fiducial cosmology. Another possibility is that there could be photo-$z$ error causing  $\phi(z)$ to be systematically biased. Both possibilities can cause shifts in $\theta_{\rm BAO}$. Using a wrong cosmology changes both $r_{\rm s} $ and $ D_{\rm A} $. 

\subsection{ The BAO damping scale }
\label{sec:DampingScale}

\begin{figure}
\begin{center}
\includegraphics[width=0.45\textwidth]{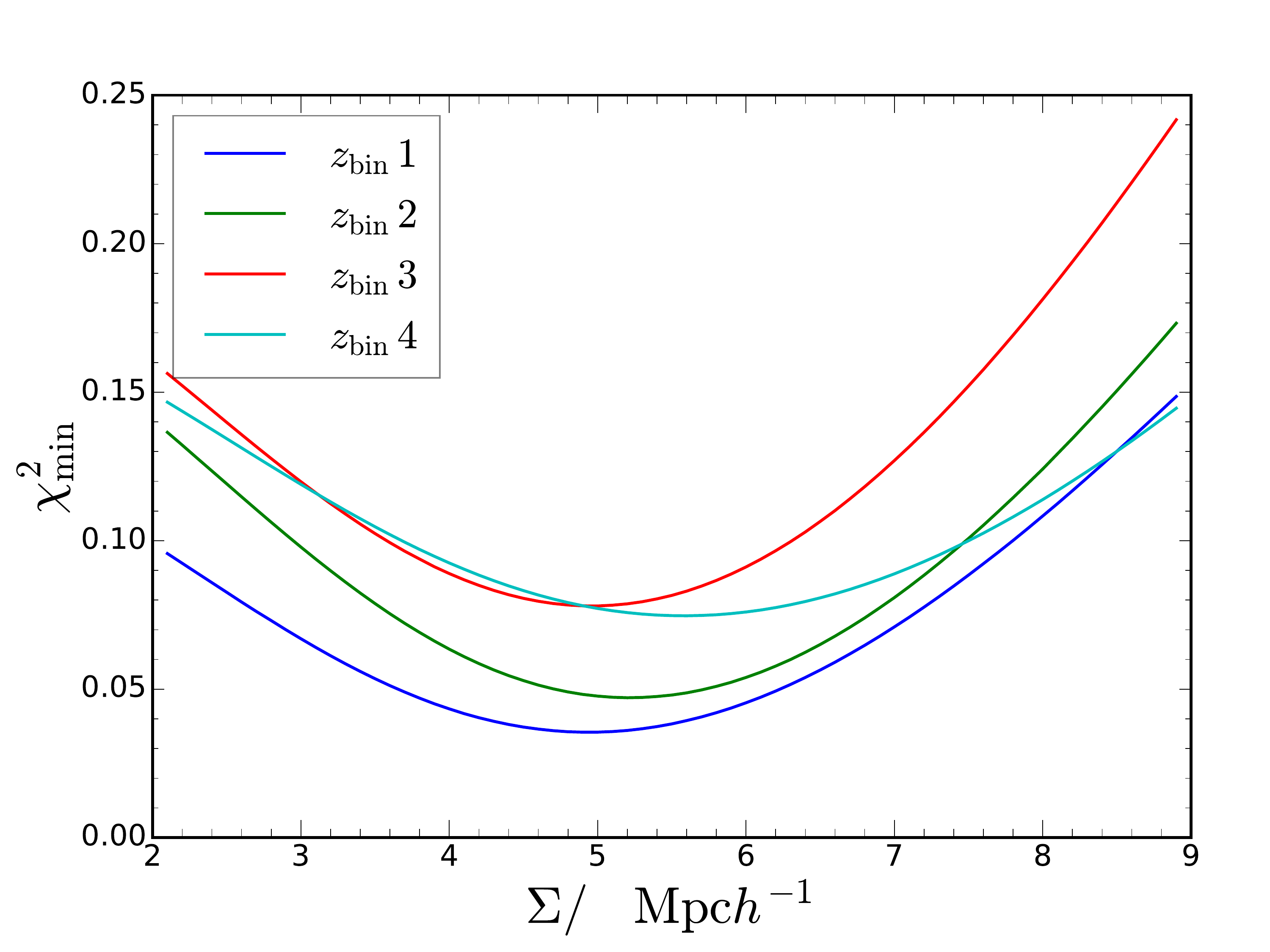}
\caption{ The minimum of the $\chi^2  $ against the BAO damping scale $\Sigma$ for the four redshift bins used.      } 
\label{fig:chi2min_dampingscale}
\end{center}
\end{figure}

To determine the correct physical damping scale, we fit the templates with only one fitting parameter $\Sigma$ to the mean $w$ of the mocks (i.e. $\alpha =1$, $B=1$, $A_i=0$).  We have considered four redshift bins and fitted to each redshift bin  separately. The minimum of the $\chi^2$  obtained with MLE is plotted against the damping scale in Fig.~\ref{fig:chi2min_dampingscale}. The best fit damping scale is in between 5 and 6 $\MpcOh$ across the redshift bins. However, we note that the highest redshift bin, bin 4, requires the largest damping. This is contrary to the expectation that the damping scale should decrease as redshift increases because the nonlinearity becomes weaker.  \change{  This is not due to the photo-$z$ distribution because of the following reason.   Although bin 4 has the largest photo-$z$ uncertainty, which can blur the BAO, it is taken into account in $ \phi(z) $ already and does not affect the more fundamental 3D damping scale. }  Throughout this work, we simply take the mean of the best fit of the 4 redshift bins, which is  $ \Sigma= 5.2 \MpcOh  $.  We have checked that the differences between using the mean damping scales and the individual best fit results in no detectable change in $ \alpha $.

\begin{figure}
\begin{center}
\includegraphics[width=0.45\textwidth]{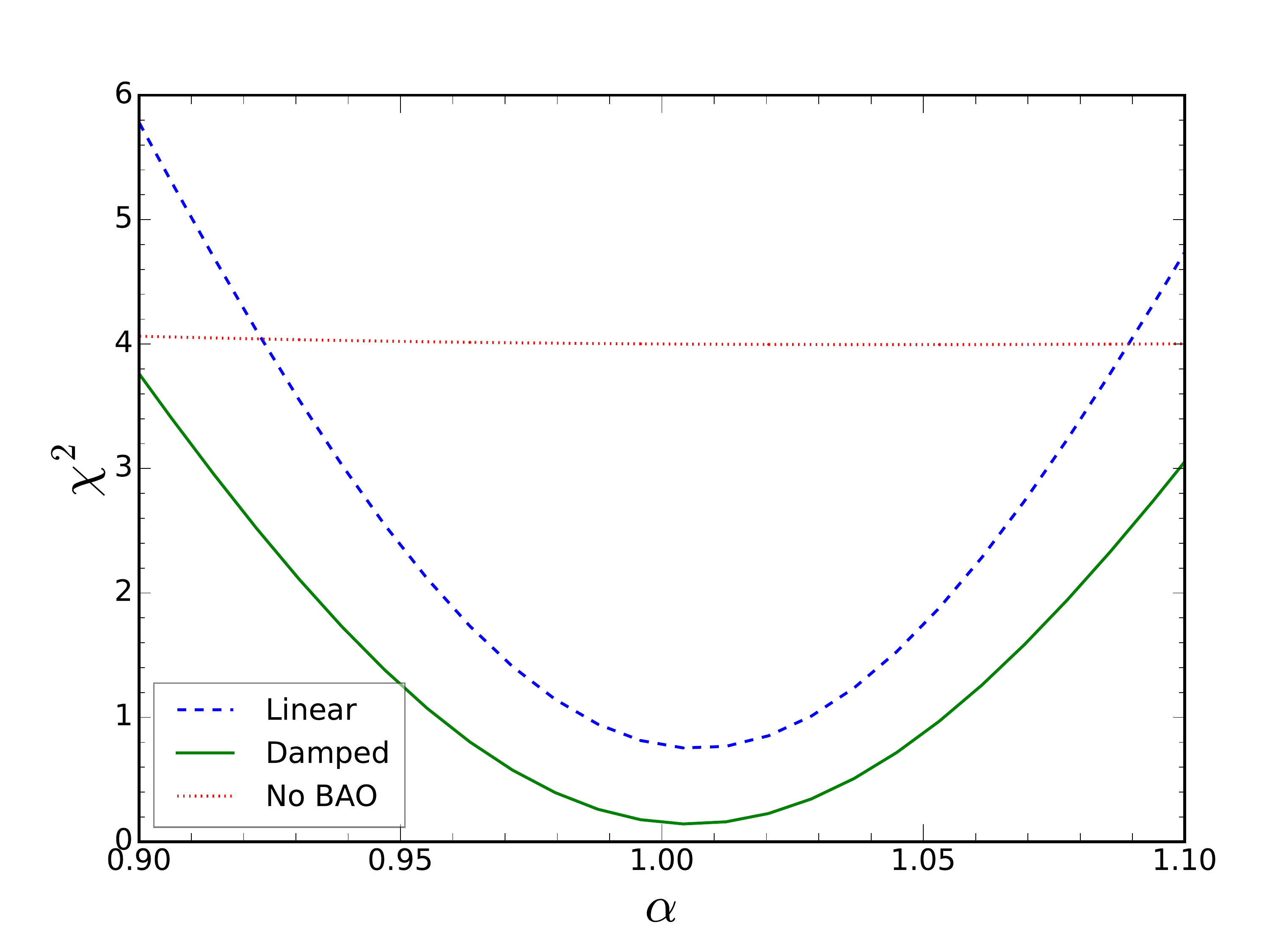}
\caption{   The $\chi^2 $ as a function of $\alpha $ when the template without damping (dashed, blue) and with the damping scale $\Sigma = 5.2 \MpcOh $ (solid, green) are used. The fit with a no-BAO template (dotted, red) is also shown.  } 
\label{fig:chi2_alpha_Lin_Damp_NoBAO}
\end{center}
\end{figure}

It is instructive to see how the fit is affected when the template without damping is used. To check this we compare the fits with $\Sigma=5.2 \MpcOh $ against the one without damping i.e.~$\Sigma=0 \MpcOh$. We jointly fit to the mean of the four bins and the results are shown in Fig.~\ref{fig:chi2_alpha_Lin_Damp_NoBAO}.  The linear case yields a larger overall value of $\chi^2  $ than the damped one. As a comparision, we also show the fit using a template without BAO, and it is clearly disfavoured compared to the BAO model. We also see that the $\chi^2 $ bound is narrower for the linear template, and hence using it we would get an artificially tighter bound on $\alpha $. On the other hand, the extreme case of no BAO, there is no bound on $\alpha$ at all.

\change{ In \citet{EisensteinSeoWhite2007} [see also \citet{SeoEisenstein_2007}], the BAO damping is modelled using the differential Lagrangian displacement field between two points at a separation of the sound horizon. Under the Zel'dovich approximation \citep{Zeldovich1970}, at $z_{\rm eff}=0.8$  the spherically averaged damping scale is  $ 5.35 \MpcOh $, which is close to our recovered value. Note that the damping scale is obtained by integrating over the linear power spectrum [Eq.~(9) in \citet{EisensteinSeoWhite2007}], and in Planck cosmology, we get $4.96 \MpcOh $ instead.  Thus in principle, we can allow the damping scale to be  a free parameter. In \citet{BAOmain}, we have tested the results obtained using different damping scales (2.6 and $7.8 \MpcOh$), consistent with the trend shown in Fig.~\ref{fig:chi2_alpha_Lin_Damp_NoBAO}, the damping scale has only small effect on the best fit, but a smaller damping scale results in a smaller error bar. Given the quality of the current data, we fix the BAO damping scale to be 5.2$\MpcOh$.  }


\subsection{ Incorrect fiducial cosmology  }
\label{sec:mismatch_cosmology}

The mock catalogs were created using the MICE cosmology. To test the cosmology dependence, we fit a template computed using the Planck cosmology \citep{Ade:2015xua}. The Planck cosmology should be closer to the current cosmology and it is used in the other analyses, such as the BOSS DR12 \citep{Ross:2016gvb}.    For the Planck cosmology, we use $\Omega_{\rm m} = 0.309$, $\sigma_8 =0.83 $, $n_{\rm s} =0.97 $, and  $h=0.676$.  From {\tt  camb} \citep{CAMB}, in the MICE cosmology, the sound horizon at the drag epoch is $153.4 \, \mathrm{Mpc} $, while in the Planck cosmology it is $147.8 \, \mathrm{Mpc} $. For the effective redshift $z_{\rm eff} = 0.8$,  $D_{\rm A}$ in MICE cosmology is larger than that in the Planck one by 3.1\%.  Thus we expect the BAO in the Planck cosmology to be smaller than that in the MICE by 4.1\% in angular scale.

\begin{figure}
\begin{center}
\includegraphics[width=\linewidth]{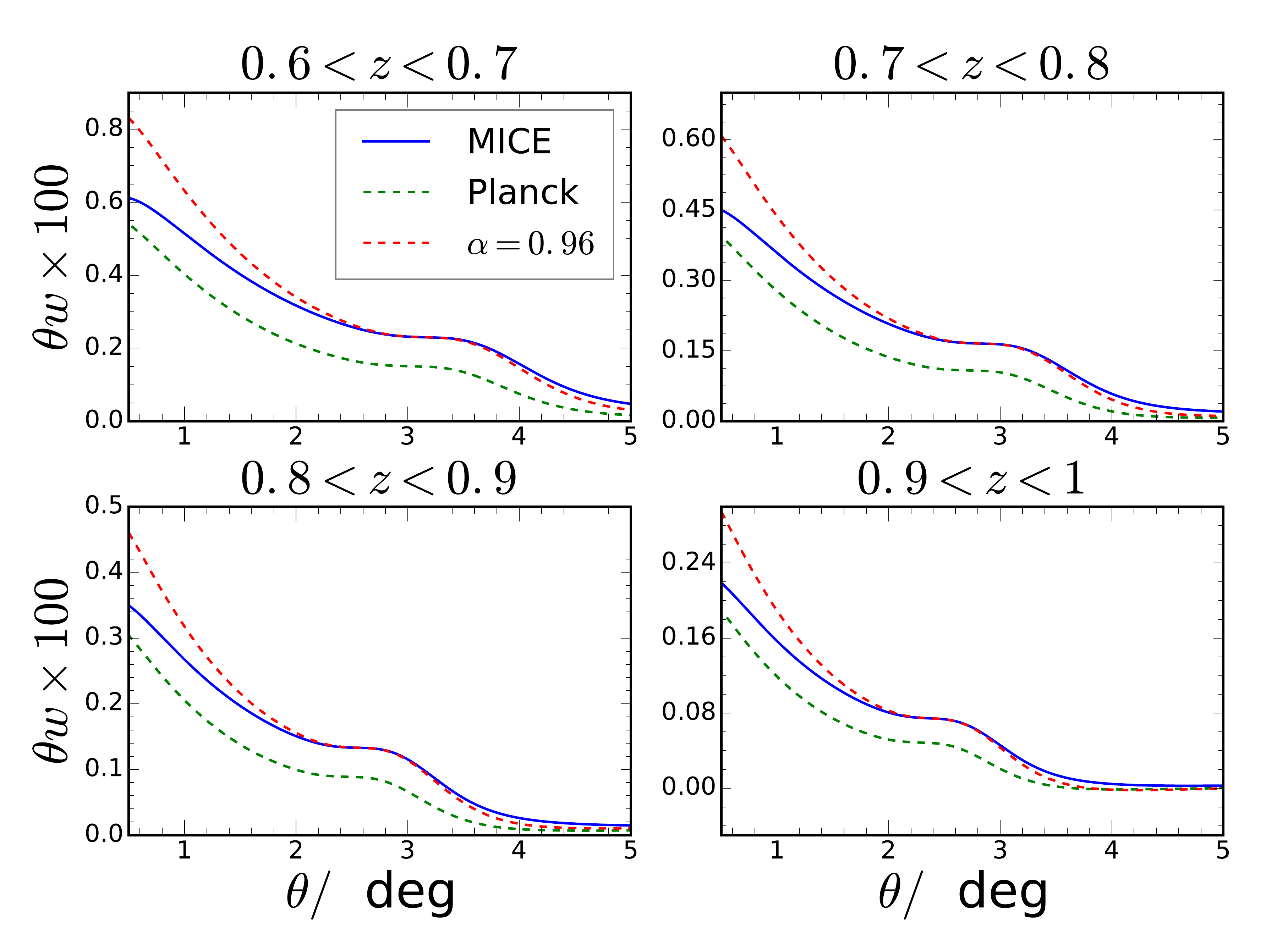}
\caption{ The angular correlation obtained using the MICE (solid, blue) and Planck (dashed, green) cosmology. The Planck correlation rescaled by  $\alpha$ = 0.96 (red)  and shifted in amplitude to match the MICE one.  }
\label{fig:BAOscale_templates}
\end{center}
\end{figure}

In Fig.~\ref{fig:BAOscale_templates}, we plot the matter angular correlation function obtained using the MICE and Planck cosmology. To check how well the rescaling parameter $\alpha $ works, we rescale the angular correlation by $ w ( \alpha \theta)$ and then match the amplitude at the BAO scale with that of the MICE one. We find that the rescaling results in a good match with the MICE one around the BAO scale. However, away from the BAO scale, the disagreement with the MICE template increases. Since we use the angular scale in the range of $[0.5^\circ,5^\circ]$ in the fitting, it is not clear if we can recover the true cosmology.  

In Table \ref{tab:MLE_MCMC_MICE_Planck}, we compare the fit obtained using the MICE template with that from the Planck cosmology one. We have displayed the fits using the mean of the mocks and the individual mocks. As the mean result is obtained by averaging over 1800 realizations, the covariance is reduced by a factor of 1800.  We have shown four decimal places as the error bars are small for the mean fit. In this very high signal-to-noise setting, there is still $\bar{\alpha} - 1 \sim  0.004$ bias in the best fit for all these estimators. It is much larger than the estimated error bar $ \sigma_\alpha \sim 0.0013$, thus the bias is systematic.  This can arise from the nonlinear effects such as the dark matter nonlinearity and galaxy bias \citep{CrocceScoccimarro_2008,PadmanabhanWhite_2009}.  When the Planck template is used, the best fit $\alpha$ is about  0.9674. Thus the ratio of the dilation parameter is  $\bar{\alpha}_{\rm Planck} / \bar{\alpha}_{\rm MICE} = 0.9634 $ and it is still smaller than the expectation 0.959 by $\sim 3 \sigma_\alpha$. Of course this high signal-to-noise case occurs if we have 1800 times the DES Y1 volume. For the present DES volume, it is still statistical error dominated as can be seen below.

Now we turn to the distribution of $\bar{\alpha }$. Since the expected  $ \bar{\alpha} \approx 0.96 $ is  not symmetric about the interval [0.8,1.2] for the case of Planck cosmology. This asymmetry can bias the distribution of $\bar{ \alpha } $. To prevent this bias, we use the detection criterion that the 1-$\sigma$ error bounds fall within the boundary [0.76, 1.16] for the case of Planck cosmology. For MCMC, the prior on $\alpha $ is changed to [0.56,1.36], although this results in no detectable significance compared to the fiducial choice [0.6,1.4].   $ \langle \bar{\alpha}  \rangle  $ shows larger variations than the mean fit case because the covariance is larger by a factor of 1800.    The ratio of the dilation parameter  $ \langle \bar{\alpha}_{\rm Planck} \rangle / \langle \bar{\alpha}_{\rm MICE} \rangle $ are 0.964, 0.965, and 0.967 for MLE, PL, and MCMC respectively. The ratio is close to what we get for the mean fit. Note that if we keep the boundary of $\alpha$ to be [0.8,1.2] for the Planck case, we get  $\langle  \bar{\alpha} \rangle = 0.970$, 0.976, and 0.979, resulting in larger difference from the mean fit results.  It is worth mentioning that this adjustment of the interval only affacts the mocks with more extreme values of the best fit.

The small difference between the fit results and  the estimation from Eq.~\eqref{eq:theta_BAO} could be because we have used the information not just about the BAO scale, but also the shape of the correlation around it.  This is related to the polynomial $A_i$ used to absorb the broad band power dependence.  We have checked that the bias in the fit reduces with the order of polynomial in $ A_i $. For example, for the case of mean fit, for $N_p=1 $, 2, and 3, $\bar{\alpha} - 1 $ are  $-0.095$, $-0.054$, and $-0.034 $ respectively.  For large range in the angles, the polynomial cannot perfectly remove the dependence. \change{  However for $N_p =3 $ the bias in the best fit $\bar{\alpha}$ is reduced to a basically negligible level given the signal-to-noise of our data, suggesting our results should be robust at least for true cosmologies between MICE and Planck. Of course one can use a template that is closer to the currently accepted cosmology, although at the risk of the confirmation bias.  }


\begin{table*}
\caption{The BAO fit obtained with MLE, PL, and MCMC for the MICE and Planck cosmology. The fits to the mean of the mocks and the individual mocks are shown.  }
\label{tab:MLE_MCMC_MICE_Planck}
  \begin{tabular}{|l|c|c|c|c|c|c|}
    \hline
    \multirow{2}{*}{} &
      \multicolumn{3}{c}{MICE} &
      \multicolumn{3}{c}{Planck} \\
      \hline 
    & best fit to &  $\langle {\bar \alpha} \rangle \pm {\rm std}({\bar \alpha})$  & frac. with ${\bar \alpha} \pm \sigma_\alpha$ & best fit to & $\langle {\bar \alpha} \rangle \pm {\rm std}({\bar \alpha})$ & frac. with ${\bar \alpha} \pm \sigma_\alpha$ \\
    & mean of mocks & (all mocks) & enclosing $\langle \bar{\alpha} \rangle $ & mean of mocks & (all mocks) & enclosing $\langle \bar{\alpha} \rangle $ \\
    \hline
MLE           & $1.0043 \pm 0.0013$    & $1.001 \pm 0.052 $   & 0.69   &  $ 0.9675 \pm 0.0014$ &   $0.965 \pm 0.050 $    &   0.70    \\
PL            & $1.0043 \pm 0.0013$    & $1.004 \pm 0.049 $   & 0.77   &  $ 0.9673 \pm 0.0014$ &   $0.969 \pm 0.048 $    &   0.78    \\
MCMC          & $1.0030 \pm 0.0012$    & $1.007 \pm 0.049 $   & 0.74   &  $ 0.9674 \pm 0.0014$ &   $0.974 \pm 0.047 $    &   0.75    \\   \hline
  \end{tabular}
\end{table*}

\subsection{ Photo-$z$ error} 

Suppose that the true photo-$z$ distribution is $\phi_0 $, but because of some photo-$z$ error $\delta \phi$,  which can arise from systematic errors in the photo-$z$ calibration, the  photo-$z$ distribution  $\phi = \phi_0 + \delta \phi$ is used.  In this section, we test how the photo-$z$ error affects the BAO fit. 

We will investigate two types of photo-$z$ errors in the templates: in one case the means of the photo-$z$ distributions are systematically shifted by 3\% in each of the four tomographic redshift bins, for the second case the standard deviations of the photo-$z$ distributions are increased  by 20\%.  While the typical relative error in the mean of the redshift distributions for the DES Y1 BAO sample is about 1\% \citep{Photoz}, 3\% is used to increase the signal-to-noise of our systematic test. The 20\% error in the spread is the upper bound of the relative error in the width of those redshift distributions \citep{Photoz}.

\begin{figure*}
\begin{center}
\includegraphics[width=0.9\linewidth]{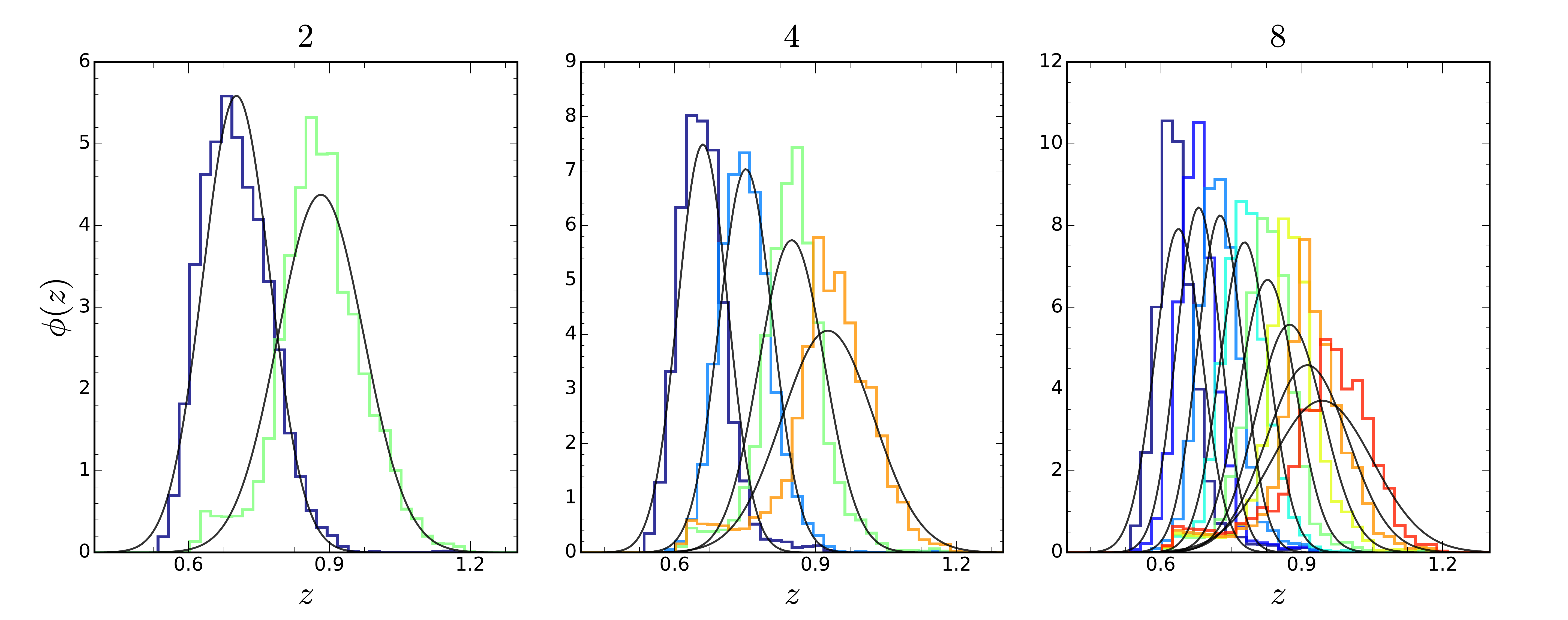}
\caption{ The photo-$z$ distribution from the mock catalog as a function of the true redshift $z$  for a number of redshift bins $N_z=2$, 4, and 8 bins respectively.   The solid black lines show the Gaussian distribution with the same mean and variance as the photo-$z$ distribution.    }
\label{fig:phi_248_compare}
\end{center}
\end{figure*}

As we can see from Fig.~\ref{fig:phi_248_compare}, the photo-$z$ distribution is  approximately Gaussian.   If $\phi_0$ is Gaussian, we parametrize  $ \delta \phi$ as variations in the mean $\mu$ and variance $\sigma^2$ by an amount of  $\delta \mu$  and $\delta \sigma^2 $   as
\beq
\label{eq:deltaphi_Gaussian}
\delta \phi(z) = \phi_0(z)  \bigg[ \frac{z - \mu}{ \sigma^2 } \delta \mu + \bigg(  \frac{(z-\mu)^2  }{ \sigma^2 } - 1   \bigg) \frac{ \delta \sigma  }{ \sigma }    \bigg]. 
\eeq
Note that $\int dz \, \delta \phi(z) = 0 $, thus  $\phi_0 + \delta \phi $ is properly normalized. However,  $\phi_0  +  \delta \phi $ is not always positive in the whole range of $z$, so it is not really a legitimate probability distribution. Nonetheless, since the variations we consider are small, the negative part is small. Furthermore, the  $\delta \mu $ term, which is odd about $\mu$, causes skewness about the mean in $ \phi + \delta \phi $. All these suggest that $\phi_0 + \delta \phi$ may not be close to Gaussian any more. If so the parameter $\delta \mu $ ($\delta \sigma $) may not be the variation of the mean (standard deviation) of $\phi_0 + \delta \phi $ relative to that of $\phi_0$.  Indeed, we find that the output variation (by direct computation) from Eq.~\eqref{eq:deltaphi_Gaussian} does not match that from the input ($\delta \mu $ and $\delta \sigma $).  Thus for the case of mean shift, we will simply translate the distribution by certain amount instead of using Eq.~\eqref{eq:deltaphi_Gaussian}. For the variation of the standard deviation, we adjust the value of $ \delta \sigma$  in  Eq.~\eqref{eq:deltaphi_Gaussian} to get the desired variation.

Because the photo-$z$ error causes variation in the mean redshift of the slice and hence $D_{\rm A}$, from Eq.~\eqref{eq:theta_BAO}, we can get a simple estimate for the shift of $\theta_{\rm BAO} $ due to a shift in $z$ by an amount of $\delta \mu $
 \beq
 \label{eq:thetaBAO_variation}
 \frac{\delta \theta_{\rm BAO} }{ \theta_{\rm BAO} } =  - \frac{ c \, \delta \mu  }{ H(z) D_{\rm A}(z)  } .
\eeq
At the effective redshift $z_{\rm eff} = 0.8 $, when $\delta \mu $ increases by 3\%,  $ \delta \theta_{\rm BAO}  /  \theta_{\rm BAO}   $ changes by $-2.4\%$.  \change{ This simple argument suggests that the leading effect comes from the systematic shift in the mean, while the variation in the standard deviation is  symmetric about the mean to the lowest order and does not shift the BAO.  } 


In Table \ref{tab:MLE_MCMC_photo_z_error}, we compare the fit results obtained using the fiducial template, the template with 3\% increase in the mean of $\phi(z)$, and the one with 20\% increase in $\sigma$.  For the fit to the mean, there is a shift of $-0.0254$, $-0.0251$, and $-0.0239$ for MLE, PL, and MCMC respectively when the mean of $\phi$ is systematically shifted by 3\%.   When  $\phi$ is systematically shifted by 3\%, we use the interval [0.776, 1.176] for $\alpha$ instead.  The shift for  $\langle \bar{ \alpha } \rangle $ are $ -0.024 $  relative to the fiducial case for all the estimators.  On the other hand, when the standard deviation of the distributions are increased by 20\%, there is no significant  systematic trend for the best fits. Although not shown explicitly here, there is about 2\% increase in $\langle \sigma_\alpha  \rangle$ for all the  estimators, signalling that the error indeed increases.

 Good agreement between the full fit results and the estimation by Eq.~\eqref{eq:thetaBAO_variation} validates the simple argument.  This is useful because it provides a convenient estimate for the accuracy of the photo-$z$ required for the BAO fit.  For example, from the redshift validation \citet{Photoz}, the mean of the photo-$z$ error after sample variance correction is about 1\%, thus the shift in the BAO position is about 0.8\%.  This is still marginal compared to other potential systematic shifts and is less than 20\% of the statistical uncertainty. For DES Year 3, the amount of data is expected to increase roughly by a factor of 3,  and so the error on $\alpha $ is expected to reduce by almost a factor of $\sqrt{3}$. Hence, the photo-$z$ uncertainty is still expected to be subdominant for Year 3.  \change{ The shift in the mean of the photo-$z$ distribution has similar effects as the parameter $\alpha$, and hence they would be degenerate with each other. It is important that the photo-$z$ distribution is calibrated in an independent means e.g.~using the external spectroscopic data.   } 
 
\begin{table*}
\caption{The BAO fit using the fiducial photo-$z$ distribution, the distribution with mean shift by 3\%, and the distribution with standard deviation increased  by 20\%.  The fits to the mean of the mocks and the individual mocks are shown.}
\label{tab:MLE_MCMC_photo_z_error}
  \begin{tabular}{|l|c|c|c|c|c|c|}
    \hline
    \multirow{2}{*}{} &
      \multicolumn{2}{c}{Fiducial $\phi(z)$} &
      \multicolumn{2}{c}{3\% increase in the  mean of $\phi(z)$} &
      \multicolumn{2}{c}{20\% increase in the std of $\phi(z)$} \\
      \hline 
    & best fit to & $\langle {\bar \alpha} \rangle \pm {\rm std}({\bar \alpha})$ & best fit to & $\langle {\bar \alpha} \rangle \pm {\rm std}({\bar \alpha})$ & best fit to & $\langle {\bar \alpha} \rangle \pm {\rm std}({\bar \alpha})$ \\
    & \ \ mean of mocks & \ \ (all mocks) & \ \ mean of mocks & \ \ (all mocks) & \ \ mean of mocks & \ \ (all mocks) \\
\hline
MLE           & $1.0043 \pm 0.0013$    & $1.001 \pm 0.052 $   & $0.9789\pm 0.0014 $  &  $0.977 \pm 0.051$ &   $1.0043 \pm 0.0012 $  &  $1.001 \pm 0.051 $    \\
LP            & $1.0043 \pm 0.0013$    & $1.004 \pm 0.049 $   & $0.9792\pm 0.0013 $  &  $0.980 \pm 0.049$ &   $1.0043 \pm 0.0013 $  &  $1.004 \pm 0.049 $    \\
MCMC          & $1.0030 \pm 0.0012$    & $1.007 \pm 0.049 $   & $0.9791\pm 0.0013 $  &  $0.983 \pm 0.049$ &   $1.0042 \pm 0.0013 $  &  $1.007 \pm 0.049 $    \\
\hline
\end{tabular}
\end{table*}


\section{Optimizing the analysis}
\label{sec:optimizing_analysis}

To derive the fit, an accurate precision matrix, i.e.~the inverse of the covariance matrix, is necessary. In \cite{BAOmain}, the workhorse covariance is derived from the mock catalogs. To get an accurate precision matrix from the mocks, we want the dimension of the covariance matrix to be small relative to the number of mocks \citep{Anderson_Statistics,Hartlap:2006kj}. Hence it is desirable to reduce the number of data bins while preserving the information content.  To this end, we consider optimizing of the number of angular bins, the number of redshift bins, and the inclusion of cross redshift bins.

\subsection{ The angular bin width}
\label{sec:AngularBinWidthOpt}

We use the following method to check the dependence of the BAO fit results on the angular bin size.  We generate a theory template with fine angular bin width ($0.01^\circ$ here). A mock data vector with coarser bin width is created by bin averaging the theory template over the bin width.  The mock data vector is then fitted using the fine template. In the analysis, we use the Gaussian covariance Eq.~\eqref{eq:Gaussian_cov_auto}.  As we see in Eq.~\eqref{eq:lnL_halfdeviation}, for the error analysis, the magnitude of the $ \chi^2 $ does not matter, and only the deviation from the best fit does.   This method  enables us to explore the likelihood  about its maximum, and hence derive the strength of the constraint. \change{ It is similar to the often used Fisher forecast \citep{Tegmark:1996bz, Dodelson_2003}. }

\begin{figure}
\begin{center}
\includegraphics[width=0.9\linewidth]{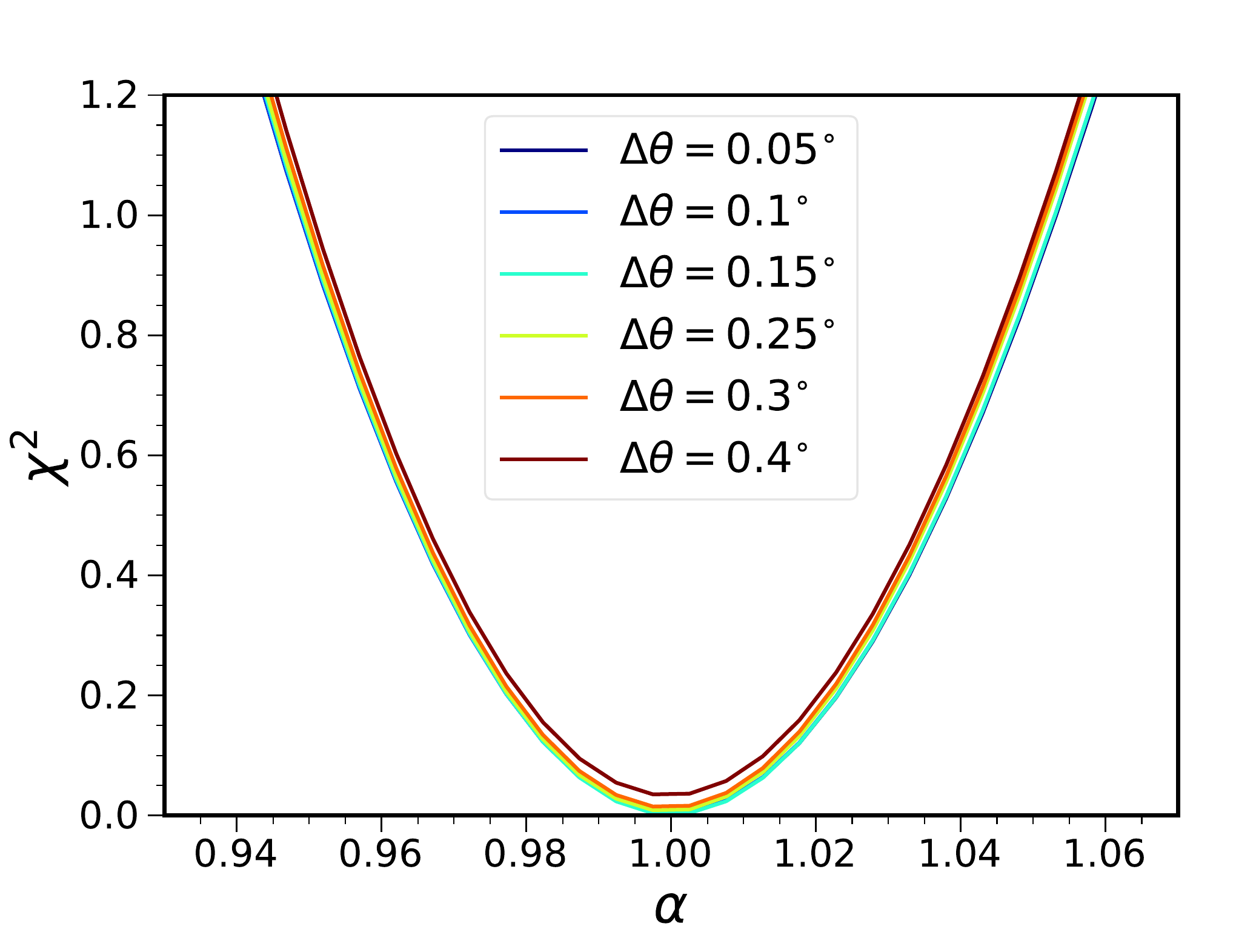}
\caption{   $ \chi^2$ as a function of $ \alpha$  for different angular bin widths.  The models with  different bin width $\Delta\theta$ yield similar results.     }
\label{fig:chi2_alpha_thetabin_NAP4}
\end{center}
\end{figure}

In Fig.~\ref{fig:chi2_alpha_thetabin_NAP4}, we show the distribution of $\chi^2 $ as a function of the dilation parameter $\alpha$ for a number of angular bin widths.  In this ideal noiseless setting, the best fit can fit the mock data extremely well, as manifested with $\chi^2 \approx 0 $ at $ \alpha =1$. The $\Delta \chi^2= 1 $ rule can give the 1-$\sigma$ constraint on $\alpha$.  
\change{ For the angular bin width $\Delta \theta = 0.05^{\circ}$, $0.1^{\circ}$,   $0.15^{\circ}$,   $0.25^{\circ}$,   $0.3^{\circ}$, and   $0.4^{\circ}$, the 1-$\sigma$ error bars are  0.0522, 0.0521, 0.0520, 0.0514, 0.0509, and 0.0503 respectively. }  The mock data with coarser bin width yields slightly smaller 1-$\sigma$ error bar  because they give a slightly less precise representation of the underlying model. 
This is often the case when a poor model is used, it gives larger $\chi^2$ (the definition of a poor model) and an artificially stringent constraint.

For BAO fitting, it is preferable to have the bin width to be fine enough so that there are a few data points in the BAO dip-peak range to delineate the BAO feature. Thus for the rest of the study, we shall stick to  $\Delta \theta =0.15 ^{\circ} $. In \cite{BAOmain}, the covariance is derived from the mock catalogs. As we find that the differences between different bin widths results are negligible,  to reduce the size of the data vector, $\Delta \theta = 0.3^{\circ } $ was adopted in \cite{BAOmain}.

\begin{table*}
  \caption{    The BAO fit obtained using varying number of redshift bins: 4 bins, 3 bins, and only 2 bins. The results from MLE, PL and MCMC are shown. }
\label{tab:number_redshiftbins}
  \begin{tabular}{|l|c|c|c|c|c|c|c|c|c|}
    \hline
    \multirow{2}{*}{} &
      \multicolumn{3}{c}{bins 1, 2, 3, \& 4} &
      \multicolumn{3}{c}{bins 2, 3, \& 4} &
      \multicolumn{3}{c}{bins 3 \& 4} \\
      \hline 
    & $\langle {\bar \alpha} \rangle \pm {\rm std}({\bar \alpha})$ & frac.~${\bar \alpha} \pm \sigma_\alpha$ &     $\langle  \sigma_\alpha \rangle   $   & $\langle {\bar \alpha} \rangle \pm {\rm std}({\bar \alpha})$ & fract.~${\bar \alpha} \pm \sigma_\alpha$ &  $\langle  \sigma_\alpha \rangle $   & $\langle {\bar \alpha} \rangle \pm {\rm std}({\bar \alpha})$ & frac.~${\bar \alpha} \pm \sigma_\alpha$ &  $\langle  \sigma_\alpha \rangle $ \\
    & &   in [0.8,1.2] & $ /\mathrm{std}( \bar{\alpha} )$  &   &  in [0.8,1.2] &  $ / \mathrm{std}( \bar{\alpha} ) $    &   &  in [0.8,1.2] & $ / \mathrm{std}( \bar{\alpha} ) $   \\
    \hline
MLE        & $1.001 \pm 0.052$    & 0.91  & 1.02   &  $ 1.001 \pm 0.053 $   &  0.87  & 1.06   &  $ 0.998 \pm 0.058$   &   0.79   &  1.11   \\
PL         & $1.004 \pm 0.049$    & 0.99  & 1.27   &  $ 1.005 \pm 0.051 $   &  0.99  & 1.32   &  $ 1.003 \pm 0.052$   &   0.99   & 1.51    \\
MCMC       & $1.007 \pm 0.049$    & 0.84  & 1.17   &  $ 1.006 \pm 0.051 $   &  0.75  & 1.22   &  $ 1.004 \pm 0.054$   &   0.63   & 1.41    \\
   \hline
  \end{tabular}
\end{table*}

\subsection{Number of photo-$z$ bins}

For a sample that relies on photo-$z$s, there can be large overlaps among the photo-$z$ distributions from different redshift bins, and thus substantial covariance between different redshift bins. 
Here we test how the constraints depend on the number of redshift bins.

Based on the assigned photo-$z$ in the mock, we divide the samples in the redshift range [0.6,1] into $N_z$ redshift bins, with equal width in $z$. For example, we have shown the photo-$z$ distribution  $\phi (z) \equiv P(z | z_{\rm photo})$ for  $N_z =2$, 4, and 8. We see that there is indeed large overlap in the photo-$z$ distribution.  We also plotted the Gaussian distribution with the same mean and variance as the photo-$z$ distribution. The photo-$z$ distribution is moderately Gaussian and the deviation from Gaussianity increases with $z$. In fact, \citet{Halogen} find that a double Gaussian distribution offers a better fit to $\phi $.

\begin{figure}
\begin{center}
\includegraphics[width=0.9\linewidth]{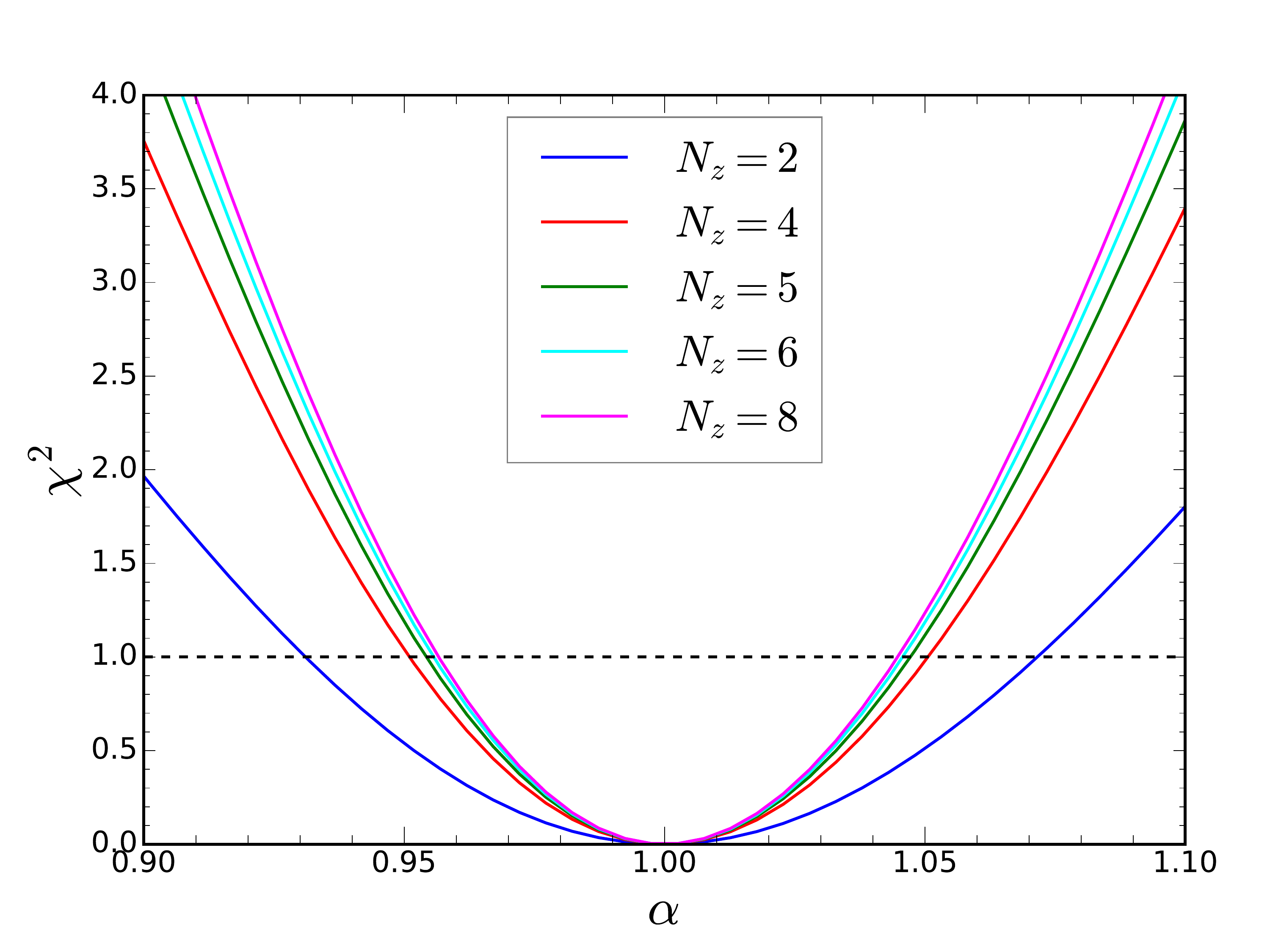}
\caption{ The $\chi^2 $ as a function of $\alpha $ for different number of redshift bins $N_z=2$, 4, 5, 6, and 8  respectively. The black dash line indicates the $\Delta \chi^2 =1$ threshold.   }
\label{fig:chi2_redshiftbincheck}
\end{center}
\end{figure}

 Similar to that in  Sec.~\ref{sec:AngularBinWidthOpt}, using the photo-$z$ distribution shown in Fig.~\ref{fig:phi_248_compare}, we generate templates with fine bin width ($\Delta \theta = 0.01^{\circ} $),  data vectors with coarser bin width ($\Delta \theta = 0.15^{\circ} $), and the Gaussian covariances. \change{ The template needs to smoothly represent the theory, and $\Delta \theta = 0.01^{\circ}  $ is sufficient. } The $\chi^2 $ as a function of $\alpha $ for different $N_z$ are displayed in Fig.~\ref{fig:chi2_redshiftbincheck}. We see that for $N_z =2 $, the data is undersampled and the 1-$\sigma$ error bar on $\alpha$ is weak (at $\Delta \chi^2 =1$, it is 0.07 ).  When $N_z$ is increased to 4, the error bar is tightened to 0.05 at $\Delta \chi^2 =1 $. Further increasing the number of redshift bins, there is little gain, and the change in 1-$\sigma$  error bar is less than $0.005 $ for $N_z = 8$. To strike a balance between retaining as much information as possible and keeping the size of the data vector small, we use $N_z = 4$ for the rest of the work and in \cite{BAOmain}.

\change{  In  \citet{BAOmain}, different combinations of redshift bins of width 0.1 in the range [0.6-1] are used to do the BAO fit.  In Table \ref{tab:number_redshiftbins} we examine how the results vary with the number of redshift bins used. We consider the fit using 4 redshift bins (bin 1, 2, 3, and 4), 3 redshift bins (bin 2, 3, and 4), and 2 redshift bins (bin 3 and 4). Recall that the redshift bins are in ascending order, e.g.~bin 1 refers to [0.6,0.7], etc. When the number of redshift bins is reduced, the constraining power of the data is expected to decrease.  The fraction of mocks with 1-$\sigma $ error bar in the range [0.8,1.2] is expected to decrease with the number of redshift bins. The trends for both the MLE and MCMC are consistent with this expectation. PL does not show any clear changes mainly because the mean and error of PL can be defined within any range, which is taken to be [0.8,1.2] here.  The mean of the best fit $\langle \bar{ \alpha} \rangle $ is essentially unchanged, but the spread $ \mathrm{std}( \bar{\alpha} ) $  increases mildly by 12\%, 6\%, and 10\% for MLE, PL, and MCMC respectively.  On the other hand, the error derived from the each realization show much larger increase. We find that $ \langle \sigma_\alpha \rangle $ increases by 21\%, 35\%, and 33\% for MLE, PL, and MCMC respectively.  We also note that $\langle \sigma_\alpha  \rangle  / \mathrm{std} ( \bar{\alpha} ) $ is closest to 1 among these fitting methods.   }

\subsection{Adding cross-correlations }

So far we considered only the auto correlation in redshift bin $w_{ii}$.  Here we test the gain on the constraint in $\alpha$ when the cross correlations among different redshift bins are included.

\begin{figure}
\begin{center}
\includegraphics[width=\linewidth]{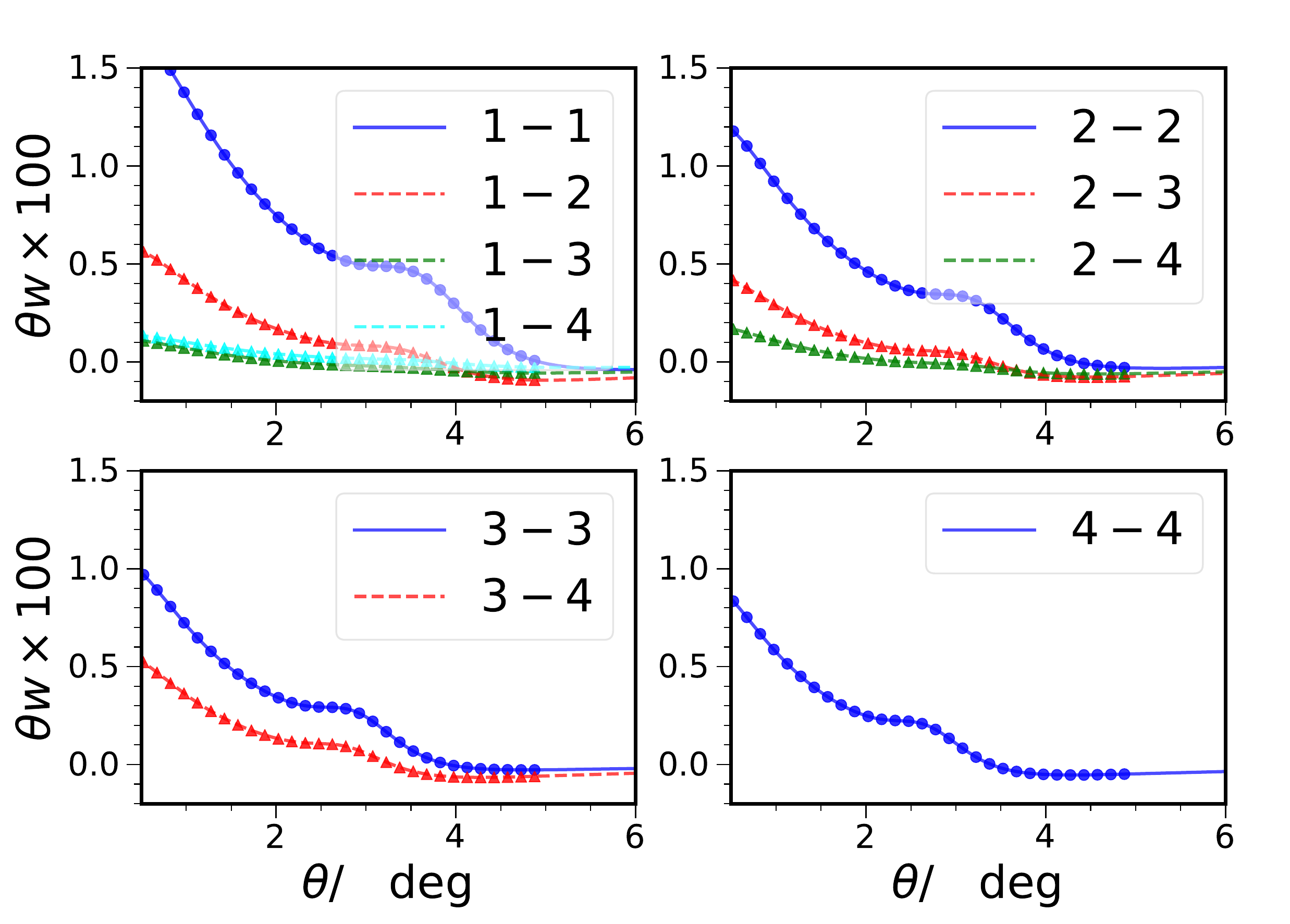}
\caption{ The auto (solid, circles, blue)  and cross correlation (dashed, trangles, other colors) between different redshift bins. } 
\label{fig:w_crossz_template_4zbins}
\end{center}
\end{figure}

 In Fig.~\ref{fig:w_crossz_template_4zbins}, the auto and cross correlation function are plotted.  We see that only the cross correlation between the adjacent redshift bins show any BAO signal, while the redshift bins that are further apart have little signal in their cross-correlation. Thus the BAO constraint improvement if any,  can only come from the cross-correlation of adjacent redshift bins. 

\begin{figure}
\begin{center}
\includegraphics[width=0.9\linewidth]{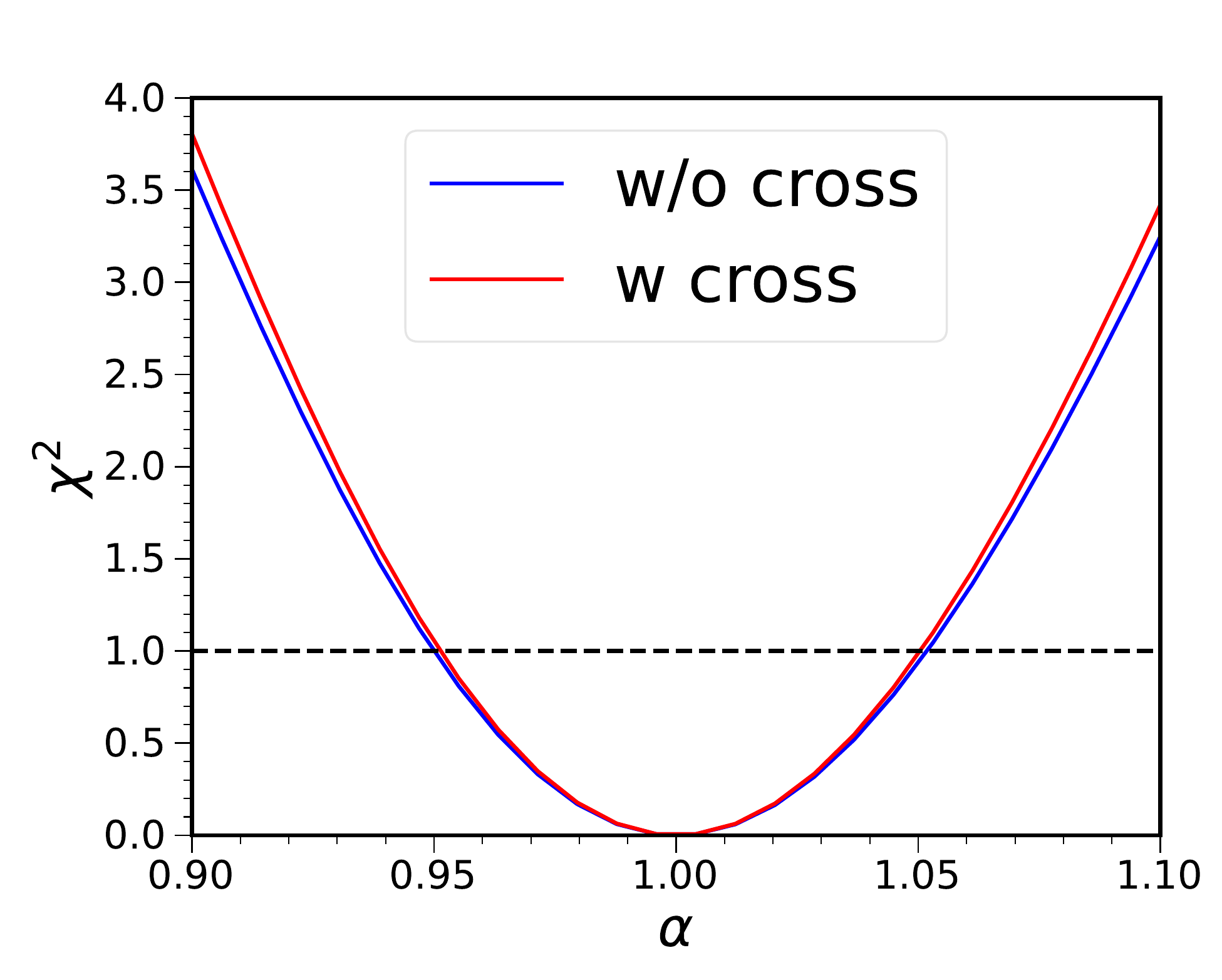}
\caption{ The $\chi^2$ for the constraint on $\alpha $ for only using the auto correlation (blue) and including also the cross correlations (red). } 
\label{fig:chi2_bestfit_zcross}
\end{center}
\end{figure}

In Fig.~\ref{fig:chi2_bestfit_zcross}, the $\chi^2$ fit  using only the auto correlation and when the cross correlations are included are compared. In the fit, for each of the $w_{ij}$, a new set of parameters $B$ and $A_a$ are introduced. Hence for our case, there are altogether $ 10 \times 4 + 1 = 41 $ free parameters. We find that the gain in the BAO constraint when the cross redshift bin correlations are included is very limited. In this estimate, Gaussian covariance is used. As it is noiseless, there is no problem of covariance matrix inversion. In practice, when the covariance is estimated from the mocks, we need to reduce the dimension of the covariance matrix.  Of course, almost all the improvement on the  $\alpha$ constraint is expect to come from the adjacent redshift bins (although this is not shown explicitly here), so in any real analysis we should include only the nearest redshift bins.  Still the dimension of the covariance matrix is increased from $4 N_{\theta}$ to $7 N_{\theta}$, where $N_{\theta}$ is the number of angular bin for each redshift, when the cross-correlations between adjacent redshift bins are included. Thus we recommend that only the auto correlations be used.

\subsection{ Testing the order of the polynomial}

\change{ As we mentioned after Eq.~\eqref{eq:BAO_template}, we test  the number of free parameters $A_i$ in the template.  In Table \ref{tab::number_free_parameters}, we show the fit results for $N_p=1$, 2, and 3 in Eq.~\eqref{eq:BAO_template}. From the limited data set it is hard to draw a solid conclusion. Overall there is no clear systematic trend with $N_p $ for these three methods, thus it shows that the fiducial order of polynomial adopted although somewhat arbitrary, it does not lead to systematics bias in the analysis. Assuringly, MLE with $N_p = 3$, which is the workhorse model adopted in \cite{BAOmain}, 
is the method that performs best in terms of being largely unbiased and recovering close to 68\% of the true answer (i.e.~the Gaussian expectation). }

\begin{table*}
\caption{The BAO fit with different number $N_p $ (=1, 2, 3) in the template Eq.~\eqref{eq:BAO_template}. The results for MLE, PL, and MCMC are compared. }
\label{tab::number_free_parameters}
  \begin{tabular}{|l|c|c|c|c|c|c|}
    \hline
    \multirow{2}{*}{} &
      \multicolumn{2}{c}{$N_p=1$} &
      \multicolumn{2}{c}{$N_p=2$} &
      \multicolumn{2}{c}{$N_p=3$} \\
      \hline 
    & $\langle {\bar \alpha} \rangle \pm {\rm std}({\bar \alpha})$ & frac. with ${\bar \alpha} \pm \sigma_\alpha$  & $\langle {\bar \alpha} \rangle \pm {\rm std}({\bar \alpha})$ & frac. with ${\bar \alpha} \pm \sigma_\alpha$ & $\langle {\bar \alpha} \rangle \pm {\rm std}({\bar \alpha})$ & frac. with ${\bar \alpha} \pm \sigma_\alpha$ \\
    & & enclosing $\langle \bar{\alpha} \rangle $  & & enclosing $ \langle \bar{\alpha} \rangle $ &  & enclosing $ \langle \bar{\alpha} \rangle $ \\
    \hline
MLE           & $0.996 \pm 0.047$    & 0.72   &  $ 0.995 \pm 0.051 $   &  0.68  &  $ 1.001 \pm 0.052$   &   0.69    \\
PL   & $0.993 \pm 0.046$    & 0.77   &  $ 1.000 \pm 0.050 $   &  0.79  &  $ 1.004 \pm 0.049$   &   0.77    \\
MCMC   & $1.005 \pm 0.046$    & 0.75   &  $ 0.994 \pm 0.050 $   &  0.71  &  $ 1.007 \pm 0.049$   &   0.74    \\
   \hline
  \end{tabular}
\end{table*}

\section{ Covariance}
\label{sec:Covariance} 

In this section, we consider the issues of covariance in more detail. We first study how the Gaussian covariance impacts the BAO fit, and then investigate how to improve the covariance derived from  a set of mock catalogs.  One of the potential issues that arises in the survey analysis pipeline is that sample properties such as the $n(z)$ and galaxy bias could be different from those used to create the mocks. We show that the correction due to these property changes can be mitigated with the help of the Gaussian covariance. We also investigate expanding the covariance matrix and precision matrix using the eigenmodes from some proxy covariance matrix. We demonstrate that this approach can substantially reduce the influence of the noise because the number of free parameters are significantly reduced, and that it can effectively mimic the effect of small changes in the sample properties.


\subsection{ BAO fit with the Gaussian covariance  }

\begin{figure*}
\begin{center}
\includegraphics[width=0.8\textwidth]{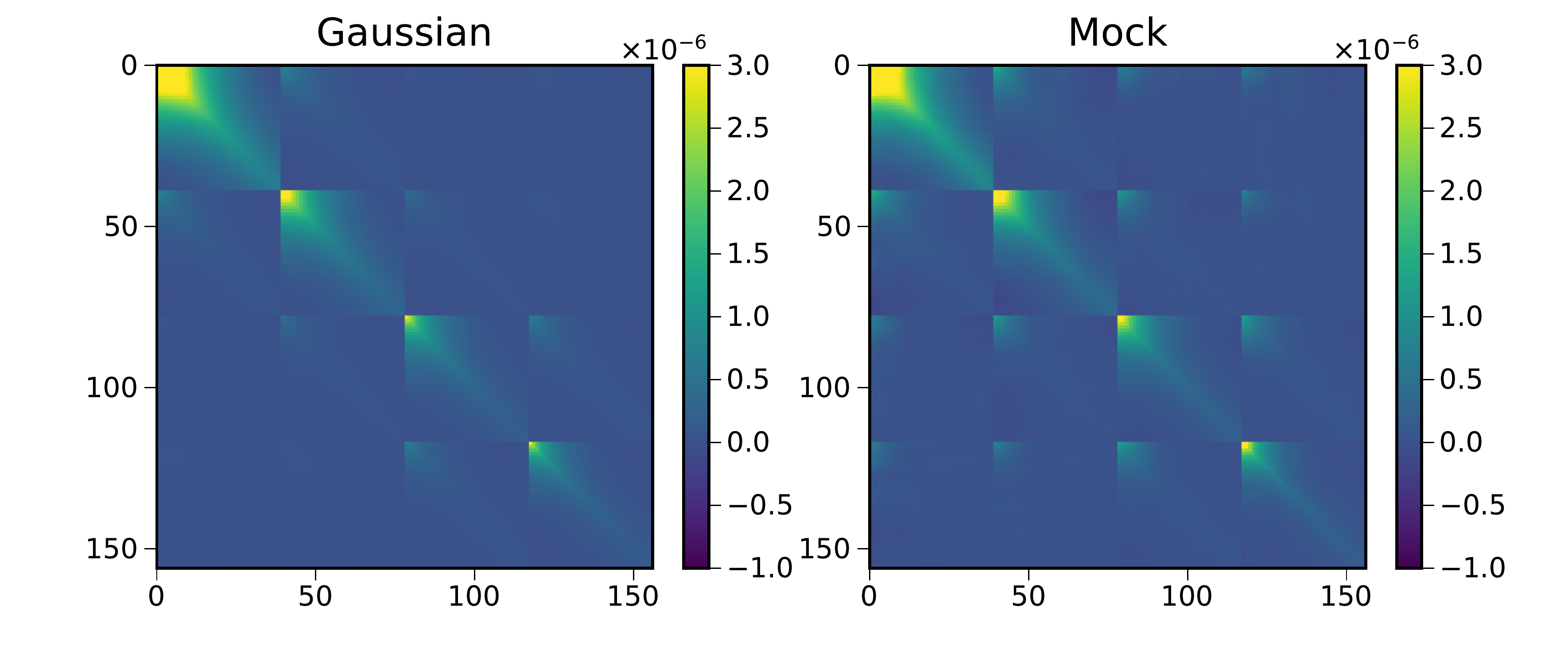} 
\caption{ The Gaussian covariance (left panel) and the mock covariance (right panel). }
\label{fig:Covmat_mock_Gaussain_color}
\end{center}
\end{figure*}

\begin{figure*}
\begin{center}
\includegraphics[width=0.8\textwidth]{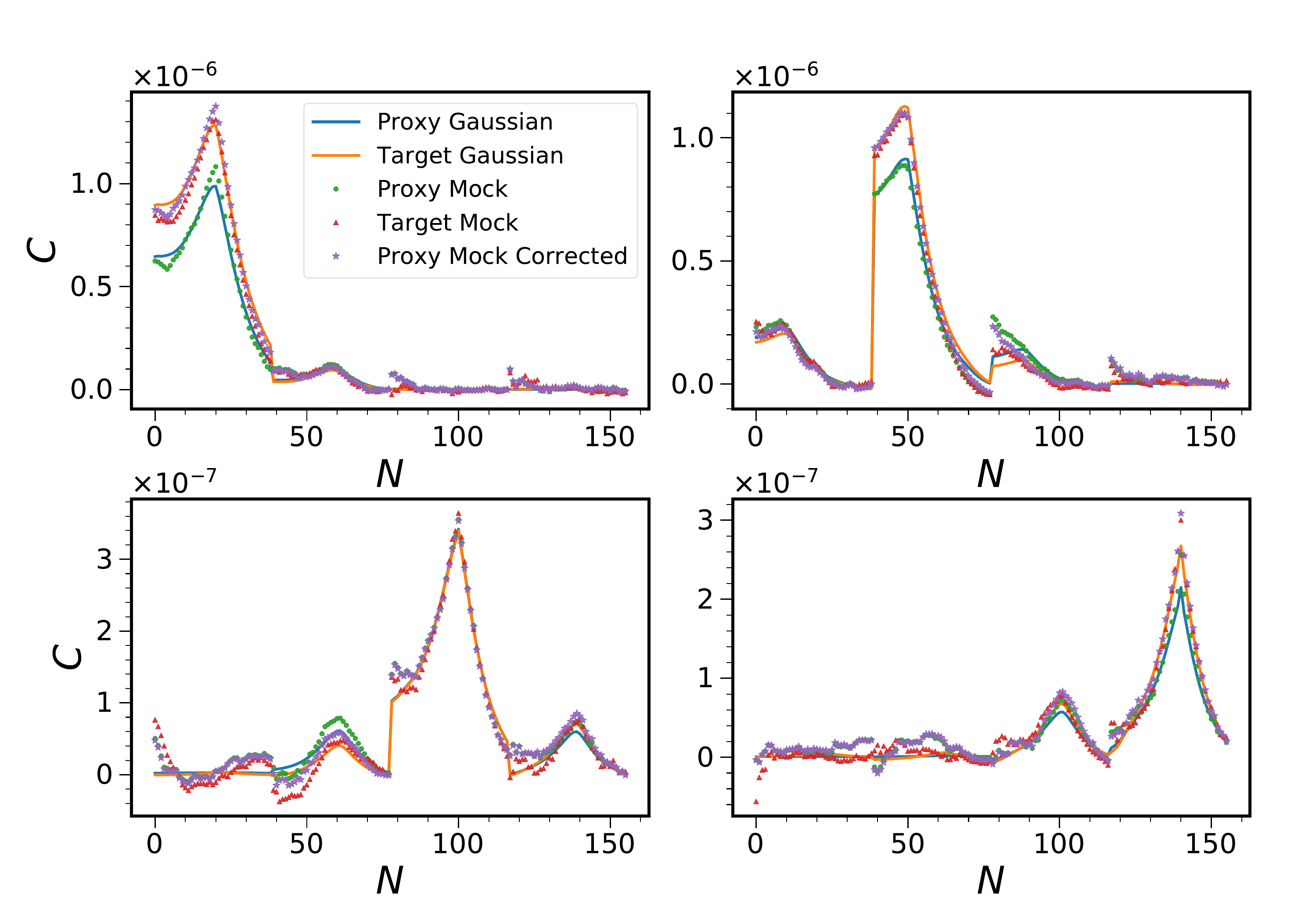}
\caption{ Four rows of the covariance matrix (the row number corresponds to the peak position) are shown. The results for Gaussian covariance (blue curve for the proxy and orange one for the target) and the mock covariance  (green circles for the proxy and red triangles for the target) are compared. The composite covariance (violet stars) obtained by combining the proxy mock results with the Gaussian correction is also displayed.  }
\label{fig:Mock_Gaussian_Corrected_covsection}
\end{center}
\end{figure*}

In Fig.~\ref{fig:Covmat_mock_Gaussain_color}, we show the covariance matrix obtained from the mock catalog and the Gaussian theory.  We arrange the data vector in  ascending order in redshift.    To see the difference more clearly, we plot four different rows of the covariance matrix in Fig.~\ref{fig:Mock_Gaussian_Corrected_covsection}. \change{ We have shown the results for two samples (proxy and target, see below). }  The symbols and lines show the mock covariance and the Gaussian covariance respectively.  From Fig.~\ref{fig:Covmat_mock_Gaussain_color} and  \ref{fig:Mock_Gaussian_Corrected_covsection} it is clear that the Gaussian covariance captures most of the features well. However the Gaussian covariance exhibits only correlation between the neighbouring redshift bins, while the mock covariances show correlation beyond the neighbouring redshift bins.



\begin{table*}
\caption{  The BAO fit results obtained using the mock and the Gaussian covariance. The results for  MLE, PL, and MCMC are shown. }
\label{tab:mock_vs_Gaussian_covmat}
\centering
\begin{tabular}{ l | c|c|c| | c|c|c  }
  \hline
  & \multicolumn{3}{c|}{Mock covariance}  & \multicolumn{3}{c|}{Gaussian covariance}   \\
  \hline
    & $ \langle\bar{\alpha}\rangle \pm \mathrm{std}( \bar{\alpha} ) $  &   $\langle  \sigma_\alpha \rangle / \mathrm{std}(\bar{\alpha} ) $   &   fraction with 1-$\sigma$                     &   $ \langle\bar{\alpha}\rangle \pm \mathrm{std}( \bar{\alpha} ) $  &   $\langle  \sigma_\alpha \rangle / \mathrm{std}(\bar{\alpha} ) $  &    fraction with 1-$\sigma$       \\
    &                                                                  &                                                                   &     enclosing $\langle \bar{\alpha} \rangle $  &                                                                     &          &  enclosing $\langle \bar{\alpha} \rangle $                                                               \\
\hline \hline
MLE           & $1.001 \pm 0.052$  & 1.02   & 0.69   &  $ 1.002 \pm 0.054 $  &  0.84      &  0.60          \\
PL            & $1.004 \pm 0.049$  & 1.27   & 0.77   &  $ 1.004 \pm 0.052 $  &  1.04      &  0.68          \\
MCMC          & $1.007 \pm 0.049$  & 1.17   & 0.74   &  $ 1.007 \pm 0.051 $  &  0.96      &  0.64          \\
\hline
\end{tabular}
\end{table*}

We summarize the BAO fit results using the mock and the Gaussian covariance in Table \ref{tab:mock_vs_Gaussian_covmat}. The distribution of $\bar{\alpha}$ is similar to that from the mock and  $\mathrm{std} (\bar{\alpha})$  is only systematically larger than that from the mock by a couple of percent. However, $\langle \sigma_\alpha \rangle$  from the Gaussian covariance is only about 80\% of  $\langle \sigma_\alpha \rangle$  derived from the mock.

\begin{figure}
\begin{center}
\includegraphics[width=0.95\linewidth]{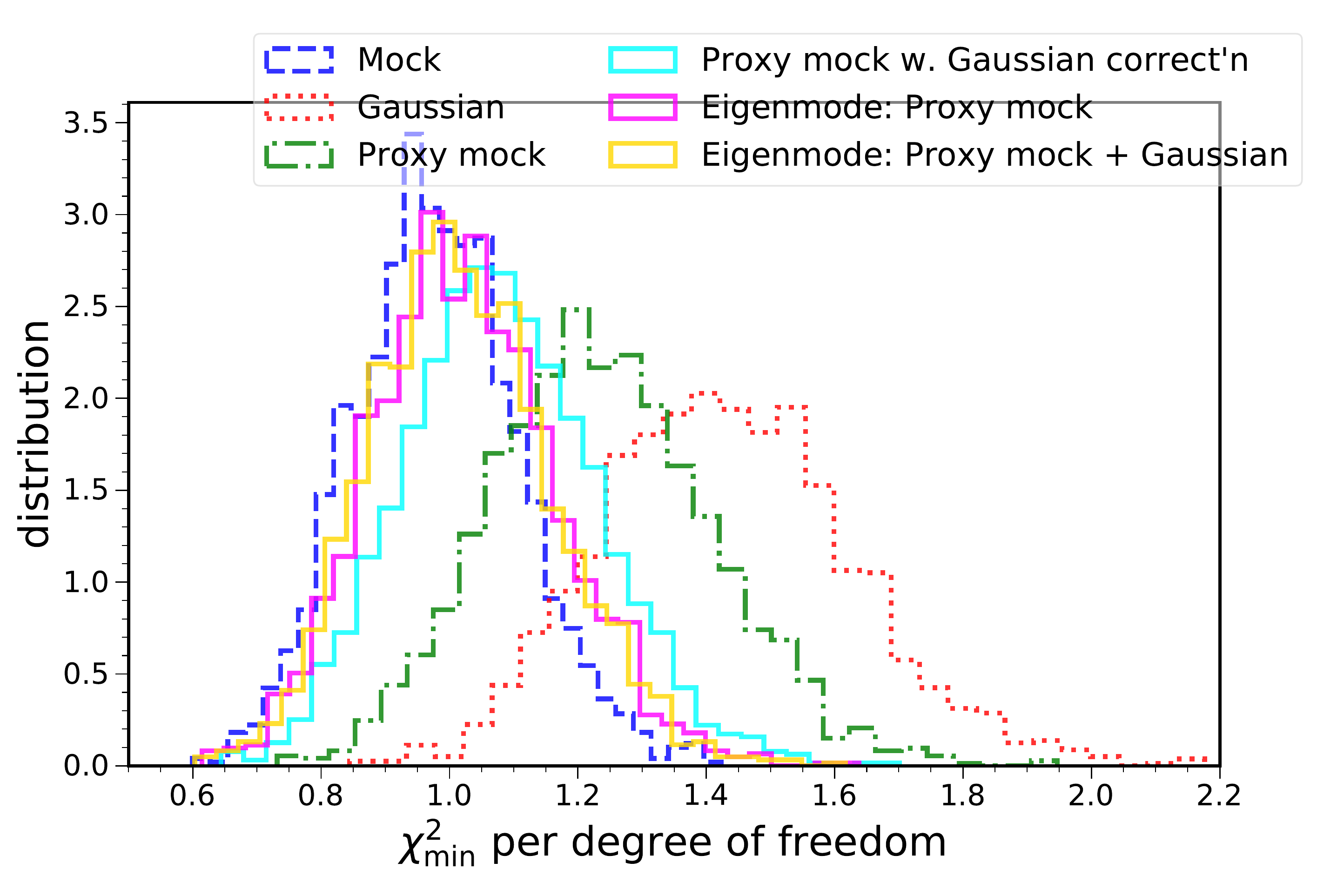}
\caption{ The distributions of the $\chi_{\rm min}^2 $ per degree of freedom for the BAO fit using various covariances. The results obtained using the target mock covariance (blue, dashed), Gaussian covariance (red, dotted), proxy  mock covariance (green, dotted-dashed), the proxy mock corrected with the Gaussian covariance (cyan, solid),  the direct eigenmode expansion on the proxy mocks (magenta, solid), and the eigenmode expansion combining the Gaussian covariance eigenmodes with the proxy mock  ones (yellow, solid). The covariance yielding the best approximation to the mock covariance is expected to give $\chi_{\rm min  }^2$ per degree of freedom closest to that obtained using the mock results.     }
\label{fig:chi2pdf_CovmatCompare}
\end{center}
\end{figure}

In Fig.~\ref{fig:chi2pdf_CovmatCompare}, we plot the $\chi^2_{\rm min} $ per degree of freedom for the BAO fit using different prescriptions for the  covariance.  The histograms are obtained by fitting to 1800 mock data vectors, and they only differ in the covariance used in the fit.  The mock covariance by construction gives the $ \chi_{\rm min}^2 $  per degree of freedom $\sim 1$, while we find that the Gaussian covariance (with the correct sample properties) gives a higher value, $\sim 1.4$. 
We will use the $\chi_{\rm min}^2$ per degree of freedom as the metric to decide which prescription of the covariance gives a better approximation to the correct mock covariance.

\subsection{  Correcting sample variation using Gaussian covariance  }
\label{sec:nz_bias_correction_Gaussian}

As mocks take time to produce, they are often created using some expected data properties. If these differ from the actual data properties, the mocks do not perfectly match the data. 
Here we investigate correcting these changes in the mock covariance matrix using the Gaussian covariance. These sample changes can result from  from variation in the bias parameter and number density of the samples. \change{ To be concrete, let us call the mock that we created using the expected properties the ``proxy mocks'', and the ones with the final correct properties  the ``target mocks''. }

As the Gaussian covariance works reasonably well, here we consider correcting the changes in the mock covariance using the Gaussian one as
\beq
C_{\rm Corrected } =  C_{\rm Proxy} + C_{\rm TargetGauss}  - C_{\rm ProxyGauss}.
\eeq
i.e.~we only correct the Gaussian part in the mock covariance, assuming that the non-Gaussian part is a small correction.

In Fig.~\ref{fig:Mock_Gaussian_Corrected_covsection}, we have plotted the  $ C_{\rm Corrected } $ and we find that the long range correlation beyond the neighbouring bins are  preserved.  In Fig.~\ref{fig:chi2pdf_CovmatCompare}, we have also plotted the $\chi^2_{\rm min}  $ per degree of freedom for the BAO fit using the proxy mock covariance with the Gaussian  correction.  It results in  values  smaller than those obtained using the proxy mock covariance or the target Gaussian covariance, and hence it signals that the corrected covariance is closer to the true one.

Overall, by combining the proxy mock covariance with the Gaussian correction, we get better agreement with the target mock covariance than using solely the proxy covariance or the Gaussian covariance.  This composite approach is expected to work when the variation of the sample is small, e.g.~the variation in $n(z) $ is small, and the non-Gaussian contribution is weak because we have not corrected the part due to the non-Gaussian covariance.   It offers a means to correct the property changes in covariance matrix without having to re-run the mocks.

\subsection{ Eigenmode expansion of the covariance matrix }

The covariance determined from mocks is susceptible to statistical noise.  In reality, we are often not interested in the covariance per se, but its inverse, the precision matrix. For example, in the likelihood, what we really need is the precision matrix.  In this case the problem is even more acute.  The noise causes bias in the  precision matrix \citep{Anderson_Statistics, Hartlap:2006kj, DodelsonSchneider2013, Percival_etal2014}.  Given a covariance matrix of size $p \times p  $ determined from $N_{\rm mock }$ mock catalogs, for it to be invertable, we must have  $ N_{\rm mock }> p$.  Even if it is invertable, the precision matrix so determined $\Psi_{\rm mock}$ is biased relative to the true precision matrix $\Psi_{\rm unbiased}$ as  \citep{Anderson_Statistics, Hartlap:2006kj}
\beq
\label{eq:precision_matrix_correction}
\Psi_{\rm unbiased} = \frac{ N_{\rm mock } - p -2  }{ N_{\rm mock } - 1 } \Psi_{\rm mock} . 
\eeq
The correction factor has been checked to work very well for the case of power spectrum \citep{Blot:2015cvj} and bispectrum \citep{Chan:2016ehg}.  Nonetheless, although the bias can be modelled and removed,  the fluctuations due to noise is unavoidable \citep{SellentinHeavens2016}. Thus it is highly desirable to reduce the number of free parameters to be determined in the covariance matrix.

There are methods that have been proposed to combine mocks with theory covariance to reduce the impact of the noise fluctuations \citep{Pope:2007vz,Taylor:2014ota,PazSanchez_2015, PearsonSamushia_2016,FriedrichEifler2017}.   Here we consider the expansion of the covariance matrix and precision matrix using the eigenmodes or the principle components  of a given covariance matrix. The effects due to noise can be mitigated because the number of free parameters is substantially reduced when the eigenmodes are given.  The study of the eigenmode of the covariance matrix is known as principle component analysis (PCA) and it is widely used in many different fields. PCA identifies the most rapidly varying direction in the data space with many variables, and hence find a  more effective way to describe the data. In the cosomological covariance context, it has been used in \cite{Scoccimarro2000, Harnois-DerapsPen_2012, MohammedSeljakVlah}.

Suppose that the eigenmodes  $\bm{v}^{(i) }$ and the  eigenvalues $\lambda^{(i) } $  of a covariance matrix $C^{\rm proxy}$ are given by 
\beq
C^{\rm proxy} \bm{v^{(i)}} = \lambda^{(i)} \bm{v}^{(i) }.  
\eeq
We can express the matrix $C^{\rm proxy}$ in terms of its eigenmodes and eigenvalues as
\beq
C_{ab}^{\rm proxy} = \sum_{i=1}^p   \lambda^{(i)} \bm{v}^{ (i) }_a \bm{v}^{ (i) }_b. 
\eeq
The precision matrix $\Psi^{\rm proxy}$ can be written as
\beq
\Psi_{ab}^{\rm proxy} =  \sum_{i=1}^p   \frac{1}{ \lambda^{(i)} }  \bm{v}^{ (i) }_a \bm{v}^{ (i) }_b. 
\eeq

We can expand a covariance matrix $C^{\rm target}$ in terms of these eigenmodes $\bm{v}^{ (i) }$, treating $ \lambda^{(i)} $ as the fitting parameters. The mode parameters can be extracted as
\beq
C^{\rm target} \bm{v}^{ (i) } \approx \kappa^{(i)}  \bm{v}^{ (i) }. 
\eeq
Here we assume that the eigenmode $\bm{v}^{ (i) } $ approximates that of $ C^{\rm target}$ well. Then we can express $C^{\rm target} $ as
\beq
\label{eq:Cmock_eigenexpand}
C^{\rm target}_{ab}  \approx \sum_{i=1}^m  \kappa^{(i)} \bm{v}^{ (i) }_a \bm{v}^{ (i) }_b. 
\eeq
Note that we use the first $m$ eigenmodes with $ \lambda^{(i)} $ ranked in descending order to approximate the covariance matrix. For the precision matrix, we have
\beq
\label{eq:Psi_eigenexpansion}
\Psi^{\rm target}_{ab}  \approx \sum_{i=1}^m  \frac{ 1}{ \kappa^{(i)} } \bm{v}^{ (i) }_a \bm{v}^{ (i) }_b. 
\eeq
For the covariance matrix, it is clear that we can safely ignore the modes with small eignevalues. However, for the inverse, the modes with small eigenvalues in fact contribute more.

The full covariance matrix of dimension $p$ has $p(p+1)/2$ independent elements, while fitting using a given basis of eigenmodes has only $p$ free parameters.  Thus, this basis cannot allow for all variations in the covariance  and they cannot fit a general symmetric matrix. The success of this method hinges on how well the given eigenmodes approximate those of the target covariance matrix.  We can obtain these eigenmodes using theory or  approximate methods, e.g.~2LPT mock in \cite{Scoccimarro2000}. In the following we use both the Gaussian theory covariance and the covariance derived from mocks with slightly different sample properties.  We consider two samples (proxy and target) which differ slightly in the $n(z)$ and bias.   The bias of these two samples differ by between a few per cent to 10\%, while the mean of the two photo-$z$ distributions differs by up to 7\%.  There are 1800 realizations for both samples. 

\begin{figure}
\begin{center}
\includegraphics[width=0.9\linewidth]{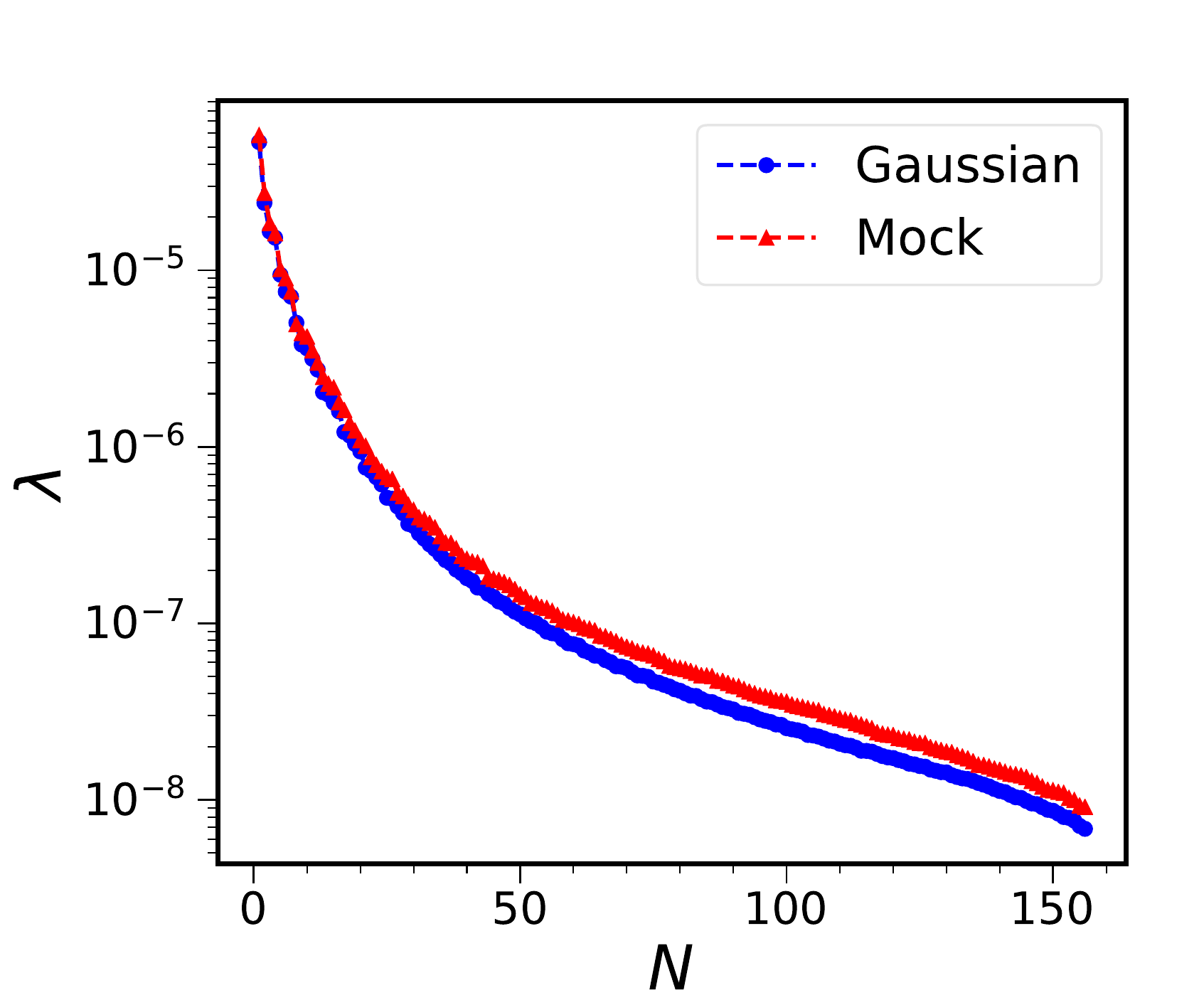}
\caption{  The eigenvalues of the covariance matrix measured from the mocks (red triangles) and the Gaussian covariance (blue circles). }
\label{fig:eigenvalues_Mock_Gaussian}
\end{center}
\end{figure}

\begin{figure}
\begin{center}
\includegraphics[width=0.9\linewidth]{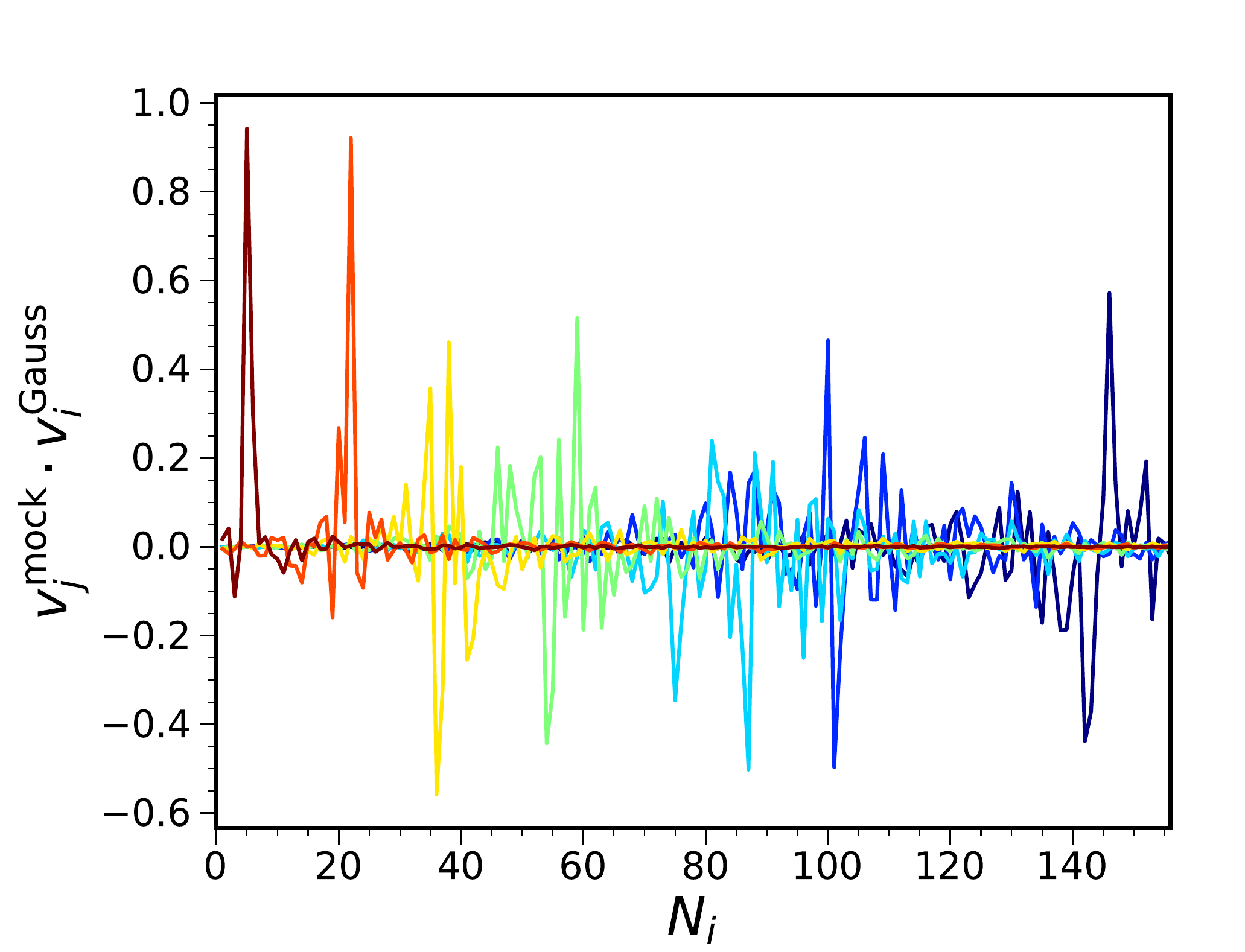}
\caption{ The dot product between the eigenmodes from the mock covariance and those from the Gaussian covariance. The eigenmodes are arranged in the same order as the eigenvalues shown in Fig.~\ref{fig:eigenvalues_Mock_Gaussian}. For each curve, the eigenmode from the mock covariance $ \bm{v}_j^{\rm mock} $ is fixed, while all the modes from the Gaussian covariance are run over.   }
\label{fig:dotproduct_MockGauss}
\end{center}
\end{figure}

In Fig.~\ref{fig:eigenvalues_Mock_Gaussian}, we first compare the eigenvalues measured from the mock and Gaussian covariance (for the target sample). We find that for the largest eigenvalues the results from the Gaussian covariance agree with those from the mocks well, while the mock ones are larger than the Gaussian ones for the relatively small ones ($\lambda \lesssim 2 \times 10^{-6} $).   \change{ We can also study the overlap between the eigenmodes from the mock covariance with those from the Gaussian one.  In Fig.~\ref{fig:dotproduct_MockGauss}, we show the dot product between the $j$th eigenmode from the mock covariance $\bm{v}_j^{\rm mock} $ and  the  $i$th eigenmode from the Gaussian covariance, $\bm{v}^{\rm  Gauss}_i $.  In this plot, for each curve $j$ is fixed while $i$ runs over all the modes. When the overlap is perfect, the dot product is 1 (or $-1$). When eigenvalue is large,  the overlap between the eigenmodes are large and well peaked.  For relatively smaller eigenvalue modes, they do not match each other well. } 
This shows that the eigenmode from the Gaussian covariance is a good approximation to that of the mock covariance only for the ones with large eigenvalues.  Since the modes with the largest eigenvalues are close to Gaussian covariance prediction, we can call them the Gaussian modes.  The fact that eigenvalues of the non-Gaussian modes from the Gaussian covariance are smaller than those from the mock covariance suggests that the Gaussian covariance underestimates the importance of the non-Gaussian modes.

\begin{figure*}
\begin{center}
\includegraphics[width=0.9\textwidth]{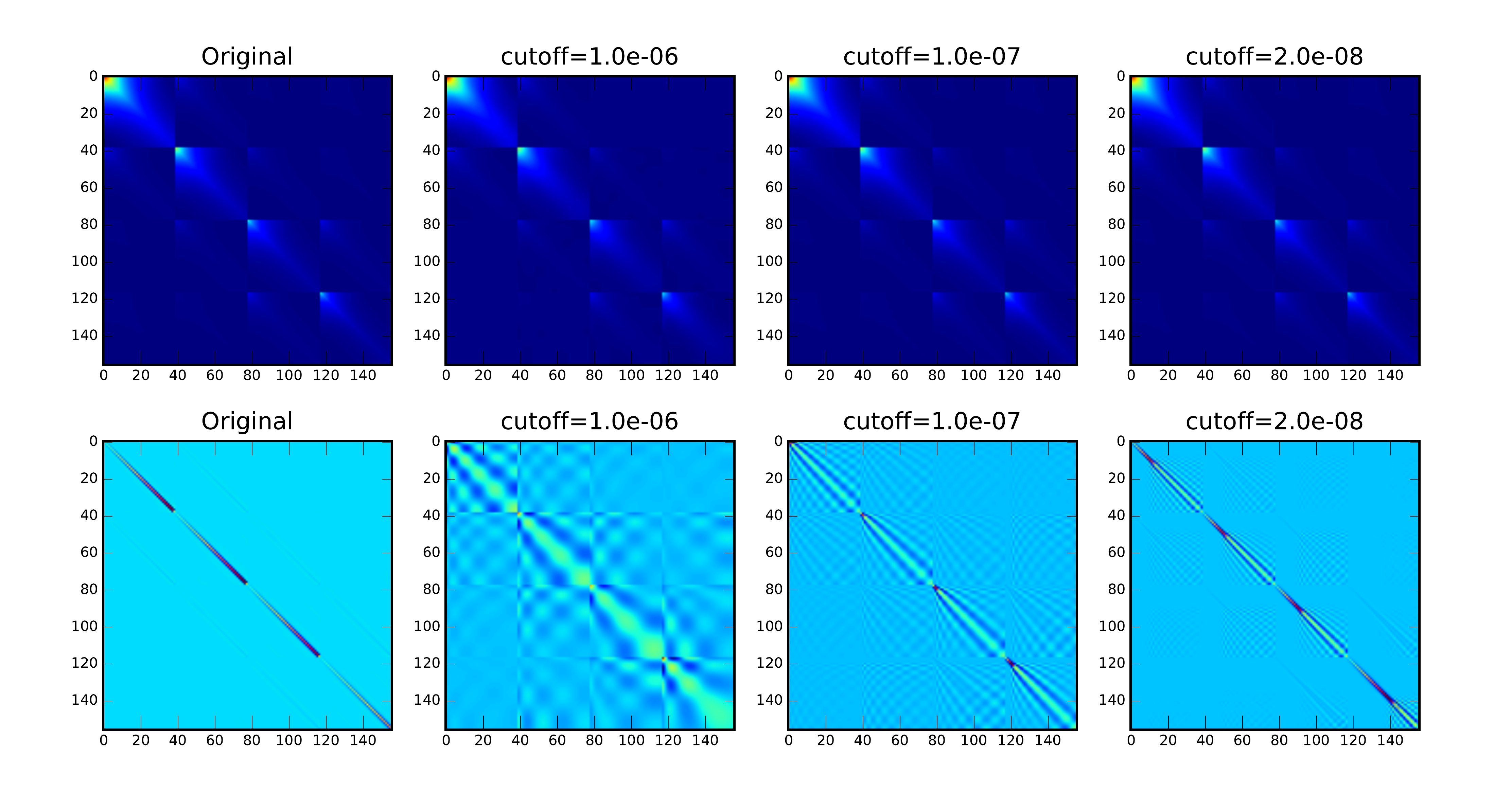}
\caption{  The covariance matrix (upper panels) and the precision matrix (lower panels) are shown. The leftmost ones  shows the original ones, and the rest are obtained using the eigenmode expansion by keeping the modes with eigenvalue larger than cutoff values  $10^{-6}$, $10^{-7}$, and  $2\times 10^{-8} $ respectively.    }
\label{fig:C_Psi_eigen_Cexpand}
\end{center}
\end{figure*}

We now check how important the non-Gaussian modes are for the covariance matrix and the precision matrix.  In Fig.~\ref{fig:C_Psi_eigen_Cexpand}, we compare the covariance matrix and the precision matrix obtained using the eigenmode expansion by keeping terms with eigenvalues larger than the cutoff values $10^{-6}$, $10^{-7}$, and  $2\times 10^{-8} $ respectively. They corresponding to the largest 19, 53, and 112 eigenvalues.   Here and for the rest of the analysis, it is useful to note that the size of the covariance matrix is $156 \times 156$.   In this exercise we have used the Gaussian covariance.  It is clear that by keeping only the terms with large eigenvalues, e.g.~larger than  $10^{-6}$ we get a reasonably good approximation to the original covariance matrix already. This is evident from Eq.~\eqref{eq:Cmock_eigenexpand}. On the other hand, we find that keeping only terms with large eigenvalues results in a poor approximation to the precision matrix.  This is not surprising because as we mentioned, the small eigenvalue modes in fact contribute more to the inverse, thus ignoring them yields an unsatisfactory approximation.  Henceforth, we will keep all the modes.

\begin{figure*}
\begin{center}
\includegraphics[width=0.9\textwidth]{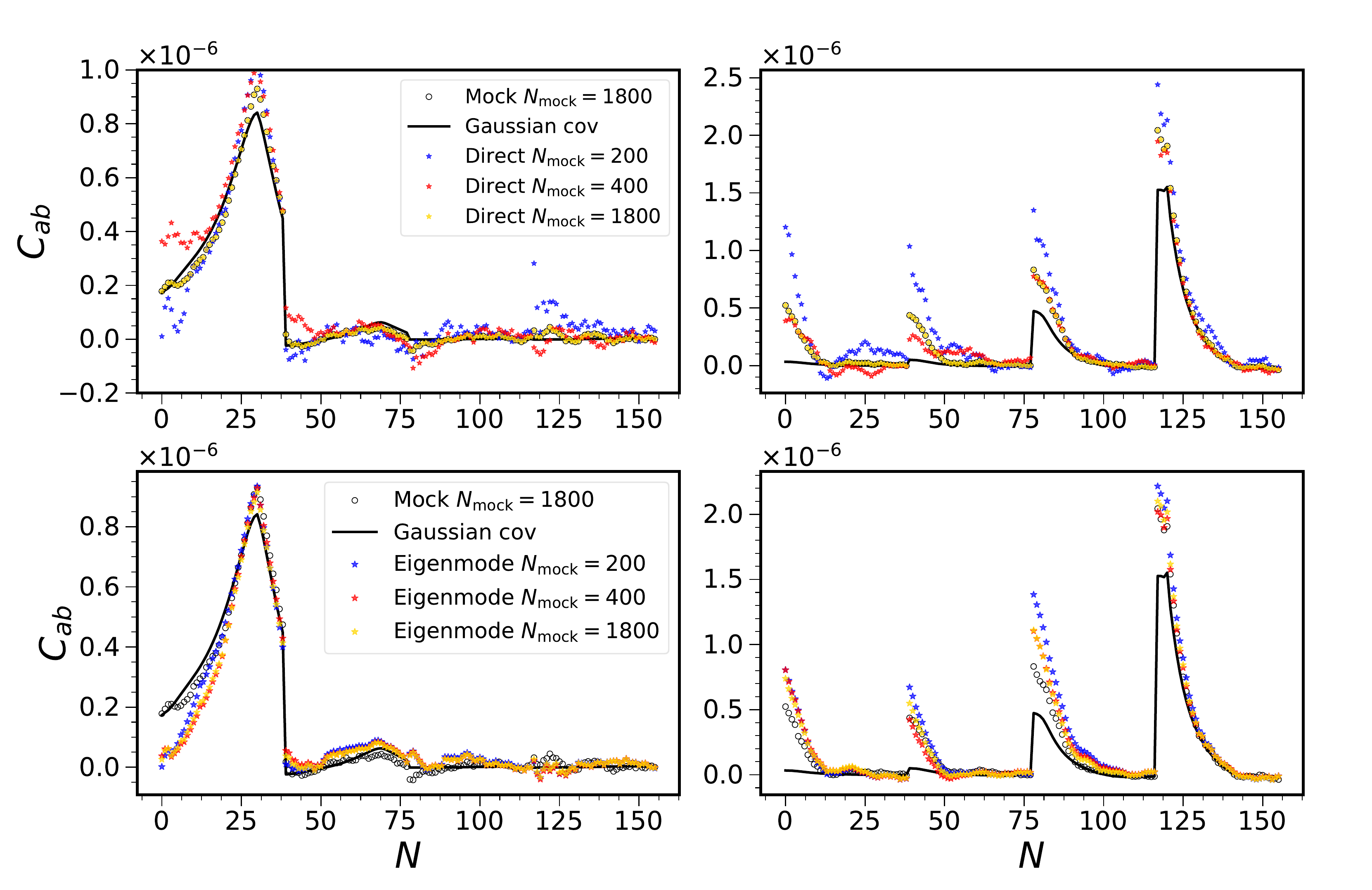}
\caption{  Two rows of the covariance matrix (left and right) obtained  from the mock (direct measurement with $N_{\rm mock} = 1800$, circles, in the upper panels coincide  with the yellow stars), Gaussian covariance (solid black lines) are shown. In the upper panels, we show the results from direct measurement (stars) using  $N_{\rm mock} = 200$  (blue), 400 (red), and 1800 (yellow). We display the corresponding results from the eigenmode expansion in the lower panels.     }
\label{fig:eigenmode_C_NumMocks_Section}
\end{center}
\end{figure*}

Because the eigenmode expansion with given basis modes has far fewer free parameters than the direct measurement, we now test how effective the eigenmode expansion is in reducing the impact of noise on the measurement of the covariance and the precision matrix. In Fig.~\ref{fig:eigenmode_C_NumMocks_Section}, we show two rows of the covariance matrices obtained using different prescriptions. The results from the mocks and the Gaussian theory are displayed. \change{  The eigenmodes are obtained from the 1800 proxy mocks. }  Using these proxy eigenmodes, we fit the ``eigenvalue'' parameters from  the target mock covariance estimated with  $N_{\rm mock}$ target mocks.  We have shown the results using $N_{\rm mock} = 200$, 400, and 1800 (all the mocks) realizations.    As expected, the direct measurement results slowly become less noisy as the number of realizations used increases.  On the other hand, the eigenmode expansion results are far less noisy than the direct measurement ones. In fact there are no marked variations among the different number of realizations shown.  However, even for $N_{\rm mock } = 1800 $, there are small deviations between the eigenmode expansion result and the direct measurement one with $N_{\rm mock } =1800$.  The systematic bias in the eigenmode expansion comes from the fact that the basis modes used are not the exact ones.

\begin{figure*}
\begin{center}
\includegraphics[width=0.9\textwidth]{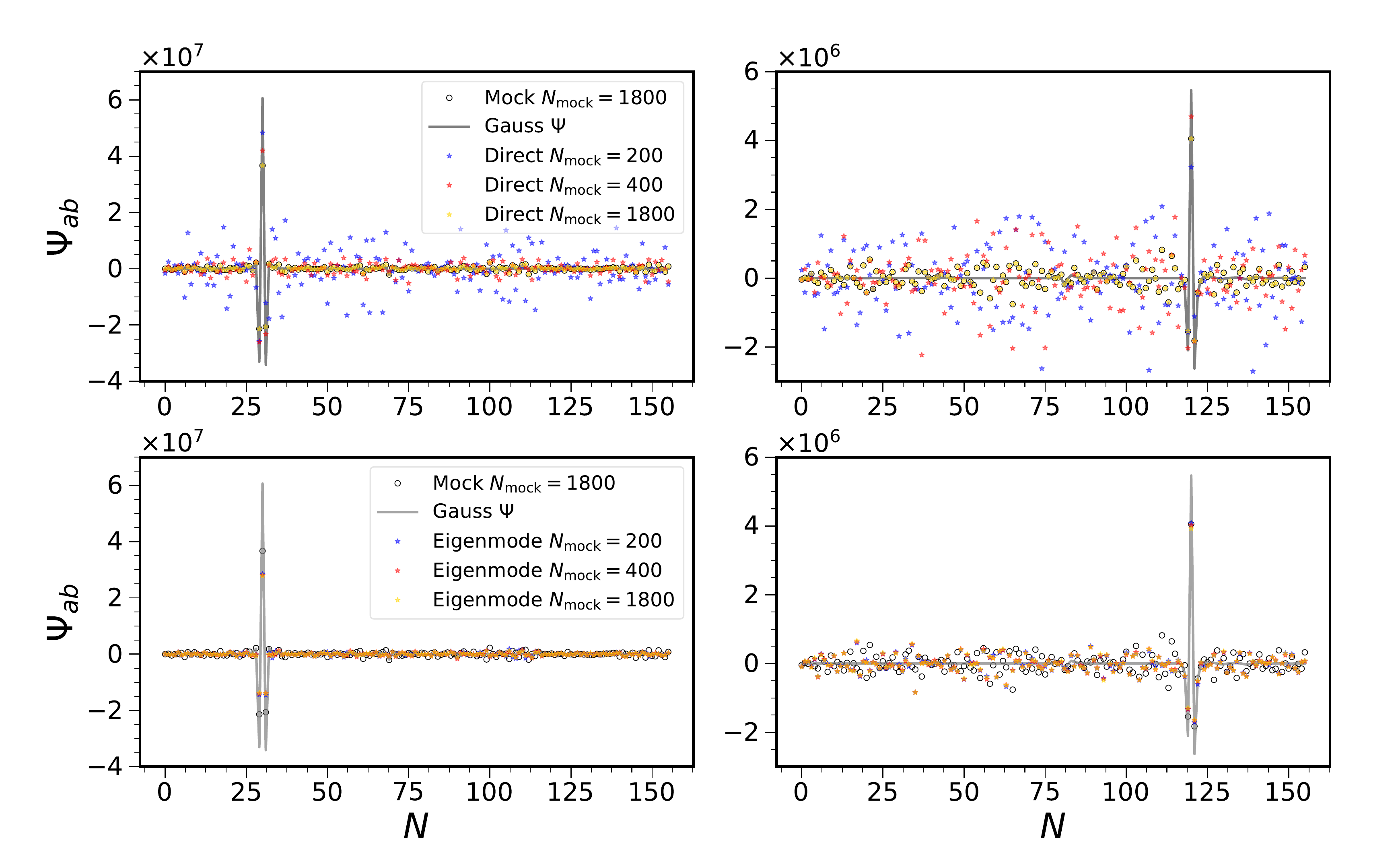}
\caption{  Two rows  of the  precision matrix obtained by inverting the covariance matrix shown in Fig.~\ref{fig:eigenmode_C_NumMocks_Section} are displayed. The symbols are the same as those in Fig.~\ref{fig:eigenmode_C_NumMocks_Section}. The upper panels are obtained from direct measurements and the lower ones are from eigenmode expansion.  }
\label{fig:eigenmode_Psi_NumMocks_Section}
\end{center}
\end{figure*}

We present the corresponding results for the precision matrix in Fig.~\ref{fig:eigenmode_Psi_NumMocks_Section}.  For $ \Psi $ from direct measurement, we have applied the bias correction factor Eq.~\eqref{eq:precision_matrix_correction}.  Similar to the case shown in Fig.~\ref{fig:eigenmode_C_NumMocks_Section}, the eigenmodes are estimated from 1800 proxy mocks, and the fitting parameters are extracted using the covariance estimated from $N_{\rm mock}$ target mocks.  While there are some improvements in the noise suppression using the eigenmode expansion for the covariance, the gain for the precision matrix is impressive. This is especially true when $N_{\rm mock}$ is not substantially larger than $p$.  For example, for the case of $N_{\rm mock}=200 $, the precision matrix from direct measurement is very noisy, while the results obtained from the eigenmode expansion is already close to the asymptotic results (the $N_{\rm mock} =1800 $ direct measurement is taken to be the asymptotic one). However, similar to the case of the covariance matrix, although the results converge quickly, they are slightly biased compared to the direct measurement with  $N_{\rm mock} =1800 $. This is again attributed to the proxy basis modes used being not exact.  In Fig.~\ref{fig:chi2pdf_CovmatCompare}, we have plotted the results obtained using the eigenmode expansion. We have used the results determined using $ N_{\rm mock} =200 $. The resultant $\chi^2_{\rm min}$ per degree of freedom is one of the closest to the true mock covariance results among all the prescriptions shown. The other one is also eigenmode expansion based, to be commented on shortly.   


We have tried even lower of number of realizations, e.g.~30.  In this case, the covariance from the direct measurement is not invertible, but the eigenmode expansion  still gives a valid result with a small amount of noise only. Thus the eigenmode expansion is much more robust to noise than the direct measurement.

If the basis modes are given by the Gaussian theory, then the convergent result \change{(i.e.~when sufficient number of $N_{\rm mock}$ are used)} gets close to the Gaussian one, and so it does not yield a good approximation to the true covariance. Alternatively we can combine the eigenmodes from the Gaussian covariance with those from the proxy mocks to form a composite covariance.  The idea is that because the large eigenvalue modes are close to the Gaussian covariance results, we can use the Gaussian theory eigenmodes for them, and this has the advantage that the sample properties can be taken into account in theory easily. For the non-Gaussian modes (with small eigenvalues), we use the basis modes determined from the proxy mock covariance.  Using the composite basis, the eigenvalue parameters are fitted to a covariance estimated from a relatively small set of target mocks as in the previous method.  This composite approach performs similar to the one using only the basis modes from the proxy mocks, as can be seen from  Fig.~\ref{fig:chi2pdf_CovmatCompare}.  If the number of proxy mocks available is not large, this hybrid approach is preferred because the influence of noise on the estimate of the proxy eigenmodes is reduced.

In summary, we find that using the eigenmode expansion, the impact of the noise on the measurement of the covariance and especially the precision matrix is significantly reduced. Thus this method enables us to substantially reduce the number of mock catalogs required.  The success of the methods relies on how well the given proxy basis modes approximate the true ones.  We find that although the expansion converges quickly, the results are slightly biased. This is due to the basis modes being not exact. One of the applications is to correct for sample changes. In this case, as we have demonstrated, the original mock samples can provide the required proxy eigenmodes.  We have demonstrated that this method performs even better than the correction using Gaussian covariance proposed in Sec.~\ref{sec:nz_bias_correction_Gaussian}.

\section{Conclusions} 
\label{sec:Conclusion}

Measurement of the BAO scale in the distribution of galaxies has been recognized as one of the most important current cosmological probes. DES already delivered encouraging BAO measurement using photometric data and future DES data, or surveys such as LSST, are expected to yield more exciting results. To meet the big improvements in the data, the analysis pipeline must be optimized and theoretical systematics must be under control.  In this paper we study the theoretical systematics and optimization of the BAO detection using the angular correlation function.  Although the fiducial setup is for DES Y1 in this paper, the techniques developed here are useful for other large scale structure survey analysis as well.

To extract the BAO from the data, some estimators are required to produce the best fit and the error bar. We have compared three common fitting methods: MLE, PL and MCMC. Among these methods, MLE yields the least bias and the error bar derived using $\Delta \chi^2 = 1 $ is closest to the Gaussian distribution (Table \ref{tab:data_pruning}). For MLE, we also find that the $\mathrm{std}( \bar{ \alpha} ) $ agrees with $\langle \sigma_\alpha \rangle $ well ($\langle \sigma_\alpha \rangle / \mathrm{std}( \bar{\alpha} )$ columns in Table \ref{tab:number_redshiftbins}). An advantage of MLE (and PL) is that it takes far less time than MCMC.

Because MLE is completely determined by the likelihood, 
it is more efficient in picking up small features such as those arising from mismatch between the template and the data. We regarded the mock catalogs whose 1-$\sigma$ interval of the best fit $\alpha$ fall outside the interval [0.8,1.2] as non-detections. Those extreme mocks are poorly fitted by our existing methodology and we have demonstrated that they cause bias in distribution of the best fit parameter in Appendix~\ref{sec:appendix}.

We  studied when there is mismatch between the template and the data either due to incorrect fiducial cosmology (Table \ref{tab:MLE_MCMC_MICE_Planck})  or the photo-$z$ error (Table \ref{tab:MLE_MCMC_photo_z_error}) whether the fitting pipeline can recover the mismatch correctly. This is especially relevant for DES Y1 as the mocks are matched to the MICE cosmology, which is quite different from the currently accepted Planck cosmology. Differences in cosmology lead to variation in both the sound horizon and the angular diameter distance, while photo-$z$ error shifts the angular diameter distance only.   We find  that only the photo-$z$ error leading to shift in the mean of the photo-$z$ distribution causes bias in the BAO fit, while reasonable variation in the width of the distribution does not produce any significant effect.  The shift in BAO obtained from direct fitting is consistent with the estimate based on its effects on  the sound horizon and/or the angular diameter distance.

A number of optimizations in the context of the DES Y1 BAO measurement were investigated: the angular bin width, the number of redshift bins, and the effect of the cross correlation. We find that the optimal binning is $\Delta \theta = 0.15^{\circ} $ (to $ 0.3^{\circ}$), and four redshift bins are sufficient. Adding the cross correlation does not enhance the BAO constraint substantially.

Sometimes the mocks are created with certain sample properties, but the finalized samples end up slightly different from those envisioned at the beginning.  We proposed to correct these sample changes in the covariance matrix using the Gaussian covariance and the eigenmode expansion method.  We show that the covariance with Gaussian covariance correction outperforms the pure Gaussian covariance or the proxy mock covariance; however, the eigenmode expansion works even better in incorporating the changes (Fig.~\ref{fig:chi2pdf_CovmatCompare}).  The success of the eigenmode expansion relies on how well the proxy eigenvectors approximate those of the target covariance.  With the basis vectors given, as the number of free parameters to be determined is significantly reduced [$p$ parameters versus $p(p+1)/2$ for direct measurement] and we demonstrate that the effect of noise is substantially mitigated relative to the direct measurements.  In this case, it does not rely on the Gaussian covariance, but a small number of target mocks are required. Thus we expect this method to work even when the covariance is highly non-Gaussian.

As this work was finished, the DES Y3 data became available and science ready. Although the depth of the Y3 data is similar to that of Y1, its area is roughly three times larger. Hence we expect that the constraint on the BAO angular scale will be tightened by a factor of $ \sim 1/\sqrt{3} $. The analysis pipeline developed here, and the lessons learned, can be easily extended to the Y3 data. Hence, we look forward to that analysis.

\section*{Acknowledgement} 

KCC acknowledges the support from the Spanish Ministerio de Economia y Competitividad grant ESP2013-48274-C3-1-P and the Juan de la Cierva fellowship.

We are grateful for the extraordinary contributions of our CTIO colleagues and the DECam Construction, Commissioning and Science Verification teams in achieving the excellent instrument and telescope conditions that have made this work possible.  The success of this project also  relies critically on the expertise and dedication of the DES Data Management group.

Funding for the DES Projects has been provided by the U.S. Department of Energy, the U.S. National Science Foundation, the Ministry of Science and Education of Spain,  the Science and Technology Facilities Council of the United Kingdom, the Higher Education Funding Council for England, the National Center for Supercomputing  Applications at the University of Illinois at Urbana-Champaign, the Kavli Institute of Cosmological Physics at the University of Chicago,  the Center for Cosmology and Astro-Particle Physics at the Ohio State University, the Mitchell Institute for Fundamental Physics and Astronomy at Texas A\&M University, Financiadora de Estudos e Projetos, Funda{\c c}{\~a}o Carlos Chagas Filho de Amparo {\`a} Pesquisa do Estado do Rio de Janeiro, Conselho Nacional de Desenvolvimento Cient{\'i}fico e Tecnol{\'o}gico and  the Minist{\'e}rio da Ci{\^e}ncia, Tecnologia e Inova{\c c}{\~a}o, the Deutsche Forschungsgemeinschaft and the Collaborating Institutions in the Dark Energy Survey.  The DES data management system is supported by the National Science Foundation under Grant Number AST-1138766.

The Collaborating Institutions are Argonne National Laboratory, the University of California at Santa Cruz, the University of Cambridge, Centro de Investigaciones En{\'e}rgeticas, Medioambientales y Tecnol{\'o}gicas-Madrid, the University of Chicago, University College London, the DES-Brazil Consortium, the University of Edinburgh, the Eidgen{\"o}ssische Technische Hochschule (ETH) Z{\"u}rich, Fermi National Accelerator Laboratory, the University of Illinois at Urbana-Champaign, the Institut de Ci{\`e}ncies de l'Espai (IEEC/CSIC), the Institut de F{\'i}sica d'Altes Energies, Lawrence Berkeley National Laboratory, the Ludwig-Maximilians Universit{\"a}t M{\"u}nchen and the associated Excellence Cluster Universe, the University of Michigan, the National Optical Astronomy Observatory, the University of Nottingham, The Ohio State University, the University of Pennsylvania, the University of Portsmouth, SLAC National Accelerator Laboratory, Stanford University, the University of Sussex, and Texas A\&M University.

This paper has gone through internal review by the DES collaboration. The DES publication number for this article is DES-2017-0306. The Fermilab preprint number is FERMILAB-PUB-17-590. We thank the anonymous referee for thoroughly reviewing this paper and providing constructive comments.

\bibliographystyle{mnras}
\bibliography{references}

\begin{thebibliography}{}
\makeatletter
\relax
\def\mn@urlcharsother{\let\do\@makeother \do\$\do\&\do\#\do\^\do\_\do\%\do\~}
\def\mn@doi{\begingroup\mn@urlcharsother \@ifnextchar [ {\mn@doi@}
  {\mn@doi@[]}}
\def\mn@doi@[#1]#2{\def\@tempa{#1}\ifx\@tempa\@empty \href
  {http://dx.doi.org/#2} {doi:#2}\else \href {http://dx.doi.org/#2} {#1}\fi
  \endgroup}
\def\mn@eprint#1#2{\mn@eprint@#1:#2::\@nil}
\def\mn@eprint@arXiv#1{\href {http://arxiv.org/abs/#1} {{\tt arXiv:#1}}}
\def\mn@eprint@dblp#1{\href {http://dblp.uni-trier.de/rec/bibtex/#1.xml}
  {dblp:#1}}
\def\mn@eprint@#1:#2:#3:#4\@nil{\def\@tempa {#1}\def\@tempb {#2}\def\@tempc
  {#3}\ifx \@tempc \@empty \let \@tempc \@tempb \let \@tempb \@tempa \fi \ifx
  \@tempb \@empty \def\@tempb {arXiv}\fi \@ifundefined
  {mn@eprint@\@tempb}{\@tempb:\@tempc}{\expandafter \expandafter \csname
  mn@eprint@\@tempb\endcsname \expandafter{\@tempc}}}

\bibitem[\protect\citeauthoryear{Abbott et~al.}{Abbott
  et~al.}{2017a}]{Abbott:2017wau}
Abbott T. M.~C.,  et~al., 2017a, {Dark Energy Survey Year 1 Results:
  Cosmological Constraints from Galaxy Clustering and Weak Lensing} (\mn@eprint
  {arXiv} {1708.01530})

\bibitem[\protect\citeauthoryear{Abbott et~al.}{Abbott et~al.}{2017b}]{BAOmain}
Abbott T. M.~C.,  et~al., 2017b, {Dark Energy Survey Year 1 Results:
  Measurement of the Baryon Acoustic Oscillation scale in the distribution of
  galaxies to redshift 1} (\mn@eprint {arXiv} {1712.06209})

\bibitem[\protect\citeauthoryear{Ade et~al.}{Ade et~al.}{2016}]{Ade:2015xua}
Ade P. A.~R.,  et~al., 2016, \mn@doi [Astron. Astrophys.]
  {10.1051/0004-6361/201525830}, 594, A13

\bibitem[\protect\citeauthoryear{{Alam} et~al.,}{{Alam}
  et~al.}{2017}]{Alam_etal2017}
{Alam} S.,  et~al., 2017, \mn@doi [MNRAS] {10.1093/mnras/stx721}, \href
  {http://adsabs.harvard.edu/abs/2017MNRAS.470.2617A} {470, 2617}

\bibitem[\protect\citeauthoryear{{Alonso}}{{Alonso}}{2012}]{Alonso_CUTE}
{Alonso} D.,  2012, {CUTE solutions for two-point correlation functions from
  large cosmological datasets} (\mn@eprint {arXiv} {1210.1833})

\bibitem[\protect\citeauthoryear{Anderson}{Anderson}{2003}]{Anderson_Statistics}
Anderson T.~W.,  2003, An Introduction to Multivariate Statistical Analysis,
  3rd Edition.
Wiley-Interscience, New York

\bibitem[\protect\citeauthoryear{{Anderson} et~al.,}{{Anderson}
  et~al.}{2014}]{Anderson_etal2014}
{Anderson} L.,  et~al., 2014, \mn@doi [\mnras] {10.1093/mnras/stu523}, \href
  {http://adsabs.harvard.edu/abs/2014MNRAS.441...24A} {441, 24}

\bibitem[\protect\citeauthoryear{{Ata} et~al.,}{{Ata}
  et~al.}{2017}]{Ata_etal2017}
{Ata} M.,  et~al., 2017, preprint, \href
  {http://adsabs.harvard.edu/abs/2017arXiv170506373A} {} (\mn@eprint {arXiv}
  {1705.06373})

\bibitem[\protect\citeauthoryear{{Aubourg} et~al.}{{Aubourg}
  et~al.}{2015}]{Aubourg:2014yra}
{Aubourg} {\'E}.,  et~al., 2015, \mn@doi [Phys. Rev.]
  {10.1103/PhysRevD.92.123516}, D92, 123516

\bibitem[\protect\citeauthoryear{{Avila}, {Murray}, {Knebe}, {Power},
  {Robotham}  \& {Garcia-Bellido}}{{Avila} et~al.}{2015}]{Avila_etal2015}
{Avila} S.,  {Murray} S.~G.,  {Knebe} A.,  {Power} C.,  {Robotham} A.~S.~G.,
  {Garcia-Bellido} J.,  2015, \mn@doi [\mnras] {10.1093/mnras/stv711}, \href
  {http://adsabs.harvard.edu/abs/2015MNRAS.450.1856A} {450, 1856}

\bibitem[\protect\citeauthoryear{Avila et~al.}{Avila et~al.}{2017}]{Halogen}
Avila S.,  et~al., 2017, {Dark Energy Survey Year 1 Results: galaxy mock
  catalogues for BAO} (\mn@eprint {arXiv} {1712.06232})

\bibitem[\protect\citeauthoryear{{Bautista} et~al.,}{{Bautista}
  et~al.}{2017}]{Bautista_etal2017}
{Bautista} J.~E.,  et~al., 2017, \mn@doi [A\&A] {10.1051/0004-6361/201730533},
  \href {http://adsabs.harvard.edu/abs/2017A%26A...603A..12B} {603, A12}

\bibitem[\protect\citeauthoryear{{Beutler} et~al.,}{{Beutler}
  et~al.}{2011}]{Beutler_etal2011}
{Beutler} F.,  et~al., 2011, \mn@doi [MNRAS]
  {10.1111/j.1365-2966.2011.19250.x}, \href
  {http://adsabs.harvard.edu/abs/2011MNRAS.416.3017B} {416, 3017}

\bibitem[\protect\citeauthoryear{Blot, Corasaniti, Amendola  \& Kitching}{Blot
  et~al.}{2016}]{Blot:2015cvj}
Blot L.,  Corasaniti P.~S.,  Amendola L.,   Kitching T.~D.,  2016, \mn@doi
  [Mon. Not. Roy. Astron. Soc.] {10.1093/mnras/stw604}, 458, 4462

\bibitem[\protect\citeauthoryear{{Bond} \& {Efstathiou}}{{Bond} \&
  {Efstathiou}}{1984}]{BondEfstathiou1984}
{Bond} J.~R.,  {Efstathiou} G.,  1984, \mn@doi [ApJL] {10.1086/184362}, \href
  {http://adsabs.harvard.edu/abs/1984ApJ...285L..45B} {285, L45}

\bibitem[\protect\citeauthoryear{{Bond} \& {Efstathiou}}{{Bond} \&
  {Efstathiou}}{1987}]{BondEfstathiou1987}
{Bond} J.~R.,  {Efstathiou} G.,  1987, \mn@doi [MNRAS]
  {10.1093/mnras/226.3.655}, \href
  {http://adsabs.harvard.edu/abs/1987MNRAS.226..655B} {226, 655}

\bibitem[\protect\citeauthoryear{Camacho et~al.}{Camacho
  et~al.}{2018}]{Clpaper}
Camacho A.,  et~al., 2018, {Dark Energy Survey Year 1 Results: BAO from
  $C_{\ell}$ } (\mn@eprint {arXiv} {---})

\bibitem[\protect\citeauthoryear{{Carnero}, {S{\'a}nchez}, {Crocce},
  {Cabr{\'e}}  \& {Gazta{\~n}aga}}{{Carnero} et~al.}{2012}]{Carnero_etal2012}
{Carnero} A.,  {S{\'a}nchez} E.,  {Crocce} M.,  {Cabr{\'e}} A.,
  {Gazta{\~n}aga} E.,  2012, \mn@doi [MNRAS]
  {10.1111/j.1365-2966.2011.19832.x}, \href
  {http://adsabs.harvard.edu/abs/2012MNRAS.419.1689C} {419, 1689}

\bibitem[\protect\citeauthoryear{Chan \& Blot}{Chan \&
  Blot}{2017}]{Chan:2016ehg}
Chan K.~C.,  Blot L.,  2017, \mn@doi [Phys. Rev.] {10.1103/PhysRevD.96.023528},
  D96, 023528

\bibitem[\protect\citeauthoryear{Chan, Moradinezhad~Dizgah  \& Nore\~na}{Chan
  et~al.}{2018}]{Chan:2017fiv}
Chan K.~C.,  Moradinezhad~Dizgah A.,   Nore\~na J.,  2018, \mn@doi [Phys. Rev.]
  {10.1103/PhysRevD.97.043532}, D97, 043532

\bibitem[\protect\citeauthoryear{Cohn}{Cohn}{2006}]{Cohn:2005ex}
Cohn J.~D.,  2006, \mn@doi [New Astron.] {10.1016/j.newast.2005.08.002}, 11,
  226

\bibitem[\protect\citeauthoryear{{Cole}, {Fisher}  \& {Weinberg}}{{Cole}
  et~al.}{1994}]{ColeFisherWeinberg1994}
{Cole} S.,  {Fisher} K.~B.,   {Weinberg} D.~H.,  1994, \mn@doi [MNRAS]
  {10.1093/mnras/267.3.785}, \href
  {http://adsabs.harvard.edu/abs/1994MNRAS.267..785C} {267, 785}

\bibitem[\protect\citeauthoryear{{Cole} et~al.,}{{Cole}
  et~al.}{2005}]{Cole_etal2005}
{Cole} S.,  et~al., 2005, \mn@doi [MNRAS] {10.1111/j.1365-2966.2005.09318.x},
  \href {http://adsabs.harvard.edu/abs/2005MNRAS.362..505C} {362, 505}

\bibitem[\protect\citeauthoryear{Cowan}{Cowan}{1998}]{GCowan}
Cowan G.,  1998, Statistical Data Analysis.
Oxford University Press

\bibitem[\protect\citeauthoryear{{Crocce} \& {Scoccimarro}}{{Crocce} \&
  {Scoccimarro}}{2008}]{CrocceScoccimarro_2008}
{Crocce} M.,  {Scoccimarro} R.,  2008, \mn@doi [\prd]
  {10.1103/PhysRevD.77.023533}, \href
  {http://adsabs.harvard.edu/abs/2008PhRvD..77b3533C} {77, 023533}

\bibitem[\protect\citeauthoryear{{Crocce}, {Cabr{\'e}}  \&
  {Gazta{\~n}aga}}{{Crocce} et~al.}{2011}]{CrocceCabreGazta_2011}
{Crocce} M.,  {Cabr{\'e}} A.,   {Gazta{\~n}aga} E.,  2011, MNRAS, 414, 329

\bibitem[\protect\citeauthoryear{Crocce et~al.}{Crocce
  et~al.}{2017}]{LSSsample}
Crocce M.,  et~al., 2017, {Dark Energy Survey Year 1 Results: Galaxy Sample for
  BAO Measurement} (\mn@eprint {arXiv} {1712.06211})

\bibitem[\protect\citeauthoryear{{Dodelson}}{{Dodelson}}{2003}]{Dodelson_2003}
{Dodelson} S.,  2003, {Modern cosmology}

\bibitem[\protect\citeauthoryear{{Dodelson} \& {Schneider}}{{Dodelson} \&
  {Schneider}}{2013}]{DodelsonSchneider2013}
{Dodelson} S.,  {Schneider} M.~D.,  2013, \mn@doi [\prd]
  {10.1103/PhysRevD.88.063537}, \href
  {http://adsabs.harvard.edu/abs/2013PhRvD..88f3537D} {88, 063537}

\bibitem[\protect\citeauthoryear{Drlica-Wagner et~al.}{Drlica-Wagner
  et~al.}{2017}]{Drlica-Wagner:2017tkk}
Drlica-Wagner A.,  et~al., 2017, {Dark Energy Survey Year 1 Results:
  Photometric Data Set for Cosmology} (\mn@eprint {arXiv} {1708.01531})

\bibitem[\protect\citeauthoryear{{Eisenstein} \& {Hu}}{{Eisenstein} \&
  {Hu}}{1998}]{EisensteinHu1998}
{Eisenstein} D.~J.,  {Hu} W.,  1998, \mn@doi [ApJ] {10.1086/305424}, \href
  {http://adsabs.harvard.edu/abs/1998ApJ...496..605E} {496, 605}

\bibitem[\protect\citeauthoryear{{Eisenstein} et~al.,}{{Eisenstein}
  et~al.}{2005}]{Eisenstein_etal2005}
{Eisenstein} D.~J.,  et~al., 2005, \mn@doi [ApJ] {10.1086/466512}, \href
  {http://adsabs.harvard.edu/abs/2005ApJ...633..560E} {633, 560}

\bibitem[\protect\citeauthoryear{{Eisenstein}, {Seo}  \& {White}}{{Eisenstein}
  et~al.}{2007}]{EisensteinSeoWhite2007}
{Eisenstein} D.~J.,  {Seo} H.-J.,   {White} M.,  2007, \mn@doi [ApJ]
  {10.1086/518755}, \href {http://adsabs.harvard.edu/abs/2007ApJ...664..660E}
  {664, 660}

\bibitem[\protect\citeauthoryear{{Estrada}, {Sefusatti}  \&
  {Frieman}}{{Estrada} et~al.}{2009}]{EstradaSefusattiFrieman2009}
{Estrada} J.,  {Sefusatti} E.,   {Frieman} J.~A.,  2009, \mn@doi [ApJ]
  {10.1088/0004-637X/692/1/265}, \href
  {http://adsabs.harvard.edu/abs/2009ApJ...692..265E} {692, 265}

\bibitem[\protect\citeauthoryear{{Foreman-Mackey}, {Hogg}, {Lang}  \&
  {Goodman}}{{Foreman-Mackey} et~al.}{2013}]{Emcee_Foreman-Mackey2013}
{Foreman-Mackey} D.,  {Hogg} D.~W.,  {Lang} D.,   {Goodman} J.,  2013, \mn@doi
  [Publications of the Astronomical Society of the Pacific] {10.1086/670067},
  \href {http://adsabs.harvard.edu/abs/2013PASP..125..306F} {125, 306}

\bibitem[\protect\citeauthoryear{{Fosalba}, {Crocce}, {Gazta{\~n}aga}  \&
  {Castander}}{{Fosalba} et~al.}{2015}]{Fosalba_etal2015}
{Fosalba} P.,  {Crocce} M.,  {Gazta{\~n}aga} E.,   {Castander} F.~J.,  2015,
  \mn@doi [\mnras] {10.1093/mnras/stv138}, \href
  {http://adsabs.harvard.edu/abs/2015MNRAS.448.2987F} {448, 2987}

\bibitem[\protect\citeauthoryear{{Friedrich} \& {Eifler}}{{Friedrich} \&
  {Eifler}}{2017}]{FriedrichEifler2017}
{Friedrich} O.,  {Eifler} T.,  2017, preprint, \href
  {http://adsabs.harvard.edu/abs/2017arXiv170307786F} {} (\mn@eprint {arXiv}
  {1703.07786})

\bibitem[\protect\citeauthoryear{Gazta\~naga et~al.}{Gazta\~naga
  et~al.}{2018}]{Photoz}
Gazta\~naga E.,  et~al., 2018, preprint (\mn@eprint {arXiv} {---})

\bibitem[\protect\citeauthoryear{Gaztanaga, Cabre  \& Hui}{Gaztanaga
  et~al.}{2009}]{Gaztanaga:2008xz}
Gaztanaga E.,  Cabre A.,   Hui L.,  2009, \mn@doi [Mon. Not. Roy. Astron. Soc.]
  {10.1111/j.1365-2966.2009.15405.x}, 399, 1663

\bibitem[\protect\citeauthoryear{Hahn, Beutler, Sinha, Berlind, Ho  \&
  Hogg}{Hahn et~al.}{2018}]{Hahn:2018zja}
Hahn C.,  Beutler F.,  Sinha M.,  Berlind A.,  Ho S.,   Hogg D.~W.,  2018

\bibitem[\protect\citeauthoryear{Hamilton}{Hamilton}{1992}]{Hamilton1992}
Hamilton A. J.~S.,  1992, ApJ, 385, L5

\bibitem[\protect\citeauthoryear{Hamilton, Rimes  \& Scoccimarro}{Hamilton
  et~al.}{2006}]{Hamilton:2005dx}
Hamilton A. J.~S.,  Rimes C.~D.,   Scoccimarro R.,  2006, \mn@doi [Mon. Not.
  Roy. Astron. Soc.] {10.1111/j.1365-2966.2006.10709.x}, 371, 1188

\bibitem[\protect\citeauthoryear{{Harnois-D{\'e}raps} \&
  {Pen}}{{Harnois-D{\'e}raps} \& {Pen}}{2012}]{Harnois-DerapsPen_2012}
{Harnois-D{\'e}raps} J.,  {Pen} U.-L.,  2012, \mn@doi [MNRAS]
  {10.1111/j.1365-2966.2012.21039.x}, \href
  {http://adsabs.harvard.edu/abs/2012MNRAS.423.2288H} {423, 2288}

\bibitem[\protect\citeauthoryear{Hartlap, Simon  \& Schneider}{Hartlap
  et~al.}{2007}]{Hartlap:2006kj}
Hartlap J.,  Simon P.,   Schneider P.,  2007, Astron. Astrophys., 464, 399

\bibitem[\protect\citeauthoryear{{Heavens}}{{Heavens}}{2009}]{Heavens2009}
{Heavens} A.,  2009, preprint, \href
  {http://adsabs.harvard.edu/abs/2009arXiv0906.0664H} {} (\mn@eprint {arXiv}
  {0906.0664})

\bibitem[\protect\citeauthoryear{{Hinshaw}, {Larson}, {Komatsu}, {Spergel},
  {Bennett}  et~al.}{{Hinshaw} et~al.}{2013}]{WMAP9Yr2013}
{Hinshaw} G.,  {Larson} D.,  {Komatsu} E.,  {Spergel} D.~N.,  {Bennett} C.~L.,
   et~al., 2013, \mn@doi [ApJS] {10.1088/0067-0049/208/2/19}, \href
  {http://adsabs.harvard.edu/abs/2013ApJS..208...19H} {208, 19}

\bibitem[\protect\citeauthoryear{{Hogg}, {Bovy}  \& {Lang}}{{Hogg}
  et~al.}{2010}]{HoggBovyLang2010}
{Hogg} D.~W.,  {Bovy} J.,   {Lang} D.,  2010, {Data analysis recipes: Fitting a
  model to data} (\mn@eprint {arXiv} {1008.4686})

\bibitem[\protect\citeauthoryear{{Hu} \& {Sugiyama}}{{Hu} \&
  {Sugiyama}}{1996}]{HuSugiyama1996}
{Hu} W.,  {Sugiyama} N.,  1996, \mn@doi [ApJ] {10.1086/177989}, \href
  {http://adsabs.harvard.edu/abs/1996ApJ...471..542H} {471, 542}

\bibitem[\protect\citeauthoryear{{Hu}, {Sugiyama}  \& {Silk}}{{Hu}
  et~al.}{1997}]{HuSugiyamaSilk1997}
{Hu} W.,  {Sugiyama} N.,   {Silk} J.,  1997, \mn@doi [Nature]
  {10.1038/386037a0}, \href {http://adsabs.harvard.edu/abs/1997Natur.386...37H}
  {386, 37}

\bibitem[\protect\citeauthoryear{{H{\"u}tsi}}{{H{\"u}tsi}}{2010}]{Hutsi2010}
{H{\"u}tsi} G.,  2010, \mn@doi [\mnras] {10.1111/j.1365-2966.2009.15824.x},
  \href {http://adsabs.harvard.edu/abs/2010MNRAS.401.2477H} {401, 2477}

\bibitem[\protect\citeauthoryear{Kaiser}{Kaiser}{1987}]{Kaiser87}
Kaiser N.,  1987, MNRAS, 227, 1

\bibitem[\protect\citeauthoryear{{Kazin} et~al.,}{{Kazin}
  et~al.}{2014}]{Kazin_etal2014}
{Kazin} E.~A.,  et~al., 2014, \mn@doi [MNRAS] {10.1093/mnras/stu778}, \href
  {http://adsabs.harvard.edu/abs/2014MNRAS.441.3524K} {441, 3524}

\bibitem[\protect\citeauthoryear{{Koo}}{{Koo}}{1985}]{Koo1985}
{Koo} D.~C.,  1985, \mn@doi [Astronomical Journal] {10.1086/113748}, \href
  {http://adsabs.harvard.edu/abs/1985AJ.....90..418K} {90, 418}

\bibitem[\protect\citeauthoryear{Krause et~al.}{Krause
  et~al.}{2017}]{Krause:2017ekm}
Krause E.,  et~al., 2017, Submitted to: Phys. Rev. D

\bibitem[\protect\citeauthoryear{{LSST Science Collaboration} et~al.,}{{LSST
  Science Collaboration} et~al.}{2009}]{LSSTScienceBook}
{LSST Science Collaboration} et~al., 2009, preprint, \href
  {http://adsabs.harvard.edu/abs/2009arXiv0912.0201L} {} (\mn@eprint {arXiv}
  {0912.0201})

\bibitem[\protect\citeauthoryear{{Lampton}, {Margon}  \& {Bowyer}}{{Lampton}
  et~al.}{1976}]{Lampton_etal1976}
{Lampton} M.,  {Margon} B.,   {Bowyer} S.,  1976, \mn@doi [\apj]
  {10.1086/154592}, \href {http://adsabs.harvard.edu/abs/1976ApJ...208..177L}
  {208, 177}

\bibitem[\protect\citeauthoryear{{Landy} \& {Szalay}}{{Landy} \&
  {Szalay}}{1993}]{LandySzalay_1993}
{Landy} S.~D.,  {Szalay} A.~S.,  1993, \mn@doi [\apj] {10.1086/172900}, \href
  {http://adsabs.harvard.edu/abs/1993ApJ...412...64L} {412, 64}

\bibitem[\protect\citeauthoryear{Lewis, Challinor  \& Lasenby}{Lewis
  et~al.}{2000}]{CAMB}
Lewis A.,  Challinor A.,   Lasenby A.,  2000, ApJ., 538, 473

\bibitem[\protect\citeauthoryear{Li, Hu  \& Takada}{Li
  et~al.}{2014}]{Li:2014sga}
Li Y.,  Hu W.,   Takada M.,  2014, \mn@doi [Phys. Rev.]
  {10.1103/PhysRevD.89.083519}, D89, 083519

\bibitem[\protect\citeauthoryear{{Linder}}{{Linder}}{2005}]{Linder2005}
{Linder} E.~V.,  2005, \mn@doi [Phys. Rev. D] {10.1103/PhysRevD.72.043529},
  \href {http://adsabs.harvard.edu/abs/2005PhRvD..72d3529L} {72, 043529}

\bibitem[\protect\citeauthoryear{{Meiksin}, {White}  \& {Peacock}}{{Meiksin}
  et~al.}{1999}]{MeiksinWhitePeacock1999}
{Meiksin} A.,  {White} M.,   {Peacock} J.~A.,  1999, \mn@doi [MNRAS]
  {10.1046/j.1365-8711.1999.02369.x}, \href
  {http://adsabs.harvard.edu/abs/1999MNRAS.304..851M} {304, 851}

\bibitem[\protect\citeauthoryear{{Mohammed}, {Seljak}  \& {Vlah}}{{Mohammed}
  et~al.}{2017}]{MohammedSeljakVlah}
{Mohammed} I.,  {Seljak} U.,   {Vlah} Z.,  2017, \mn@doi [MNRAS]
  {10.1093/mnras/stw3196}, \href
  {http://adsabs.harvard.edu/abs/2017MNRAS.466..780M} {466, 780}

\bibitem[\protect\citeauthoryear{{Padmanabhan} \& {White}}{{Padmanabhan} \&
  {White}}{2009}]{PadmanabhanWhite_2009}
{Padmanabhan} N.,  {White} M.,  2009, \mn@doi [\prd]
  {10.1103/PhysRevD.80.063508}, \href
  {http://adsabs.harvard.edu/abs/2009PhRvD..80f3508P} {80, 063508}

\bibitem[\protect\citeauthoryear{{Padmanabhan} et~al.,}{{Padmanabhan}
  et~al.}{2007}]{Padmanabhan_etal2007}
{Padmanabhan} N.,  et~al., 2007, \mn@doi [MNRAS]
  {10.1111/j.1365-2966.2007.11593.x}, \href
  {http://adsabs.harvard.edu/abs/2007MNRAS.378..852P} {378, 852}

\bibitem[\protect\citeauthoryear{{Paz} \& {S{\'a}nchez}}{{Paz} \&
  {S{\'a}nchez}}{2015}]{PazSanchez_2015}
{Paz} D.~J.,  {S{\'a}nchez} A.~G.,  2015, \mn@doi [\mnras]
  {10.1093/mnras/stv2259}, \href
  {http://adsabs.harvard.edu/abs/2015MNRAS.454.4326P} {454, 4326}

\bibitem[\protect\citeauthoryear{{Pearson} \& {Samushia}}{{Pearson} \&
  {Samushia}}{2016}]{PearsonSamushia_2016}
{Pearson} D.~W.,  {Samushia} L.,  2016, \mn@doi [MNRAS] {10.1093/mnras/stw062},
  \href {http://adsabs.harvard.edu/abs/2016MNRAS.457..993P} {457, 993}

\bibitem[\protect\citeauthoryear{Peebles}{Peebles}{1980}]{Peebles}
Peebles P.~J.~E.,  1980, The Large-Scale Structure of the Universe.
Princeton University Press, New Jersey

\bibitem[\protect\citeauthoryear{{Peebles} \& {Groth}}{{Peebles} \&
  {Groth}}{1975}]{PeeblesGroth1975}
{Peebles} P.~J.~E.,  {Groth} E.~J.,  1975, \mn@doi [ApJ] {10.1086/153390},
  \href {http://adsabs.harvard.edu/abs/1975ApJ...196....1P} {196, 1}

\bibitem[\protect\citeauthoryear{{Peebles} \& {Yu}}{{Peebles} \&
  {Yu}}{1970}]{PeeblesYu1970}
{Peebles} P.~J.~E.,  {Yu} J.~T.,  1970, \mn@doi [ApJ] {10.1086/150713}, \href
  {http://adsabs.harvard.edu/abs/1970ApJ...162..815P} {162, 815}

\bibitem[\protect\citeauthoryear{{Percival} et~al.,}{{Percival}
  et~al.}{2010}]{Percival_etal2010}
{Percival} W.~J.,  et~al., 2010, \mn@doi [MNRAS]
  {10.1111/j.1365-2966.2009.15812.x}, \href
  {http://adsabs.harvard.edu/abs/2010MNRAS.401.2148P} {401, 2148}

\bibitem[\protect\citeauthoryear{{Percival} et~al.,}{{Percival}
  et~al.}{2014}]{Percival_etal2014}
{Percival} W.~J.,  et~al., 2014, \mn@doi [\mnras] {10.1093/mnras/stu112}, \href
  {http://adsabs.harvard.edu/abs/2014MNRAS.439.2531P} {439, 2531}

\bibitem[\protect\citeauthoryear{Pope \& Szapudi}{Pope \&
  Szapudi}{2008}]{Pope:2007vz}
Pope A.~C.,  Szapudi I.,  2008, \mn@doi [Mon. Not. Roy. Astron. Soc.]
  {10.1111/j.1365-2966.2008.13561.x}, 389, 766

\bibitem[\protect\citeauthoryear{Press, Teukolsky, Vetterling  \&
  Flannery}{Press et~al.}{2007}]{Press:2007:NRE:1403886}
Press W.~H.,  Teukolsky S.~A.,  Vetterling W.~T.,   Flannery B.~P.,  2007,
  Numerical Recipes 3rd Edition: The Art of Scientific Computing, 3 edn.
Cambridge University Press, New York, NY, USA

\bibitem[\protect\citeauthoryear{{Ross}, {Percival}, {Crocce}, {Cabr{\'e}}  \&
  {Gazta{\~n}aga}}{{Ross} et~al.}{2011}]{2011MNRAS.415.2193R}
{Ross} A.~J.,  {Percival} W.~J.,  {Crocce} M.,  {Cabr{\'e}} A.,
  {Gazta{\~n}aga} E.,  2011, \mn@doi [MNRAS]
  {10.1111/j.1365-2966.2011.18843.x}, \href
  {http://adsabs.harvard.edu/abs/2011MNRAS.415.2193R} {415, 2193}

\bibitem[\protect\citeauthoryear{{Ross}, {Samushia}, {Howlett}, {Percival},
  {Burden}  \& {Manera}}{{Ross} et~al.}{2015}]{Ross_etal2015}
{Ross} A.~J.,  {Samushia} L.,  {Howlett} C.,  {Percival} W.~J.,  {Burden} A.,
  {Manera} M.,  2015, \mn@doi [MNRAS] {10.1093/mnras/stv154}, \href
  {http://adsabs.harvard.edu/abs/2015MNRAS.449..835R} {449, 835}

\bibitem[\protect\citeauthoryear{Ross et~al.}{Ross
  et~al.}{2017a}]{Ross:2016gvb}
Ross A.~J.,  et~al., 2017a, \mn@doi [Mon. Not. Roy. Astron. Soc.]
  {10.1093/mnras/stw2372}, 464, 1168

\bibitem[\protect\citeauthoryear{Ross et~al.}{Ross
  et~al.}{2017b}]{Ross:2017emc}
Ross A.~J.,  et~al., 2017b, \mn@doi [Mon. Not. Roy. Astron. Soc.]
  {10.1093/mnras/stx2120}, 472, 4456

\bibitem[\protect\citeauthoryear{{Salazar-Albornoz} et~al.,}{{Salazar-Albornoz}
  et~al.}{2017}]{Salazar-Albornoz:2016psd}
{Salazar-Albornoz} S.,  et~al., 2017, \mn@doi [MNRAS] {10.1093/mnras/stx633},
  \href {http://adsabs.harvard.edu/abs/2017MNRAS.468.2938S} {468, 2938}

\bibitem[\protect\citeauthoryear{Sanchez, Baugh  \& Angulo}{Sanchez
  et~al.}{2008}]{Sanchez:2008iw}
Sanchez A.~G.,  Baugh C.~M.,   Angulo R.,  2008, \mn@doi [Mon. Not. Roy.
  Astron. Soc.] {10.1111/j.1365-2966.2008.13769.x}, 390, 1470

\bibitem[\protect\citeauthoryear{{Scoccimarro}}{{Scoccimarro}}{2000}]{Scoccimarro2000}
{Scoccimarro} R.,  2000, \apj, 544, 597

\bibitem[\protect\citeauthoryear{{Sellentin} \& {Heavens}}{{Sellentin} \&
  {Heavens}}{2016}]{SellentinHeavens2016}
{Sellentin} E.,  {Heavens} A.~F.,  2016, \mn@doi [\mnras]
  {10.1093/mnrasl/slv190}, \href
  {http://adsabs.harvard.edu/abs/2016MNRAS.456L.132S} {456, L132}

\bibitem[\protect\citeauthoryear{{Seo} \& {Eisenstein}}{{Seo} \&
  {Eisenstein}}{2007}]{SeoEisenstein_2007}
{Seo} H.-J.,  {Eisenstein} D.~J.,  2007, \mn@doi [\apj] {10.1086/519549}, \href
  {http://adsabs.harvard.edu/abs/2007ApJ...665...14S} {665, 14}

\bibitem[\protect\citeauthoryear{{Seo} et~al.,}{{Seo}
  et~al.}{2012}]{Seo_etal2012}
{Seo} H.-J.,  et~al., 2012, \mn@doi [\apj] {10.1088/0004-637X/761/1/13}, \href
  {http://adsabs.harvard.edu/abs/2012ApJ...761...13S} {761, 13}

\bibitem[\protect\citeauthoryear{Smith, Scoccimarro  \& Sheth}{Smith
  et~al.}{2008}]{Smith:2007gi}
Smith R.~E.,  Scoccimarro R.,   Sheth R.~K.,  2008, \mn@doi [Phys. Rev.]
  {10.1103/PhysRevD.77.043525}, D77, 043525

\bibitem[\protect\citeauthoryear{{Sunyaev} \& {Zeldovich}}{{Sunyaev} \&
  {Zeldovich}}{1970}]{SunyaevZeldovich1970}
{Sunyaev} R.~A.,  {Zeldovich} Y.~B.,  1970, \mn@doi [Astrophysics and Space
  Science] {10.1007/BF00653471}, \href
  {http://adsabs.harvard.edu/abs/1970Ap%26SS...7....3S} {7, 3}

\bibitem[\protect\citeauthoryear{Takada \& Hu}{Takada \&
  Hu}{2013}]{Takada:2013bfn}
Takada M.,  Hu W.,  2013, \mn@doi [Phys. Rev.] {10.1103/PhysRevD.87.123504},
  D87, 123504

\bibitem[\protect\citeauthoryear{Taylor \& Joachimi}{Taylor \&
  Joachimi}{2014}]{Taylor:2014ota}
Taylor A.,  Joachimi B.,  2014, \mn@doi [Mon. Not. Roy. Astron. Soc.]
  {10.1093/mnras/stu996}, 442, 2728

\bibitem[\protect\citeauthoryear{Tegmark, Taylor  \& Heavens}{Tegmark
  et~al.}{1997}]{Tegmark:1996bz}
Tegmark M.,  Taylor A.,   Heavens A.,  1997, \mn@doi [Astrophys. J.]
  {10.1086/303939}, 480, 22

\bibitem[\protect\citeauthoryear{Trotta}{Trotta}{2017}]{Trotta:2017wnx}
Trotta R.,  2017, {Bayesian Methods in Cosmology} (\mn@eprint {arXiv}
  {1701.01467})

\bibitem[\protect\citeauthoryear{{Weinberg}, {Mortonson}, {Eisenstein},
  {Hirata}, {Riess}  \& {Rozo}}{{Weinberg}
  et~al.}{2013}]{WeinbergMortonson_etal2013}
{Weinberg} D.~H.,  {Mortonson} M.~J.,  {Eisenstein} D.~J.,  {Hirata} C.,
  {Riess} A.~G.,   {Rozo} E.,  2013, \mn@doi [\physrep]
  {10.1016/j.physrep.2013.05.001}, \href
  {http://adsabs.harvard.edu/abs/2013PhR...530...87W} {530, 87}

\bibitem[\protect\citeauthoryear{Zel'dovich}{Zel'dovich}{1970}]{Zeldovich1970}
Zel'dovich Y.~B.,  1970, A\&A, 5, 84

\bibitem[\protect\citeauthoryear{{de Simoni} et~al.,}{{de Simoni}
  et~al.}{2013}]{deSimoni_etal2013}
{de Simoni} F.,  et~al., 2013, \mn@doi [\mnras] {10.1093/mnras/stt1496}, \href
  {http://adsabs.harvard.edu/abs/2013MNRAS.435.3017D} {435, 3017}

\makeatother
\end{thebibliography}

\appendix

\section{ Comparison of the estimators (almost) without pruning  } 
\label{sec:appendix}

\begin{table*}
  \caption{The BAO fit using  MLE, PL, and MCMC.  Only  those mocks whose 1-$\sigma$ interval $\bar{\alpha }  \pm \sigma_\alpha  $ fall within the interval [0.6,1.8] are considered.  }
\label{tab:data_almost_no_pruning}
  \begin{tabular}{|l|c|c|c|c|}
    \hline
    &  ${\bar \alpha} \pm \sigma_\alpha$ in [0.6,1.4]  &  $\langle {\bar \alpha} \rangle \pm {\rm std}({\bar \alpha})$  & frac. with ${\bar \alpha} \pm \sigma_\alpha$ &       $ \langle \sigma_\alpha  \rangle / \mathrm{std}( \bar{\alpha}  )    $  \\
    &  (frac. selected)  &    & enclosing $\langle \bar{\alpha} \rangle $ &        \\
    \hline
 MLE           & 0.96      &  $ 0.971 \pm 0.074 $ &  0.65    &  0.75  \\
PL             & 0.98      &  $0.975 \pm 0.069 $  &  0.83    &  1.34  \\
MCMC           & 1.00      &  $0.990 \pm 0.078 $  &  0.71    &  0.92  \\
    \hline
  \end{tabular}
\end{table*}

\change{ In the main text, we presented the results with the pruning criterion that the best fit $\bar{\alpha}$  with its 1-$\sigma$ error bar must fall within the interval $[0.8,1.2]$. Although the interval [0.8,1.2] seems reasonable, this choice is to some degree arbitrary. In this appendix, we show the results when a much more generous interval [0.6,1.4] is used. I.e.~for MLE and MCMC, we require that the best fit with its 1-$\sigma$ interval falls within [0.6,1.4]. For PL, it is computed in this interval. 
}

\change{ In Table \ref{tab:data_almost_no_pruning}, we show the results  obtained with these three estimators. The results should be compared to Table \ref{tab:data_pruning}. For MLE and MCMC, the fraction of mocks satisfying this criterion has increased relative to the cases shown in  Table \ref{tab:data_pruning}, and it is almost 100\% for MCMC.  For PL it does not change significantly.    The extreme mocks  not only increase  $ \mathrm{std}( \bar{\alpha} )$, they also result in   appreciable bias in the mean of the MLE and PL estimators. However, the bias of the MCMC estimator is more mild, and  $| \langle \bar{\alpha} \rangle - 1 | = 0.001 $ only.   The fractions of time that the 1-$\sigma$ interval enclosing $\langle \bar{\alpha}  \rangle $  are only 3\% from the Gaussian expectation for both MLE and MCMC.   Furthermore,  $ \langle \sigma_\alpha  \rangle / \mathrm{std}( \bar{\alpha}  ) = 0.92 $ for MCMC. From these statistics, we find  that when the pruning is weak, the MCMC is the most desirable estimator based on the criteria outlined in Sec.~\ref{sec:selection_criterion}.    }

\begin{figure}
\begin{center}
\includegraphics[width=\linewidth]{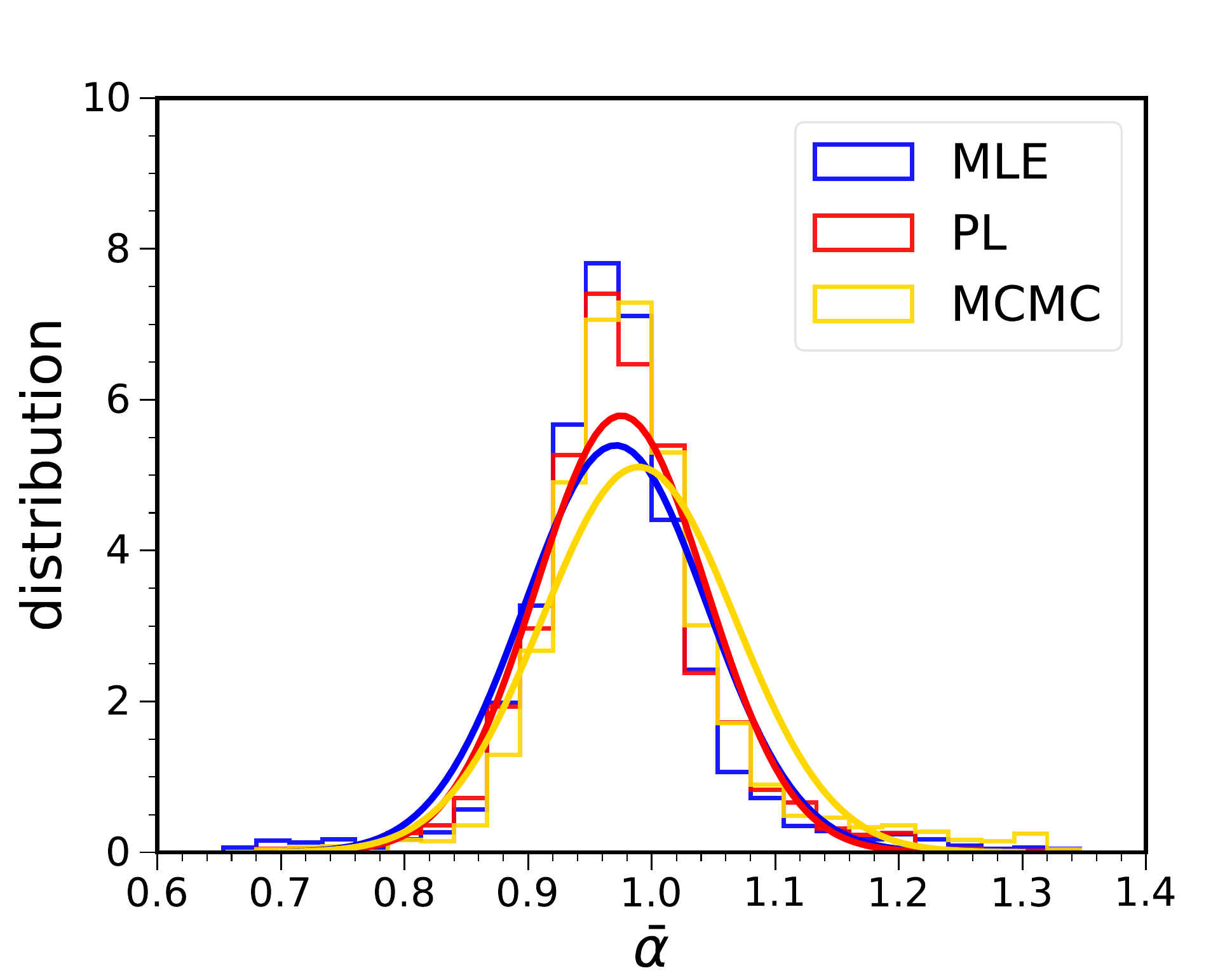}
\caption{ Similar to Fig.~\ref{fig:alpha_distribution_NAP4}, except the interval of $\alpha$ is limited to [0.6,1.4] instead of  [0.8,1.2]. }
\label{fig:alpha_distribution_NAP4_wide_alpha}
\end{center}
\end{figure}

\begin{figure}
\begin{center}
\includegraphics[width=\linewidth]{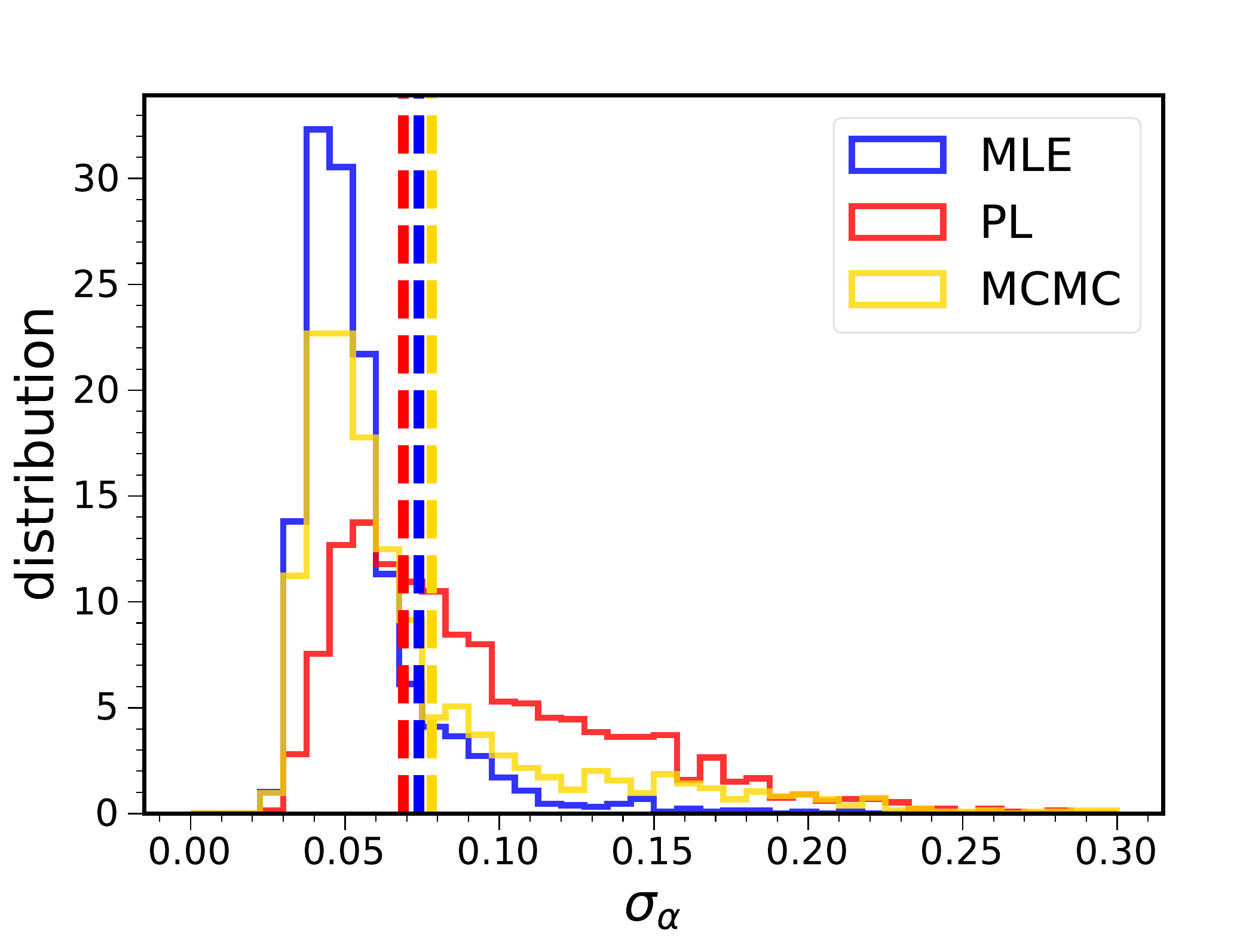}
\caption{   Similar to Fig.~\ref{fig:error_distribution_NAP4}, except the interval of $\alpha$ is limited to [0.6,1.4] instead of  [0.8,1.2].  }
\label{fig:error_distribution_NAP4_wide_alpha}
\end{center}
\end{figure}

\begin{figure}
\begin{center}
\includegraphics[width=\linewidth]{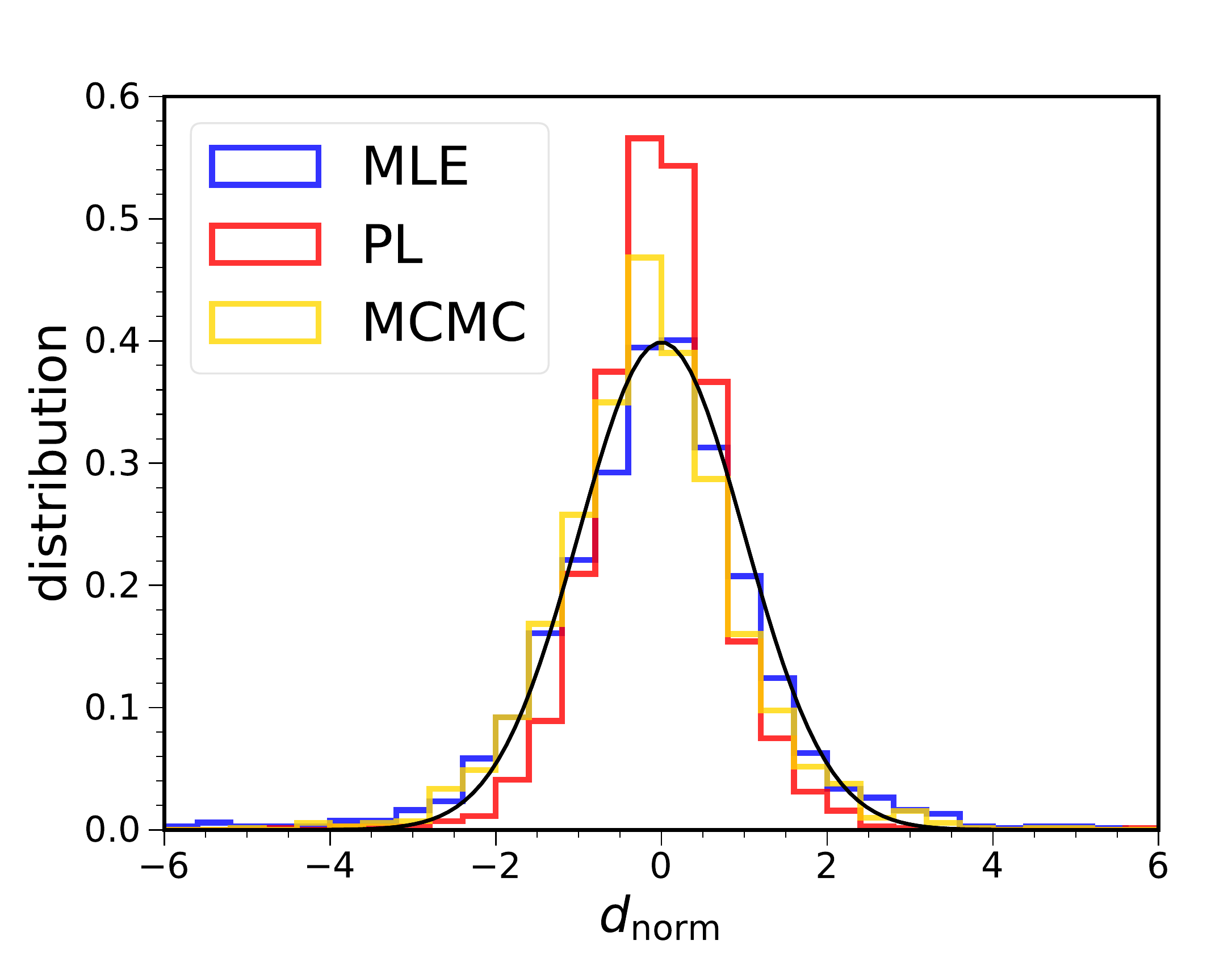}
\caption{  Similar to Fig.~\ref{fig:dnorm_fiducial}, except the interval of $\alpha$ is limited to [0.6,1.4] instead of  [0.8,1.2].    }
\label{fig:dnorm_fiducial_wide_alpha}
\end{center}
\end{figure}

\change{  In Fig.~\ref{fig:alpha_distribution_NAP4_wide_alpha}, we plot the distributions of the best fit $ \bar{\alpha} $. The Gaussian distributions with the same mean and variance as the histograms are also shown.  Because of the strong tail in the histogram, the variance is not a good proxy for its width, a Gaussian distribution with the same variance as the histogram does not agree with the histogram.  From the Gaussian distribution, we also see that only the MCMC one is close to be unbiased. In contrast the distributions in Fig.~\ref{fig:alpha_distribution_NAP4} are much closer to the Gaussian distributions.   The distributions of the error are shown in Fig.~\ref{fig:error_distribution_NAP4_wide_alpha}. In comparison with Fig.~\ref{fig:error_distribution_NAP4_wide_alpha}, the derived errors have much stronger tails when the interval [0.6,1.4] is adopted.  In particular, PL and MCMC show appreciable increase in the tail of the distribution. We plot the distributions  of $d_{\rm norm } $ in Fig.~\ref{fig:dnorm_fiducial_wide_alpha}. Besides thicker tails than those in Fig.~\ref{fig:dnorm_fiducial}, we also note that the distribution for MCMC is skewed aside and the PL case shows large deviation from the unit Gaussian distribution.  }

\change{ When the allowed interval is enlarged, the mocks with $\bar{\alpha} $  far from $\alpha = 1 $  are included. They are often  badly fitted by our model and  cause strong tails in the distribution of the best fit parameters. We show that these strong tails cause significant deviation from the Gaussian distribution. They not only enlarge the derived error bars but also cause bias in the mean although it is relatively mild for the MCMC. Those extreme mocks are poorly handled by our existing methodology, and little BAO information can be extracted from them.  Therefore we choose a smaller interval in the main text to get rid of those cases.      }

\bsp	
\label{lastpage}
\end{document}